\documentclass[aps,twocolumn,prl,amsmath,amssymb,superscriptaddress,floats]{revtex4}

\usepackage{amsmath}	%allows the use of mathematical symbols
\usepackage{gensymb}
\usepackage{hyperref}
\hypersetup{colorlinks=true}
\usepackage{upgreek}
\usepackage{graphics}
\usepackage{hyperref}
\usepackage{epsfig}
\usepackage{color}
\usepackage{bm}
\usepackage{float}
\usepackage{ulem} % \sout
\usepackage{amssymb,mathrsfs,amsmath}

\usepackage{url}
\usepackage{multirow}

\begin{document}
%\begin{CJK*}{GBK}{kai}

\title{Triple-meron crystal in high-spin Kitaev magnets}

\author{Ken Chen}
\affiliation{School of Physical Science and Technology $\&$ Key Laboratory for Magnetism and Magnetic Materials of the MoE, Lanzhou University, Lanzhou 730000, China}
\affiliation{Lanzhou Center for Theoretical Physics and Key Laboratory of Theoretical Physics of Gansu Province, Lanzhou University, Lanzhou 730000, China.}
\author{Qiang Luo}
\affiliation{College of Science, Nanjing University of Aeronautics and Astronautis, Nanjing, 211106, China}
\affiliation{Department of Physics, University of Toronto, Toronto, Ontario M5S 1A7, Canada}
\author{Zongsheng Zhou}
\affiliation{School of Physical Science and Technology $\&$ Key Laboratory for Magnetism and Magnetic Materials of the MoE, Lanzhou University, Lanzhou 730000, China}
\affiliation{Lanzhou Center for Theoretical Physics and Key Laboratory of Theoretical Physics of Gansu Province, Lanzhou University, Lanzhou 730000, China.}
\author{Saisai He}
\affiliation{School of Physical Science and Technology $\&$ Key Laboratory for Magnetism and Magnetic Materials of the MoE, Lanzhou University, Lanzhou 730000, China}
\affiliation{Lanzhou Center for Theoretical Physics and Key Laboratory of Theoretical Physics of Gansu Province, Lanzhou University, Lanzhou 730000, China.}
\author{Bin Xi}
\email[]{xibin@yzu.edu.cn}
\affiliation{College of Physics Science and Technology, Yangzhou University, Yangzhou 225002, China}
\author{Chenglong Jia}
\affiliation{School of Physical Science and Technology $\&$ Key Laboratory for Magnetism and Magnetic Materials of the MoE, Lanzhou University, Lanzhou 730000, China}
\affiliation{Lanzhou Center for Theoretical Physics and Key Laboratory of Theoretical Physics of Gansu Province, Lanzhou University, Lanzhou 730000, China.}
\author{Hong-Gang Luo}
\affiliation{School of Physical Science and Technology $\&$ Key Laboratory for Magnetism and Magnetic Materials of the MoE, Lanzhou University, Lanzhou 730000, China}
\affiliation{Lanzhou Center for Theoretical Physics and Key Laboratory of Theoretical Physics of Gansu Province, Lanzhou University, Lanzhou 730000, China.}
\affiliation{Beijing Computational Science Research Center, Beijing 100084, China}
\author{Jize Zhao}
\email[]{zhaojz@lzu.edu.cn}
\affiliation{School of Physical Science and Technology $\&$ Key Laboratory for Magnetism and Magnetic Materials of the MoE, Lanzhou University, Lanzhou 730000, China}
\affiliation{Lanzhou Center for Theoretical Physics and Key Laboratory of Theoretical Physics of Gansu Province, Lanzhou University, Lanzhou 730000, China.}

\date{\today}

\begin{abstract}
Spin textures with nontrivial topology hold great promise in future spintronics applications since
they are robust against local deformations. The meron, as one of such spin textures,
is widely believed to appear in pairs due to its topological equivalence to a half skyrmion. Motivated by recent progresses in
high-spin Kitaev magnets, here we investigate numerically a classical Kitaev-$\Gamma$ model with a single-ion anisotropy.
An exotic spin texture including three merons is discovered. Such a state  
features a peculiar property with an odd number of merons in one magnetic unit cell and it can induce the topological Hall effect.
Therefore, these merons cannot be dissociated from skyrmions as reported in the literature and
a general mechanism for such a deconfinement phenomenon calls for further studies.
Our work demonstrates that high-spin Kitaev magnets can host robust unconventional spin textures and
thus they offer a versatile platform not only for exploring exotic states in spintronics
but also for understanding the deconfinement mechanism in the condensed-matter physics and the field theory.
\end{abstract}	

\pacs{}
%\keywords{}

\maketitle

\textit{Introduction}.\;--
Topological spin textures~(TSTs), which have attracted enormous attention in condensed matter physics,
can be described by the homotopy theory\cite{Braun2012} with a non-trivial mapping $\pi_n(S^m)$,
where $n$ and $m$ are the corresponding degrees of freedom in the real and the parameter spaces, respectively.
A topological charge $Q$ thus counts the number of times of the real space configuration covering the parameter space.
Among these TSTs, the skyrmions ($\pi_2(S^2)$), originally proposed in the particle physics\cite{Skyrme1962},
have become the focus of recent research in magnetism due to their potential applications 
in spintronics\cite{NagaosaTokura2013, FertRC2017, HellmanHTetal2017, Zhou2018, BogdanovPanagopoulos2020, GobelMT2021}.
Experimentally, they have already been successfully discovered in noncentrosymmetric
chiral magnets~\cite{MuhlbauerBJetal2009, YuOKetal2010, SekiYIT2012, NayakKMetal2017, FujishiroKNetal2019}, 
magnetic heterostructures~\cite{HeinzeBMetal2011, RommingHMetal2013, TokunagaYRetal2015, MoreauMRetal2016}
and centrosymmetric magnets~\cite{KurumajiNHetal2019, HirschbergerNGetal2019, KhanhNYetal2020}.
In a skyrmion, $Q$ is an integer which is given by\cite{Braun2012}
\begin{eqnarray}
	Q &=& \frac{1}{4\pi}\int {\bf{S}}\cdot \left( \frac{\partial {\bf{S}}}{\partial x} \times \frac{\partial {\bf{S}}}{\partial y}\right) dxdy
\label{continue-Q}
\end{eqnarray}
where ${\bf{S}=\bf{S}}(x,y)$ is the unit spin vector at the position $(x,y)$. In addition, recent studies show that,
given a sufficiently large effective easy-plane anisotropy,
a skyrmion can be transferred into a meron-antimeron pair\cite{LinSB2015, YuKTetal2018, GaoJIetal2019, GaoRGetal2020}.
Such meron and antimeron carry one half skyrmion topological charge, and thus are treated as half skyrmions.
From the spin configurations, the most distinct difference between skyrmions and merons
is that at the perimeter the spins of merons lie within the plane while those of skyrmions point out of the plane.

\begin{figure}[htpb]
    \includegraphics[width=0.42\textwidth]{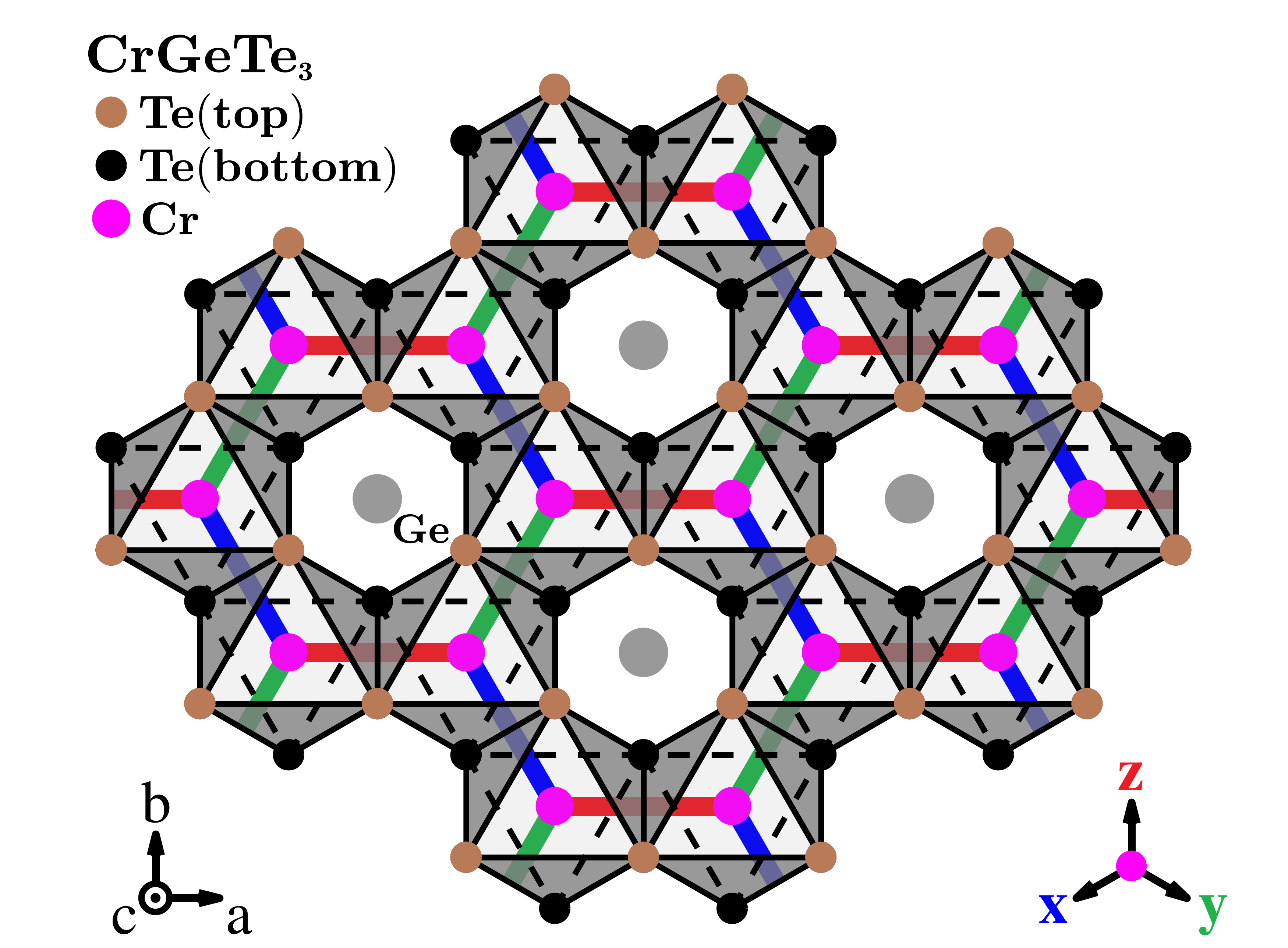}
\caption{Top view of the crystal structure for the monolayer $\rm CrGeTe_3$. The honeycomb network is formed by the
Cr atoms which are at the center of octahedrons. These Cr atoms carry effective spin 3/2.
The $\mathbf{abc}$ and the $\mathbf{xyz}$ coordinate systems are shown on the bottom left
and bottom right, respectively. The two-dimensional honeycomb lattice lies in the $\rm\bf{ab}$ plane
and $\rm \bf{c}$ is perpendicular to the plane. The $\mathbf{xyz}$ coordinate systems are along the three Cr-Te bonds.
In the $\rm \bf{abc}$ coordinate system, $\rm \mathbf{x,y}$ and $\rm \mathbf{z}$ are given by
$\mathbf{x} =\left[-\sqrt{2}/2,-\sqrt{6}/6,\sqrt{3}/3\right]$, $\mathbf{y} = \left[\sqrt{2}/2 , -\sqrt{6}/6 , \sqrt{3}/3\right]$, 
$\mathbf{z} = \left[0,\sqrt{6}/3,\sqrt{3}/3\right]$, respectively. The Kitaev interactions on the bonds in blue, 
green and red correspond to $x, y$ and $z$ Ising type, respectively.}	
\label{STRUCT}	
\end{figure}

\begin{figure*}[htpb]
    \begin{minipage}[t]{0.46\textwidth}	
	\includegraphics[width=\textwidth, scale=0.96]{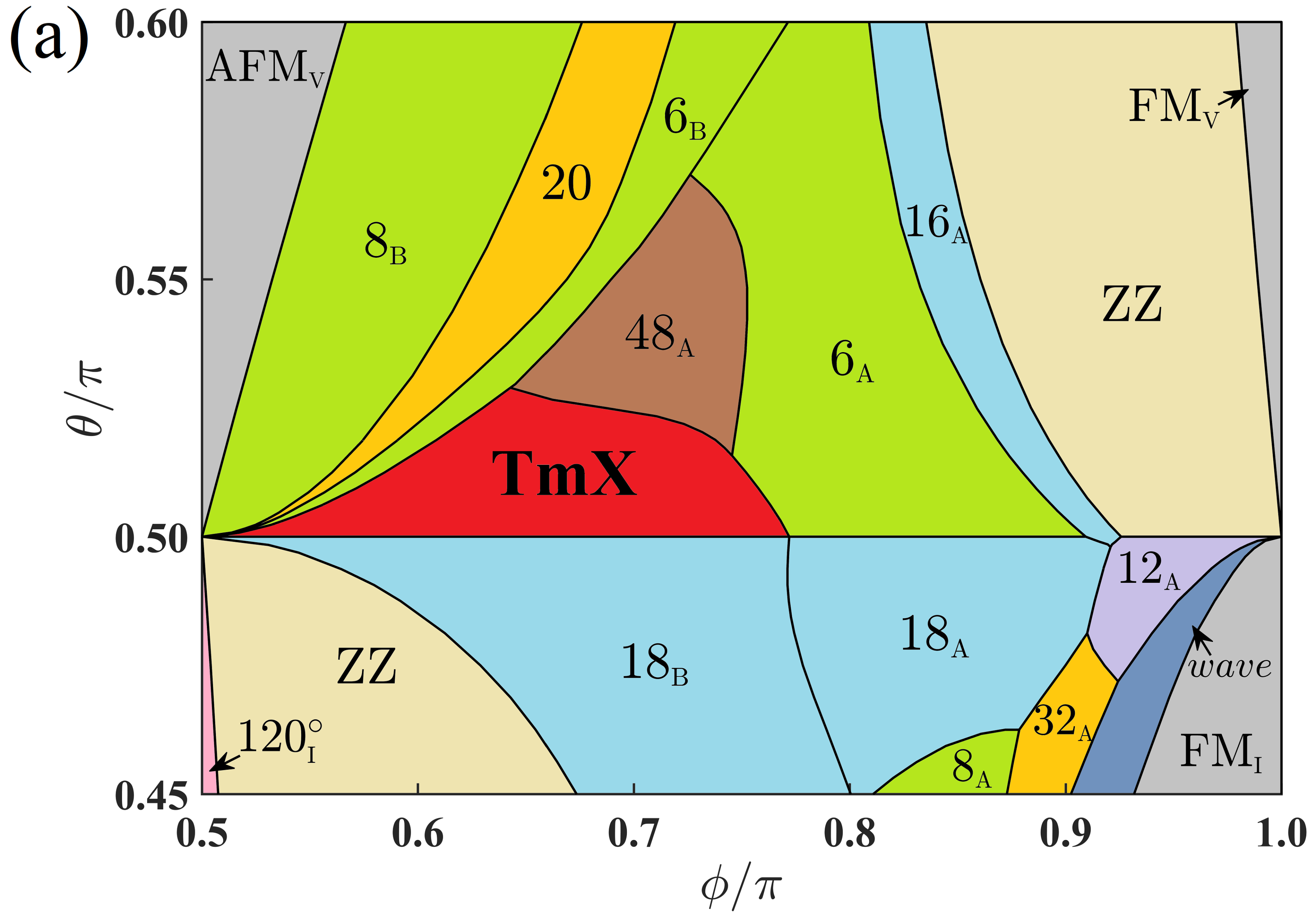}
     \end{minipage}
     \hspace{8mm}	
     \begin{minipage}[t]{0.42\textwidth}	
        \includegraphics[width=\textwidth, scale=0.96]{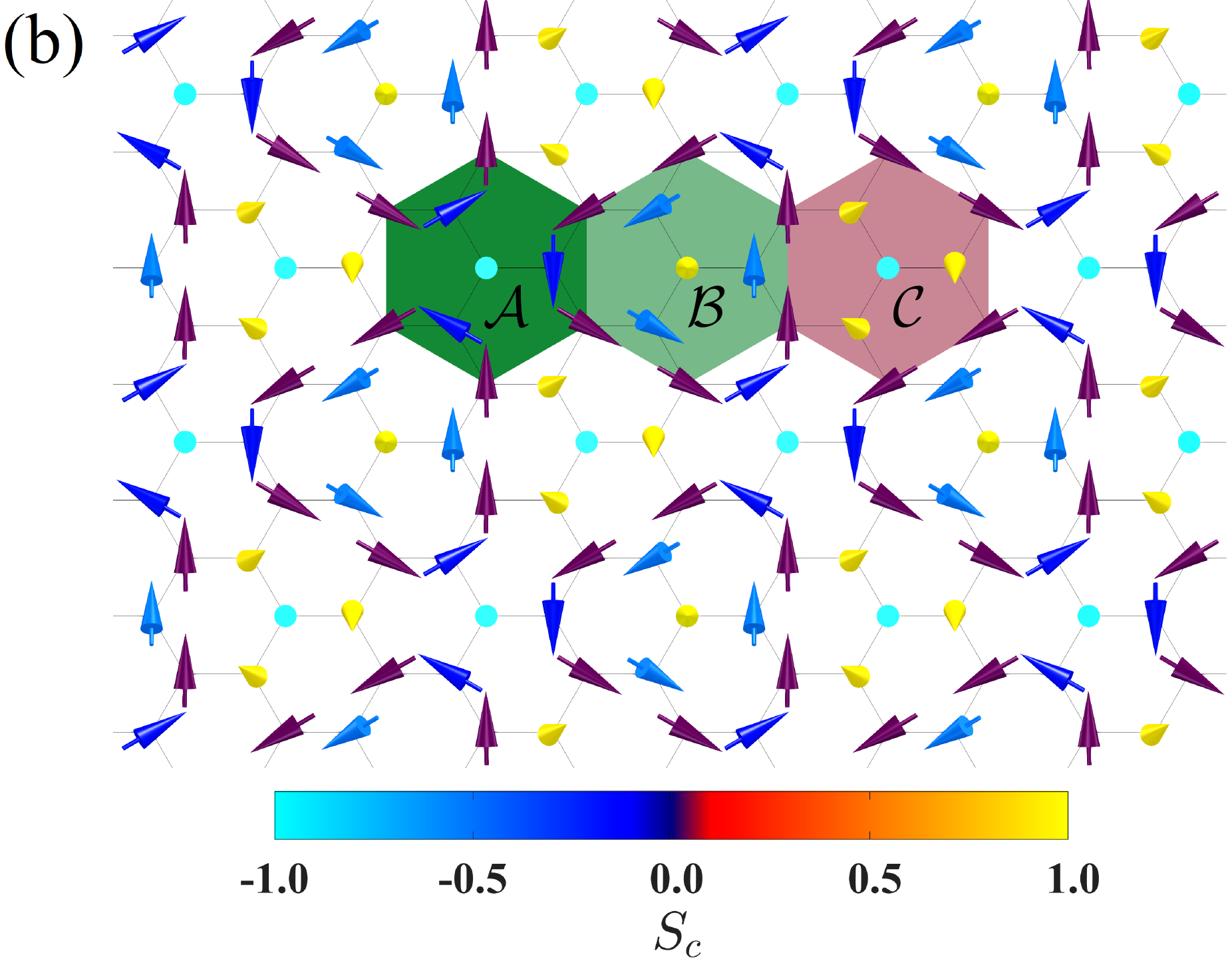}
     \end{minipage}
\caption{(a) The ground-state phase diagram of the classical $\rm K\Gamma A$ model. The parameters $K, 
\Gamma$ and $A$ are parameterized by $\left(\theta, \phi\right)$. Here we restrict $\phi\in[0.5\pi,1.0\pi], \theta\in[0.45\pi,0.60\pi]$. 
FM, AFM, ZZ, $\rm 120^\circ$ and $wave$ mean ferromagnetism, 
antiferromagnetism, zigzag, $\rm 120^\circ$ and wave-like magnetic orders, respectively.
The subscript $\mathbf{I}$ and $\mathbf{V}$ indicate the spins lie in the plane or point out of the plane, respectively.
The numbers 6, 8, 12, 16, 18, 20, 32 and 48 mark the number of spins in one MUC and the subscripts $\rm A$ and $\rm B$
are used to distinct those phases marked by the same number~\cite{Supp}. 
The phase marked by TmX is the key findings of this work and details are discussed in the main text.
(b) The typical ground-state spin configuration in the TmX phase where $\theta=0.515\pi, \phi=0.68\pi$. 
The three (anti)merons are marked by $\mathcal{A}$, $\mathcal{B}$ and $\mathcal{C}$, respectively. 
The topological charges for (anti)merons $\mathcal{A}$, $\mathcal{B}$ and $\mathcal{C}$ are 1.055231, 
0.472850, and -0.528081. Their summation gives 1.0.}
\label{phaseDiagram}
\end{figure*}

On the other hand, ever since the discovery of the exact solution of the Kitaev honeycomb model\cite{Kitaev2006},
great efforts have been made to synthesize materials to realize such a model. These materials, so called Kitaev magnets,
are two-dimensional transition-metal compounds with strong spin-orbit coupling~\cite{JackeliKhaliullin2009}.
The promising candidates include $\rm Na_2IrO_3$, $\rm Li_2IrO_3$ and
$\rm \alpha$-$\rm RuCl_3$ \cite{SinghMRetal2012, PlumbCSetal2014}.
In these compounds, honeycomb-located cations are surrounded by the edge-sharing octahedral anions.
Strong spin-orbit coupling entangles spin and orbital degrees of freedom together,
resulting in an effective spin-1/2 model with the Heisenberg ($J$) and Kitaev ($KS_i^\gamma S_j^\gamma$) interactions.
Further research suggests that a bond-dependent symmetric off-diagonal $\Gamma$ and/or $\Gamma'$
interactions should also be taken into account\cite{RauLK2014, KatukuriNYetal2014, WangDYetal2017}.
This $\rm{JK\Gamma\Gamma^\prime}$ model and its relevance have been proposed\cite{MaksimovChernyshev2020}
to explain the possible quantum spin liquids observed in experiments.

However, the Kitaev interaction is not limited to spin-1/2 compounds. Recent numerical and theoretical analysis\cite{XuFXetal2018, LeeUWetal2020, StavropoulosPLetal2021}
suggest that it may also exist in the van der Waals materials $\rm CrI_3$, $\rm CrGeTe_3$ and $\rm CrSiTe_3$.
Interestingly, the crystal structures of these compounds bear a strong resemblance to those of the
honeycomb iridates but Cr owns an effective $S = 3/2$ spin. Similar to its spin-1/2 counterparts these compounds can also
be described by the $\rm JK\Gamma\Gamma^\prime$ model but possibly with an additional
single-ion anisotropy ($A$).
Although the sought-after quantum spin liquids are not preferable in these high-spin Kitaev magnets\cite{ZhouCLetal2021},
stable TSTs originating from frustration~\cite{OkuboCK2012, LeonovMostovoy2015, HayamiOM2017, KharkovSM2017, WangSLetal2021} are highly desirable. 
This calls for extensive and immediate studies to search for the TSTs
in these high-spin Kitaev magnets. However, so far, most works focus
simply on the magnetic orders\cite{ChernKLetal2020, LiuSRetal2020, RayyanLK2021}
but relevant topological phenomena are rarely touched. We fill in this 
gap by investigating the classical $\rm K\Gamma{A}$ model numerically.
A triple-meron crystal~(TmX) with three (anti)merons in one magnetic unit cell~(MUC) is discovered,
suggesting the mechanism for generating (anti)merons is beyond our present knowledge.
Furthermore, we show that such a TmX can cause the topological Hall effect~(THE)\cite{Hamamoto2015} which 
is absent in the meron-antimeron crystals\cite{HayamiOM2021}.

\textit{Model}.\;--
The crystal structures\cite{HuangCNetal2017, GongLLetal2017, XingCOetal2017, XuFXetal2018} of the Cr-based compounds mentioned
above are sketched in Fig.~\ref{STRUCT}.
Due to their similarity, here we exemplify them by the $\rm {CrGeTe_3}$.
In $\rm {CrGeTe_3}$ six Te atoms are at the corners of an octahedron and the Cr atom sits at the center.
These edge-shared octahedrons form honeycomb layers stacking along the $\rm \textbf{c}$ axis, which are
stabilized by a weak interlayer van der Waals coupling.
Due to the symmetry of the crystal, two coordinate systems\cite{XuFXetal2018}, $\rm \bf{abc}$ and $\rm \bf{xyz}$
are involved in our later discussion~(Fig.~\ref{STRUCT}).
In principle, $J, K, \Gamma, \Gamma^\prime$ and $A$ terms are all allowed by the lattice symmetry.
However, for a given compound, some terms may be zero.  For example, the density-functional theory calculation shows that
in $\rm CrGeTe_3$ the Heisenberg term $J$ becomes zero at some compressive strain\cite{XuFKetal2020}.
Here, for simplicity, we consider the $\rm K\Gamma A$ model, which is given by
\begin{equation}
	\mathcal{H} = \sum_{\langle i,j\rangle_\gamma} KS_i^{\gamma}S_j^\gamma+\Gamma\left(S_i^\alpha S_j^\beta+S_i^\beta S_j^\alpha\right)+ A\sum_i \left(\mathbf{S}_i\cdot\bf{c}\right)^2,
    \label{HAM}
\end{equation}
where $\langle{i,j}\rangle$ means that the summation runs over all nearest neighbors.
$S_i^\gamma = \mathbf{S}_i\cdot\vec{\gamma}(\vec{\gamma}=\bf{x,y,z})$, e.g., the $\gamma$-component of the spin vector at site $i$
associated with the $\gamma$ bonds on the honeycomb lattice and $\alpha, \beta$ are other two components.
This Hamiltonian can be parameterized as $K = \textrm{sin}\theta\cos\phi$, $\Gamma = \sin\theta\sin\phi$ and $A=\cos\theta$,
with $\mathbf{\theta}\in[0,\pi]$ and $\mathbf{\phi}\in[0,2\pi)$. Now the model~(\ref{HAM}) depends on only two parameters $\theta$ and $\phi$.
Furthermore, one can see that the transformation
$\mathbf{\phi}\rightarrow\mathbf{\phi+\pi}$ is equivalent to flipping all the spins in one sublattice.
Therefore, we only need to consider the parameters in the region $\phi \in [0,\pi)$.

To get the ground-state phase diagram, the Hamiltonian~(\ref{HAM}) is studied by the parallel-tempering Monte Carlo
simulations\cite{HukushimaNemoto1996, MetropolisRRetal1953, MiyatakeYKetal1986, JanssenAV2016} in combination with other
numerical techniques~(Supplementary Material~(SM)\cite{Supp} Sec.~I). In our Monte Carlo simulations, the maximum size is $2\times36\times36$
and the temperature ranges from 0.005 to 0.2. The ground state is then obtained by optimizing the Monte Carlo data.
Various sizes are used to confirm we obtain the correct MUC~(SM\cite{Supp} Sec.~II).
The ground-state energy, the spin configurations and the static spin structure factor
are calculated to map out the phase diagram. In the following, all the figures are plotted in the $\rm \textbf{abc}$ coordinate system.

\textit{Results}.\;--
In Fig.~\ref{phaseDiagram}(a), we present the ground-state phase diagram of 
the $\rm K\Gamma{A}$ model in the region $\theta\in[0.45\pi,0.60\pi]$, $\phi\in[0.5\pi,1.0\pi]$~(see SM\cite{Supp} Sec.~VI for the full phase diagram).
Of all these phases, the one in red marked by $\rm{TmX}$ is the most exciting discovery in our work.
In the following, we will focus on this phase and present the details.
For simplicity, we choose one representative point in the TmX phase, i.e., $\theta =0.515\pi, \phi=0.68\pi$ to
demonstrate its unusual properties. In Fig.~\ref{phaseDiagram}(b), we
plot the ground-state spin configuration at the representative point. One can see there are 18 spins in one MUC.
Moreover, these 18 spins are divided into three hexagons with their edges shared, which are marked by $\mathcal{A}$, $\mathcal{B}$ and $\mathcal{C}$, respectively.
After a simple inspection one finds the spin at the core of each hexagon points out of the plane,
while the edge spins lie almost in the plane.
This is the characteristic feature of (anti)merons, indicating that there are three (anti)merons in one MUC.
This finding is very distinct from our previous knowledge that the meron and antimeron should emerge in pairs\cite{LinSB2015,YuKTetal2018, GaoJIetal2019, GaoRGetal2020}.
In a common sense, the meron and antimeron with a topological charge $\mp{1/2}$
can be mapped\cite{LinSB2015} onto the southern and northern hemispheres, respectively, and wrap them once.
Hence, considering a large enough easy-plane interaction, such as an in-plane anisotropy or the dipole-dipole interaction,
a skyrmion could be dissociated into two halves from the equator, transforming into a meron-antimeron pair\cite{LinSB2015}.
However, as shown in Fig.~\ref{phaseDiagram}(b), there are three Bloch-type (anti)merons.
Each one is composed of ten spins, including one polarized core spin, its three nearest neighbours~(NNs) and six next-nearest neighbours~(NNNs).
All the core spins locate on the same sublattice, and are completely polarized 
along the $\rm \mathbf{c}$ axis~(down for $\mathcal{A}$ and $\mathcal{C}$, up for $\mathcal{B}$).
The three NNs of each (anti)meron locate on the other sublattice and have the same polar angle. In particular,
in $\mathcal{B}$ and $\mathcal{C}$ they are on the opposite hemisphere to the core spin.
At the edges, the equator is formed by six NNNs locating on the same sublattice as the core spin.
All NNNs lie almost in the $\mathbf{ab}$ plane. These characters suggest that $\mathcal{B}$ and $\mathcal{C}$ are of the AFM type.
We want to mention that in our model FM merons are readily available by the transformation $\phi\rightarrow\phi+\pi$,
equivalent to flipping all spins on one sublattice.

One could notice that, if swirling around one meron in any direction, its NN and NNN spins rotate in the same direction
while the core spins in $\mathcal{B}$ and $\mathcal{C}$ are antiparallel.
In addition, for one loop around the meron, the spins in $\mathcal{A}$ rotate $4\pi$, wrapping the hemisphere twice.
By contrast, the spins wrap the hemisphere once in $\mathcal{B}$ and $\mathcal{C}$, 
suggesting the topological charge of $\mathcal{A}$ should be twice of those of $\mathcal{B}$ and $\mathcal{C}$.
In order to confirm our analysis, we calculate the topological charges of the three 
merons at the representative parameter point following the definition given by Berg and L\"uscher\cite{BergLuscher1981}~(SM\cite{Supp} Sec.~III),
which turn out to be $Q_\mathcal{A}=1.055231, Q_\mathcal{B}=0.472850$ and $Q_\mathcal{C}=-0.528081$,
indicating the existence of a meron ($Q_\mathcal{C}$) and antimeron ($Q_\mathcal{B}$) pair and a high-$Q$ antimeron ($Q_\mathcal{A}$).
Particularly, the high-$Q$ antimeron is topologically equivalent to a half high-$Q$ antiskyrmion\cite{ZhangZhouEzawa2016}.
It is known theoretically that the topological charge for a ideal (high-$Q$) (anti)meron should be exactly $n/2$ with $n$ an integer.
However, in the lattice system, the three (anti)merons share the outmost spins as the equator.
When frustrations disturb the outmost edge slightly off the $\mathbf{ab}$ plane, we can expect that the topological charge of each (anti)meron
may deviate from its theoretical value accordingly. Since the increase or decrease of solid angles should occur synchronously
and one can expect that the summation of $Q_\mathcal{A}, Q_\mathcal{B}$ and $Q_\mathcal{C}$ should be an integer.
Indeed, this is what we have observed from our results, i.e., it is 1.0. Particularly, in the TmX phase there are points with their 
$Q_\mathcal{A}, Q_\mathcal{B}$ and $Q_\mathcal{C}$ agreeing well with their theoretical values and 
the summation giving 1.0 as well. 

The crystals formed by magnetic spin textures with a nonzero $Q$ often exhibit nontrivial transport 
properties. The prototypical example is the THE\cite{Hamamoto2015}. 
However, it has been shown~\cite{HayamiOM2021} that there is no THE in the meron-antimeron crystals due to the vanishing net $Q$.  
Thus, the TmX provides an opportunity to explore the THE in the meron crystals. To show this, we study the double-exchange model, 
which describes itinerant electrons coupled with the spin texture\cite{Hamamoto2015},
\begin{eqnarray}
    H &=& t\sum_{ij}c^\dagger_{i\sigma} c_{j\sigma} -J\sum_{i\sigma\sigma'}\mathbf{S}_{i}\cdot c^\dagger_{i\sigma} \boldsymbol{\sigma} c_{i\sigma'},
\label{HDE} 	
\end{eqnarray}
where $c^\dagger_{i\sigma}$ ($c_{i\sigma}$) is the creation (annihilation) operator of electrons with the spin $\sigma$ at the site $i$. 
$t$ is the hopping integral between two nearest neighbors $i$ and $j$. $J$ is the coupling strength between the electron 
and on-site magnetization $\mathbf{S}_i$, and $\boldsymbol{\sigma}$ denotes the Pauli matrix. 
Since each unit cell of the TmX carries a finite $Q$, it provides a finite magnetic flux and leads to the THE.

The $n$th band structure $E_n\left(\mathbf{q}\right)$ and the corresponding eigenvector $\phi_n(\mathbf{q})$ can be easily obtained by 
diagonalizing the Hamiltonian (\ref{HDE}) in the strong coupling limit\cite{Hamamoto2015} (see the SM\cite{Supp} Sec.~IV for details). 
We plot the band structure along the high-symmetry lines in Fig~\ref{THE}(a). One may notice that there is a Dirac-cone 
structure with a certain large gap in the vicinity of $\Gamma$ point between the bands $n = 6$ and $n = 7$. 
With the $E_n\left({\mathbf{q}}\right)$ and $\phi_n(\mathbf{q})$ available, some important quantities such as the Berry curvature, 
the Chern number $\mathcal{C}_n$ and the topological Hall conductance $\sigma_{xy}$ are readily obtained. 
In Fig~\ref{THE}(b), we plot $\sigma_{xy}$ given by the Kubo formula\cite{Hamamoto2015} (see SM\cite{Supp} Sec.~IV for other quantities),
\begin{eqnarray}
	\sigma_{xy} & & = - \frac{2\pi i e^2}{h\Xi} \sum_{\mathbf{q}}\sum_{n,m(\ne n)} f(E_n(\mathbf{q}))\times  \\
& & \frac{\langle \phi_n(\mathbf{q})| \frac{\partial H} {\partial q_x} | m(\mathbf{q})\rangle
		\langle \phi_m(\mathbf{q})| \frac{\partial H}{\partial q_y} |\phi_n(\mathbf{q})\rangle-(n\leftrightarrow m)}{(E_n(\mathbf{q})-E_m(\mathbf{q}))^2}  \nonumber
\end{eqnarray}
where $\Xi$ is the size of the system and $f$ is the Fermi distribution function. 
One may notice that at zero temperature the Hall conductance is quantized to be $-1$
in the band gap between $n=6$ and $n=7$ where $E_f \in [-0.73,-0.61]$, which indicates a finite total Chern number, 
i.e., $\sum_{E_n<E_f}\mathcal{C}_n=\sigma_{xy}/(e^2/h)$.

\begin{figure}[bt]
	\begin{tabular}{c}
		\begin{minipage}[t]{0.5\textwidth}
			\includegraphics[width=0.9\textwidth, clip]{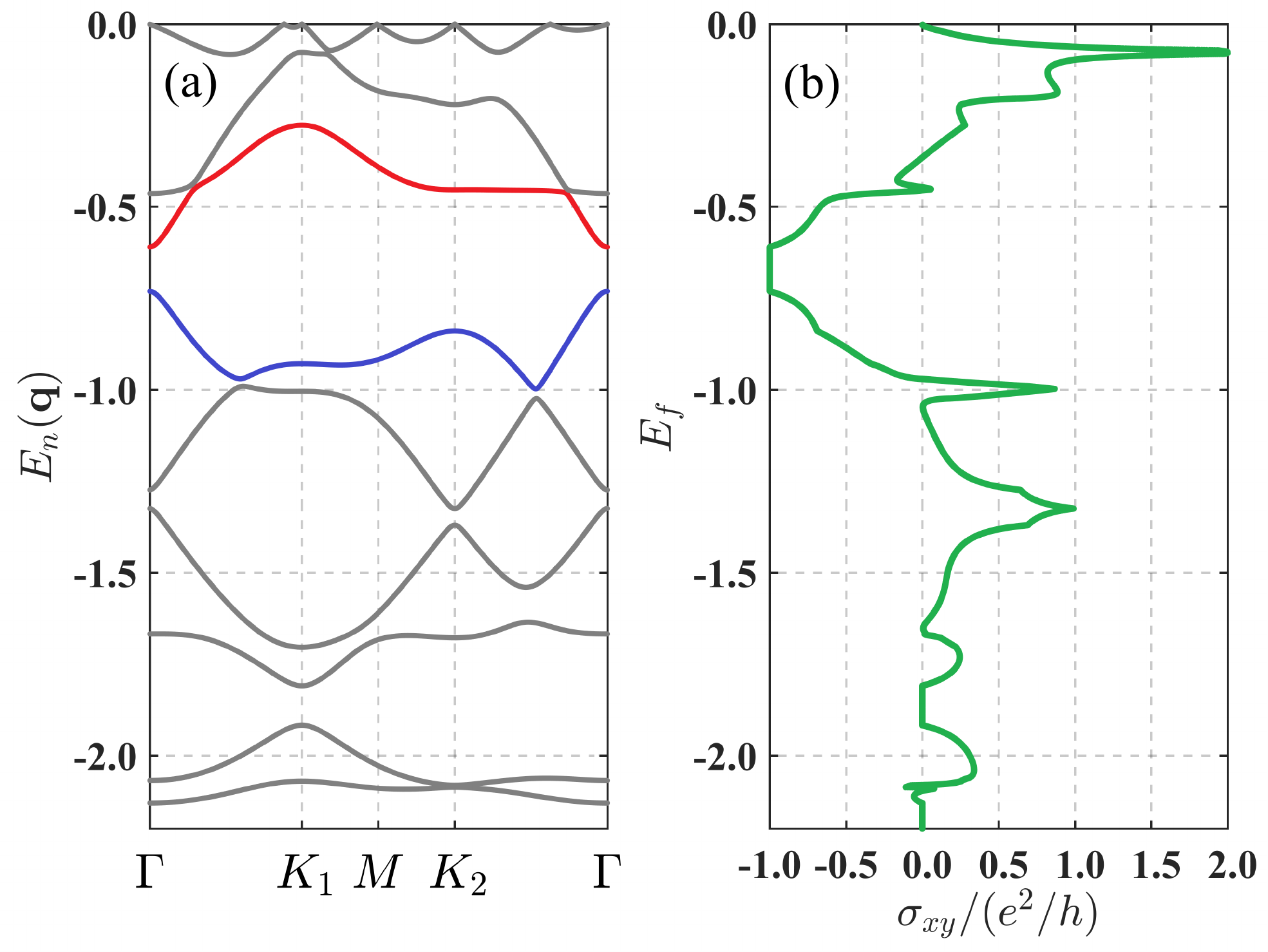}
		\end{minipage}
	\end{tabular}
	\vspace{-0.5cm}
	\caption{(a) Band structures of the model (\ref{HDE}) along the high-symmetry lines in the first Brillouin zone.
	The TmX spin configuration is obtained at $\theta=0.515\pi, \phi=0.656\pi$. $n\left(=1,2,\cdots\right)$ is the band index from 
	the bottom to the top. The Chern number $\mathcal{C}_n$ is $0$ for $n = 1, 2, 3, 5$, and $\mathcal{C}_4 = 1$, $\mathcal{C}_6 = -2$, 
	$\mathcal{C}_7 = 2$. Note that the two lowest bands are separated by a tiny gap.
	         (b) The Hall conductance $\sigma_{xy}$ at zero temperature is shown as a function of Fermi energy $E_f$.
		 The quantized conductance locates in the band gap between the 6th~(blue) and the 7th~(red) bands.}
	\label{THE}
\end{figure}

\begin{figure}[bt]
	\begin{tabular}{c}
		\begin{minipage}[t]{0.45\textwidth}
			\includegraphics[width=1.0\textwidth, clip]{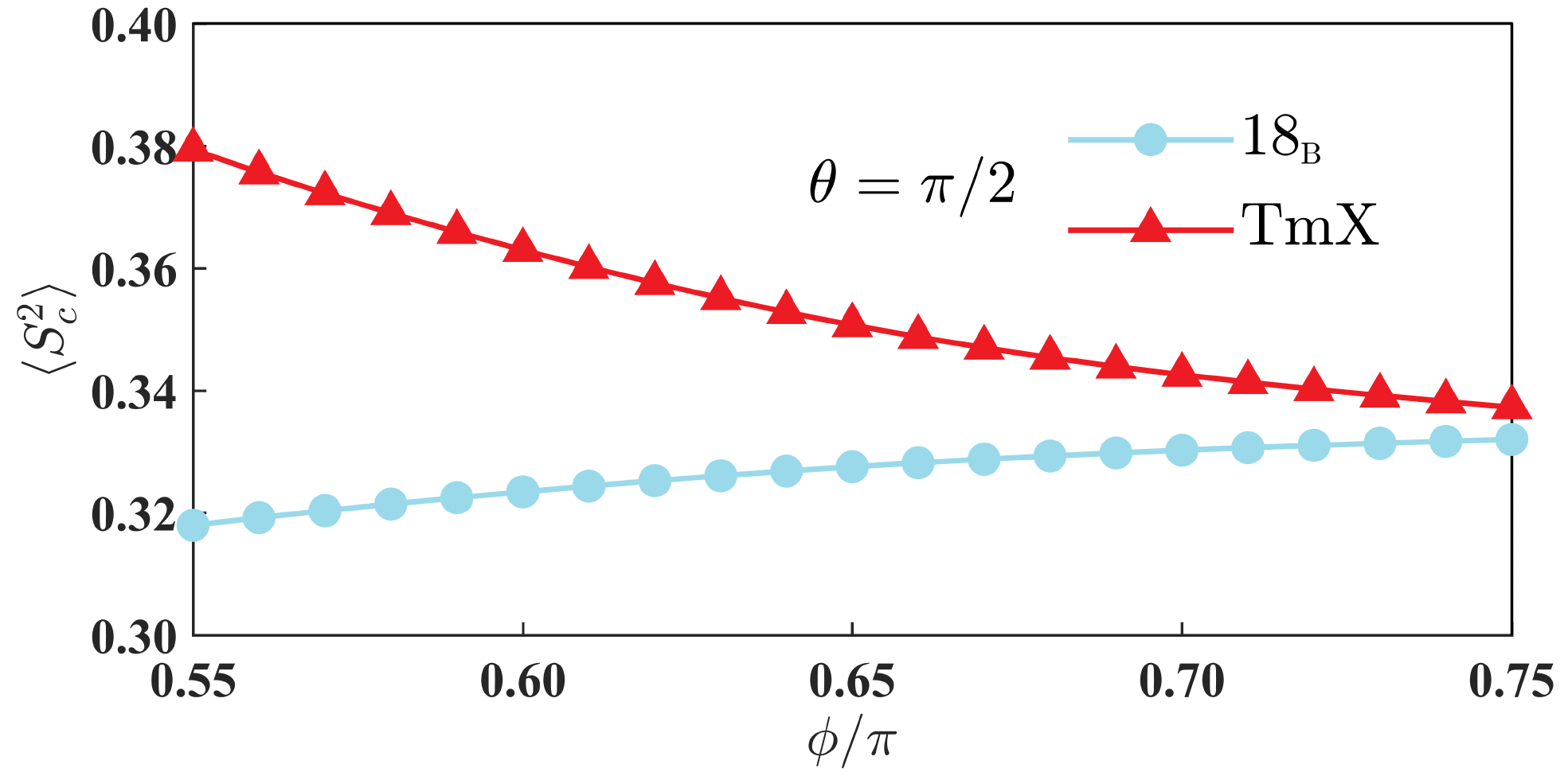}
		\end{minipage}
	\end{tabular}
	\vspace{-0.5cm}
	\caption{$\langle{S_c^2}\rangle=\sum{\left(\mathbf{S}_i\cdot\mathbf{c}\right)^2}/N$ for the TmX and $18_B$ states is
		shown as a function of $\phi$. $\theta=\pi/2$ is fixed which separates the TmX phase and the $18_B$ phase.}
	\label{SC2}
\end{figure}

As shown in Fig.~\ref{phaseDiagram}, the $\rm{TmX}$ phase and the $18_{\rm B}$ phase are separated
by the line $\theta=\pi/2$, where $A=0$. In both phases, there are 18 spins in one MUC.
On the line $\theta=\pi/2~(A=0)$, these two states are degenerate~(SM\cite{Supp} Sec.~I).
One may notice that in the TmX phase $A<0$, which is thus of the easy-axis type.
In the literature, several mechanisms have been proposed to stabilize merons, 
such as the dipole-dipole interaction~\cite{ZhouEzawa2014}, the easy-plane anisotropy~\cite{LinSB2015, HellmanHTetal2017}
as well as their interplay~\cite{AugustinJEetal2021},
the relative phase shifts among multiple helical spin density waves\cite{HayamiOM2021}, 
and the interplay between the biquadratic interaction and the Dzyaloshinskii-Moriya interaction\cite{HayamiYambe2021}.
One question naturally arises as to why the TmX phase is stable under the easy-axis anisotropy.
Actually, the TmX state is rooted in the $\Gamma$ model, which has highly degenerate ground states~\cite{RousochatzakisPerkins2017}. 
Its degeneracy is partly lifted by the Kitaev term, resulting in the degenerate TmX phase and 18$_\mathrm{B}$ phase (see SM\cite{Supp} Sec. I and III).
Moreover, as shown in Fig.~\ref{SC2}, $\langle S_c^2\rangle=\sum_i\left(\mathbf{S}_i\cdot\mathbf{c}\right)^2/N$
is different in these two states. Therefore, in the presence of a finite $A$, the degeneracy is lifted.
Particularly, $\langle{S_c^2}\rangle$ is larger in the TmX state, and hence when $A<0$ the TmX state is energetically favored.

\textit{Summary and Outlook}.\;--
In this work, we report our discovery of the TmX state in the high-spin Kitaev magnets and 
the topological effect led by such a state is demonstrated.
Although the Kitaev interaction was first proposed in the spin-1/2 Kitaev honeycomb model\cite{Kitaev2006},
recent studies show that it may also exist in high-spin magnets.
This offers great opportunities to study TSTs in such high-spin Kitaev magnets.
Our discovery of the TmX state represents a fantastic progress in this direction and laid
the foundation for the forthcoming Kitaev spintronics.
On the other hand, the merons, as the classical solutions of Yang-Mills equation, were proposed as the mechanism for the quark confinement.
Owing to their one half topological charge, they can only exist in pairs\cite{Actor1979, Ezawa2011, PhatakLH2012}.
This assessment is also supported by numerous studies in magnets\cite{LinSB2015,YuKTetal2018, GaoJIetal2019, GaoRGetal2020}  and 
photonic systems\cite{GuoXGetal2020}.
Our findings are obviously beyond this paradigm and thus extend the scope of the deconfinement\cite{SenthilVBetal2004},
which leads to our conjecture that merons with a half-integer $Q$ should be in pairs while those with an integer $Q$ can appear solely. 
Further studies, with Kitaev magnets as a feasible starting point, are necessary to explore such a deconfinement phenomenon.

\begin{acknowledgments}
We are grateful to Shi-Zeng Lin, Yan Zhou, Meng Xiao and Yalei Lu for helpful discussions.
This work is supported by the National Natural Science Foundation of
China (Grant Nos. 11874188, 12047501, 11774300, 11834005, 91963201, 12174164).
The computational resource is partly supported by the Supercomputing Center of Lanzhou University.
\end{acknowledgments}

\clearpage

\onecolumngrid

%%%%%%%%%% Merge with supplemental materials %%%%%%%%%%
%%%%%%%%%% Prefix a "S" to all equations, figures, tables and reset the counter %%%%%%%%%%
\newpage

\newcounter{equationSM}
\newcounter{figureSM}
\newcounter{tableSM}
\stepcounter{equationSM}
\setcounter{equation}{0}
\setcounter{figure}{0}
\setcounter{table}{0}
\setcounter{page}{1}
\makeatletter
\renewcommand{\theequation}{\textsc{sm}-\arabic{equation}}
\renewcommand{\thefigure}{\textsc{sm}-\arabic{figure}}
\renewcommand{\thetable}{\textsc{sm}-\arabic{table}}

\begin{center}
	\Large{\bf{Supplementary Material on ``Triple-meron crystal in high-spin Kitaev magnets''}}
\end{center}
%\author{Ken Chen, Qiang Luo, Zongsheng Zhou, Saisai He, Bin Xi, Chenglong Jia, Hong-Gang Luo, Jize Zhao}
%\begin{abstract}
   In this Supplementary Material, we present more details of the numerical methods, the corresponding results and the complete phase diagram.
The energy optimization method and how to use it to determine the phase transition points are shown in Sec.~I.
In Sec.~II we show how to determine the magnetic unit cell numerically. 
In Sec.~III, we explain the origin of the TmX phase.
In Sec. ~IV, the details for calculating the topological charges and the topological Hall effect caused by the TmX are provided.
In Sec.~V, we show results from the atomistic dynamics simulations and in Sec.~VI we provide the complete phase diagram. The phase diagram is
determined by the ground-state energy, the spin configurations and the static spin structure factors.
%\end{abstract}
\maketitle
%\date{}

\begin{center}
\textsf{\textbf{Sec. I: The ground-state energy and phase-transition points by the Energy Optimization Method}}
\end{center}
\setcounter{subsection}{1}

~Based on the spin configuration obtained by the Monte Carlo simulation,
we could further use an energy optimization method to target the ground-state energy with a controllable precision.
In some cases, we can derive an analytic expression for the ground-state energy.
Here, we take the TmX as an example to illustrate this method.

\begin{figure}[ht]
   \includegraphics[width=0.7\columnwidth]{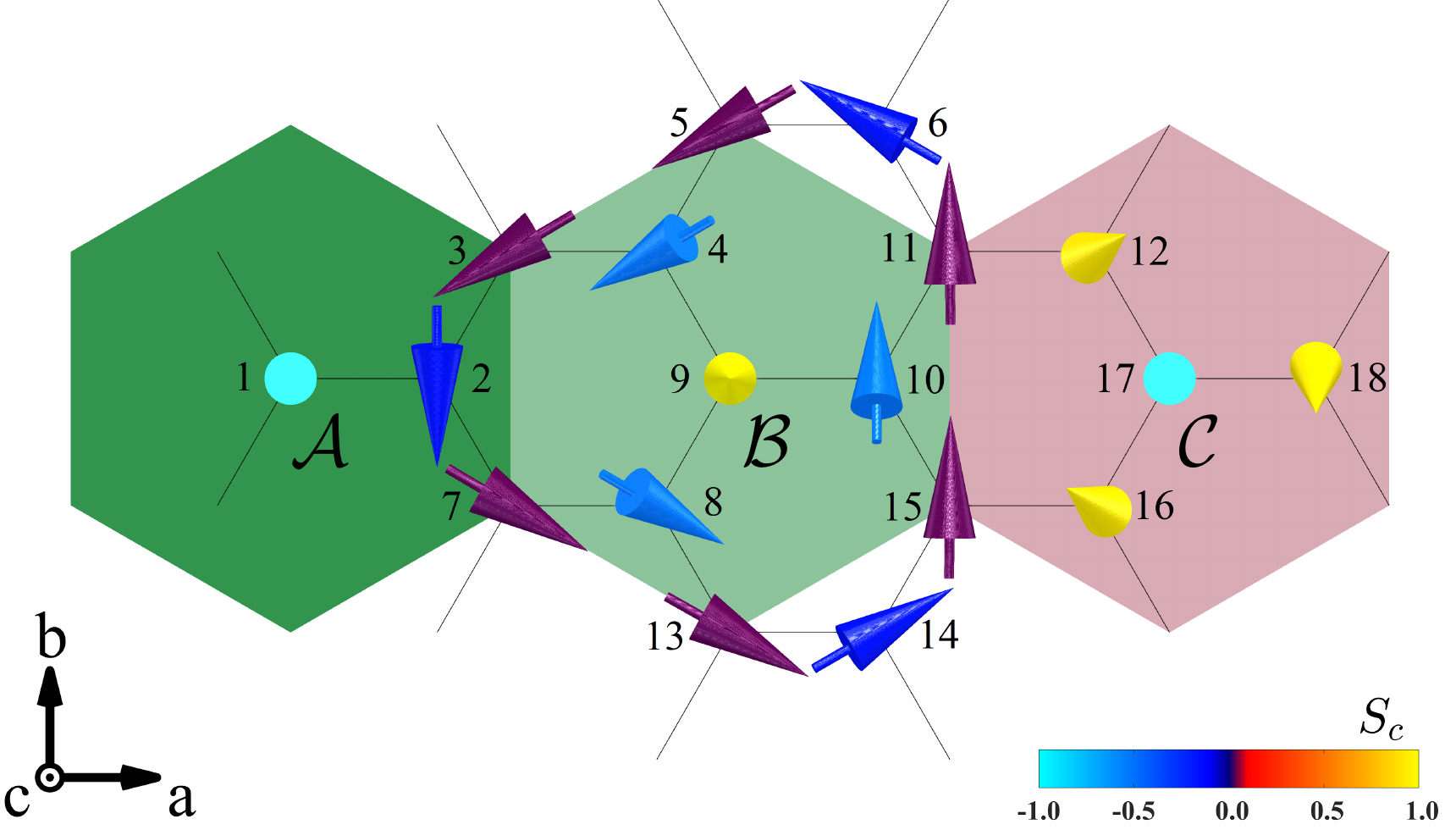}
\caption{The spin configuration of the TmX state in one magnetic unit cell. The numbers labelling lattice sites range from 1 to 18.
The spins $\mathbf{S}_i(\vartheta_i, \varphi_i)$ are defined in the $\rm \bf{abc}$ coordinate system.
    The color gives the $S_c$, while the in-plane orientation gives the azimuthal angle $\varphi$.}
    \label{FIG:TmX18UC}
\end{figure}

The paradigmatic spin configuration in the TmX state, which includes three kinds of merons, is shown in Fig.~2 of the main text.
As presented in Fig.~\ref{FIG:TmX18UC}, there are eighteen spins in one magnetic unit cell.
For each meron, the core spin and its three NNs
can be written as $(\mathbf{S}_c|\mathbf{S}_i \mathbf{S}_j \mathbf{S}_k)$ for short.

In the $\bf{abc}$ coordinate system, the classical spin at site $i$ can be parameterized as
\begin{equation}\label{Mclassicalspin3}
\mathbf{S}_i = S\left(\sin{\vartheta_i}\cos{\varphi_i}, \sin{\vartheta_i}\sin{\varphi_i}, \cos{\vartheta_i}\right),
\end{equation}
where $\vartheta_i \in [0, \pi]$ and $\varphi_i \in [0, 2\pi)$.
Pertaining to the core spin and its NNs in each meron, the only variable parameter is the polar angle $\vartheta_i$.
For merons centered at site 1, 9, and 17, the auxiliary angles are found to be
\begin{eqnarray}\label{EQ:SpinMeronA}
\textrm{meron $\mathcal{A}$}:\quad
\left(\begin{array}{c|ccc}
	\mathbf{S}_{1}      & \mathbf{S}_{6}    &  \mathbf{S}_{14}      &   \mathbf{S}_{2}      \\
    \hline
    \pi                         & \vartheta_{1}             &  \vartheta_{1}                &   \vartheta_{1}               \\
    -                       & 5\pi/6                    &  \pi/6                        &   3\pi/2                      \\
\end{array}\right),
\end{eqnarray}
\begin{eqnarray}\label{EQ:SpinMeronB}
\textrm{meron $\mathcal{B}$}:\quad
\left(\begin{array}{c|ccc}
	\mathbf{S}_{9}      & \mathbf{S}_{10}   &  \mathbf{S}_{4}       &   \mathbf{S}_{8}      \\
    \hline
    0                           & \vartheta_{2}             &  \vartheta_{2}                &   \vartheta_{2}               \\
    -                      & \pi/2                     &  7\pi/6                       &   11\pi/6                     \\
\end{array}\right)
\end{eqnarray}
and
\begin{eqnarray}\label{EQ:SpinMeronC}
\textrm{meron $\mathcal{C}$}:\quad
\left(\begin{array}{c|ccc}
	\mathbf{S}_{17}     & \mathbf{S}_{16}   &  \mathbf{S}_{12}      &   \mathbf{S}_{18}     \\
    \hline
    \pi                         & \vartheta_{3}             &  \vartheta_{3}                &   \vartheta_{3}               \\
    -                      & 5\pi/6                    &  \pi/6                        &   3\pi/2                      \\
\end{array}\right).
\end{eqnarray}
Here, elements in the second and third rows of Eqs.~\eqref{EQ:SpinMeronA}-\eqref{EQ:SpinMeronC} are the polar angles $\vartheta_i$ and azimuthal angles $\varphi_i$, respectively.
The other six spins rely on two angles ($\vartheta_0$, $\varphi_0$), subjecting to the following relation
\begin{eqnarray}\label{EQ:ItinerantSpin}
\left(\begin{array}{cc|cc|cc}
	\mathbf{S}_{3}    & \mathbf{S}_{5}    & \mathbf{S}_{13}      & \mathbf{S}_{7}     & \mathbf{S}_{11}    & \mathbf{S}_{15}    \\
    \hline
    \vartheta_{0}                  & \vartheta_{0}                  & \vartheta_{0}                  & \vartheta_{0}                  & \vartheta_{0}       & \vartheta_{0}     \\
    \frac{5\pi}{3}\!-\!\varphi_0   & \frac{2\pi}{3}\!+\!\varphi_0   & \frac{\pi}{3}\!-\!\varphi_0   & \frac{4\pi}{3}\!+\!\varphi_0   & \pi\!-\!\varphi_0   & \varphi_0         \\
\end{array}\right).
\end{eqnarray}

Hence, the Hamiltonian $\mathcal{H}$ could be recast as a multivariable function
$\mathcal{F}(\vartheta_0, \varphi_0; \vartheta_1, \vartheta_2, \vartheta_3)$
which depends on five variational parameters,
and the ground-state energy is determined by minimizing $\mathcal{F}$ in the allowed parameter spaces.
We emphasize that proper trial angles $(\{\vartheta_i\}, \varphi_0)$ guided by the Monte Carlo simulation are essential to reach the global minima of $\mathcal{F}$.

As shown in the Fig.~\ref{PDM}, the classical $\rm K\Gamma{A}$ model has a rich phase diagram which contains scores of magnetically ordered phases.
Here, five of them are conventional magnetic orders which also frequently appear in other spin models.
In the limit of ferromagnetic (FM) Kitaev point, there are two FM phases termed FM$\rm{_I}$ phase and FM$\rm{_V}$ phase.
In the former the spins lie in the $\rm \mathbf{ab}$ plane while in the latter the spins point out of the plane.
Meanwhile, in the vicinity of antiferromagnetic (AFM) $\Gamma$ limit,
there are an AFM$\rm{_V}$ phase for a negative $A$, and a zigzag~(ZZ) phase and an in-plane 120$\rm^{\circ}_I$ phase in the case of positive $A$.
We note that the ZZ phase lies in the $\rm \mathbf{ac}$ plane with a tilted angle $\alpha_{\tau}$ away from the $\rm \mathbf{a}$ axis.
The ground-state energy $E_g$ and the polarization directions in these phases can be given analytically, which are shown in Tab.~\ref{TabSM-1}.

\begin{table}[th!]
\caption{\label{TabSM-1}
The ground-state energy $E_g$ and polarization directions in the FM$\rm_I$, FM$\rm_V$, AFM$\rm_V$, ZZ and 120$\rm^\circ_I$ phases.}
% \begin{ruledtabular}
\begin{tabular}{ l | l | c}
\colrule\colrule
	Phases                               &  \hspace{1cm}  $E_g$        & polarization direction   \\
\colrule
FM$_{\rm I}$         & $E_g = \frac{S^2}{2}\left(K-\Gamma\right)$                 & $\parallel \rm\mathbf{ab}$   \\
FM$\rm_V$         & $E_g = S^2\left(A+K/2+\Gamma\right)$                 & $\perp \rm\mathbf{ab}$   \\
AFM$\rm_V$         & $E_g = S^2\left(A-K/2-\Gamma\right)$                 & $\perp \rm\mathbf{ab}$   \\
ZZ         & $E_g = -\frac{S^2}{12}\left[\sqrt{32(\Gamma-K)^2 + (2K+7\Gamma+6A)^2} + (3\Gamma-6A) \right]$                 & $\tan(2\alpha_{\tau}) = \frac{4\sqrt2(\Gamma-K)}{2K+7\Gamma+6A}$    \\
120$\rm^{\circ}_I$         & $E_g = -S^2\left(K/2-\Gamma\right)$                 & $\parallel\rm\mathbf{ab}$    \\
\colrule\colrule
\end{tabular}
% \end{ruledtabular}
\end{table}

\begin{figure}[!ht]
\centering
\includegraphics[width=0.6\columnwidth, clip]{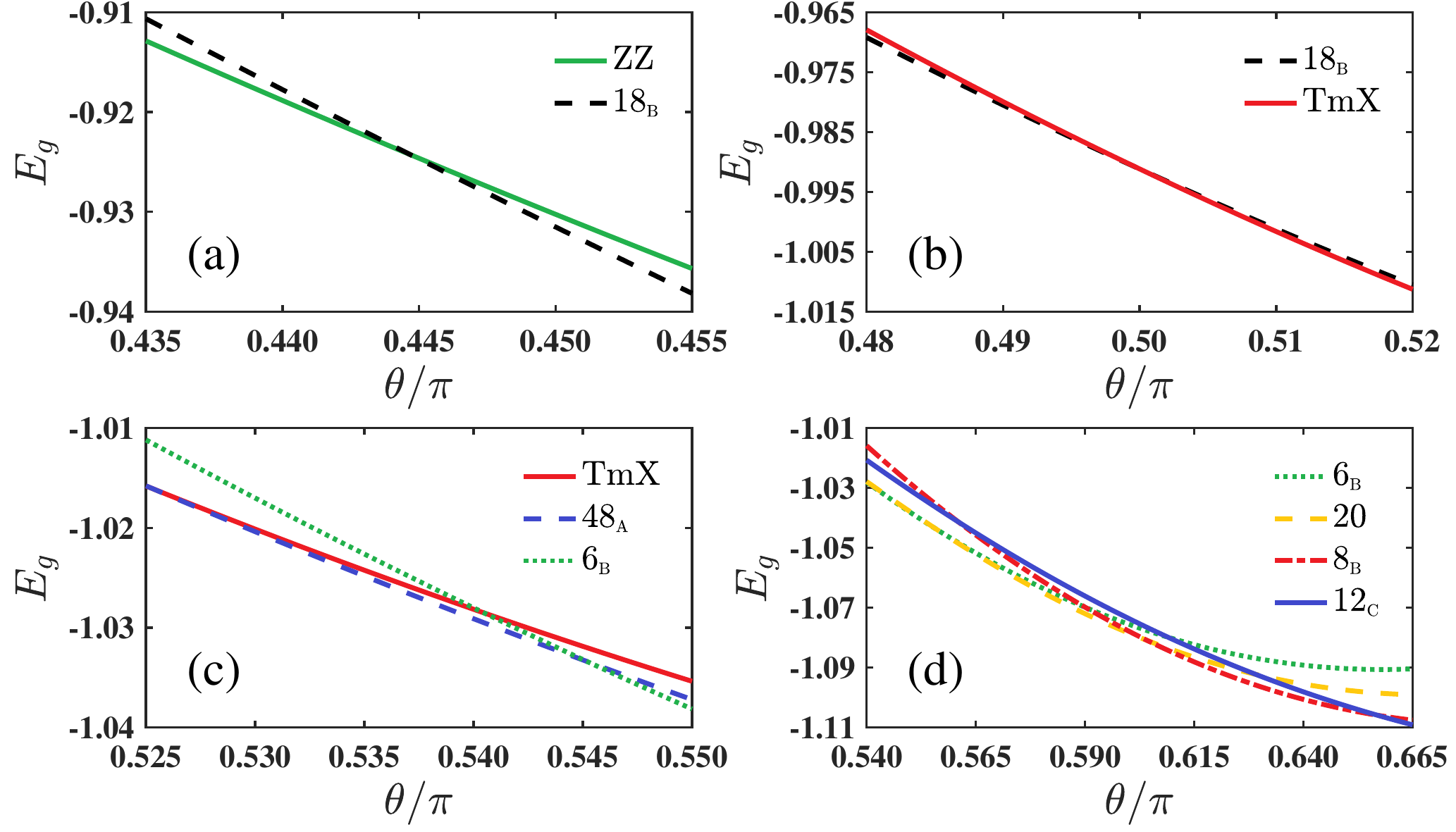}\\
\caption{Level crossings of the ground-state energy $E_g$ along the line of $\phi = 0.68\pi$.
    For the benefit of visualization, the level crossings are split into four panels which indicate the transitions
    (a) ZZ--18$\rm_B$, (b) 18$\rm_B$--TmX, (c) TmX--48$\rm_A$ and 48$\rm_A$--6$\rm_B$,
and (d) 6$\rm_B$--20, 20--8$\rm_B$, and 8$\rm_B$--12$\rm_C$, respectively. Particularly, the transition between
the 18$\rm_B$ and TmX occurs at $\theta=\pi/2$.
}\label{FIGSM-2}
\end{figure}

In the $\rm K\Gamma{A}$ model, when $K < 0$ and $\Gamma > 0$ the system is highly frustrated and
many magnetic orders with a large unit cell could be stabilized.
To illustrate it, we start by performing the Monte Carlo simulations
along the line of $\phi = 0.68\pi$
and tune the polar angle $\theta$ from $0.25\pi$ (ZZ phase) to $0.7\pi$ (12$\rm_C$ phase).
In between, we find six distinct phases, 18$\rm_B$, TmX, 48$\rm_A$, 6$\rm_B$, 20 and 8$\rm_B$, with the increase of $\theta$.
While closed forms of their ground-state energy are not available, the expression $\mathcal{F}$ that only relies on a few variational
parameters could be obtained by the energy optimization method.
The ground-state energy is thus targeted by minimizing $\mathcal{F}$ w.r.t. the variational parameters in the neighboring of optimal trial values.
The energy of these phases is shown in Fig.~\ref{FIGSM-2}, and the transition points are identified by the level-crossing points
between neighboring phases.

\begin{center}
\textsf{\textbf{Sec. II: Details for determing the unit cell of the TmX phase}}
\end{center}
\setcounter{subsection}{2}
\begin{figure}
   \includegraphics[width=0.35\columnwidth]{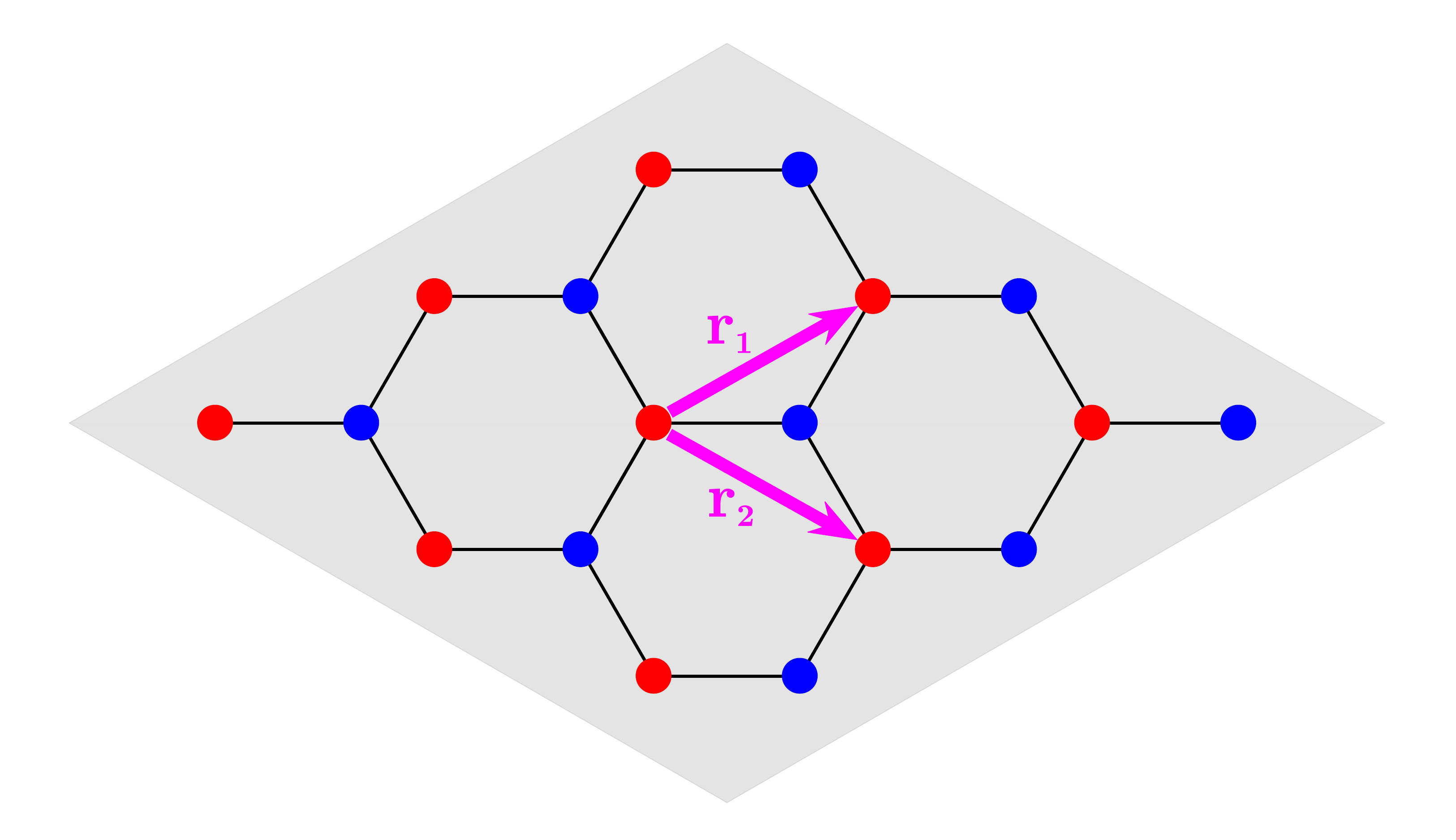} 
   \caption{Primitive lattice vectors of the honeycomb lattice.}	
   \label{PLV}	
\end{figure}
\begin{figure}[ht]
   \includegraphics[width=0.58\columnwidth, clip]{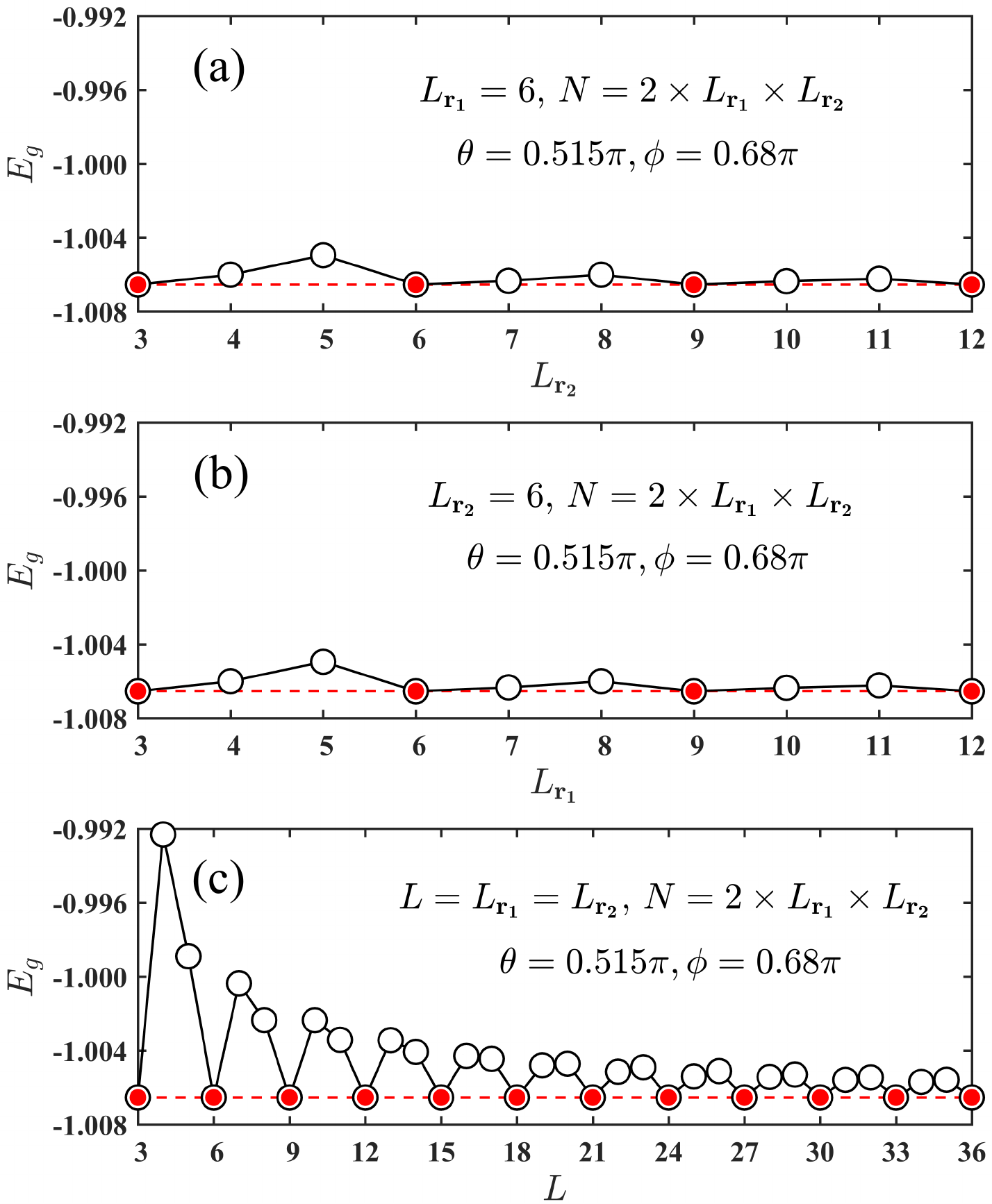}
      \caption{The procedure for determing the magnetic unit cell is illustrated. $\mathbf{r}_1, \mathbf{r}_2$ are given in Fig. \ref{PLV}.}
   \label{FSMUC}
\end{figure}
Since the magnetic unit cell of each phase is unknown aforehand, we need to determine their magnetic
unit cell numerically. In Monte carlo simulations, this can be done by varying the cluster sizes and compare their
ground-state energy. Here we exemplify this procedure by applying it to the TmX phase. Again we choose
$\theta=0.515\pi, \phi=0.68\pi$ as the representative point. If the cluster does not match the magnetic unit cell,
we will end with a higher energy than that of the true ground state in the Monte Carlo simulations.
In such a case, its ground state energy may approach the true one as the size increases.
Therefore we need to adjust the cluster size and compare the energy to determine the magnetic unit cell, 
which is shown in Fig.~\ref{FSMUC} with primitive lattice vetors $\mathbf{r}_1, \mathbf{r}_2$ given in Fig.~\ref{PLV}.
In Fig.~\ref{FSMUC}(a), we fix $L_{\mathbf{r}_1}=6$ and vary $L_{\mathbf{r}_2}$ from $3$ to $12$. 
We find that when $L_{\mathbf{r}_2}=3n$ with $n$ an integer we have the same lowest
energy $E_g = -1.0065447701$ while the energy is higher for $L_{\mathbf{r}_2}=3n+1$ and $3n+2$.
Moreover, the difference decreases as $n$ increases. This is well consistent with
our expectation. In Fig.~\ref{FSMUC}(b), we fix $L_{\mathbf{r}_2}=6$ and vary $L_{\mathbf{r}_1}$ 
from $3$ to $12$. We can find the same results as that from Fig.~\ref{FSMUC}(a). 
In Fig.~\ref{FSMUC}(c), we fix the ratio $L_{\mathbf{r}_1}/L_{\mathbf{r}_2}=1$ and vary $L_{\mathbf{r}_1}$ from 3 to 36.
The lowest energy $E_g$ is at $L_{\mathbf{r}_1}=3n$ and it is independent of $n$. Moreover, it is the same as that 
in Fig.~\ref{FSMUC}(a) and Fig.~\ref{FSMUC}(b).
These results obviously tell us the magnetic unit cell of the TmX phase has $2\times3\times3=18$ spins.

We want to mention that to determine the unit cell of a magnetic ordered phase, this procedure is necessary.
We also apply the same method to determine other phases in the phase diagram.

\begin{center}
\textsf{\textbf{Sec. III: Some formal analysis on the TmX phase}}
\end{center}
\setcounter{subsection}{3}
Once we know the magnetic unit cell and the corresponding magnetic order, we can do further analysis.
As demonstrated in Sec.~I, the variational function $\mathcal{F}(\vartheta_0,\varphi_0;\vartheta_1,\vartheta_2,\vartheta_3)$ of $E_g$ in the TmX phase can be written as
\begin{displaymath}
	\begin{aligned}	
\mathcal{F}(\vartheta_0,\varphi_0;\vartheta_1,\vartheta_2,\vartheta_3)=&3A-2\Gamma\cos\vartheta_{1}+2\Gamma\cos\vartheta_{2}-2\Gamma\cos\vartheta_{3}-K\cos\vartheta_{1}+K\cos\vartheta_{2}-K\cos\vartheta_{3}\\
&+3A\left(\cos\vartheta_{1}\right)^{2}+3A\left(\cos\vartheta_{2}\right)^{2}+3A\left(\cos\vartheta_{3}\right)^{2}+4\Gamma\cos\vartheta_{0}\cos\vartheta_{1}+4\Gamma\cos\vartheta_{0}\cos\vartheta_{2}	
\\&+4\Gamma\cos\vartheta_{0}\cos\vartheta_{3}+2K\cos\vartheta_{0}\cos\vartheta_{1}	+2K\cos\vartheta_{0}\cos\vartheta_{2}+2K\cos\vartheta_{0}\cos\vartheta_{3}
\\&-\sqrt{2}\Gamma\sin\vartheta_{1}-\sqrt{2}\Gamma\sin\vartheta_{2}-\sqrt{2}\Gamma\sin\vartheta_{3}+\sqrt{2}K\sin\vartheta_{1}+6A\left(\cos\vartheta_{0}\right)^{2}		
\\&+\sqrt{2}K\sin\vartheta_{2}+\sqrt{2}K\sin\vartheta_{3}-5\Gamma\sin\varphi_{0}\sin\vartheta_{0}\sin\vartheta_{1}	
-4\Gamma\sin\varphi_{0}\sin\vartheta_{0}\sin\vartheta_{2}
\\&+\Gamma\sin\varphi_{0}\sin\vartheta_{0}\sin\vartheta_{3}
-K\sin\varphi_{0}\sin\vartheta_{0}\sin\vartheta_{1}+K\sin\varphi_{0}\sin\vartheta_{0}\sin\vartheta_{2}
\\&+2K\sin\varphi_{0}\sin\vartheta_{0}\sin\vartheta_{3}-\sqrt{2}\Gamma\cos\vartheta_{0}\sin\vartheta_{1}			
+\sqrt{2}\Gamma\cos\vartheta_{0}\sin\vartheta_{2}
\\&-\sqrt{2}\Gamma\cos\vartheta_{0}\sin\vartheta_{3}			
+\sqrt{2}K\cos\vartheta_{0}\sin\vartheta_{1}-\sqrt{2}K\cos\vartheta_{0}\sin\vartheta_{2}			
\\&+\sqrt{2}K\cos\vartheta_{0}\sin\vartheta_{3}-\sqrt{6}\Gamma\cos\varphi_{0}\cos\vartheta_{1}\sin\vartheta_{0}
+\sqrt{6}\Gamma\cos\varphi_{0}\cos\vartheta_{2}\sin\vartheta_{0}
\\&+\sqrt{6}K\cos\varphi_{0}\cos\vartheta_{1}\sin\vartheta_{0}			
-\sqrt{6}K\cos\varphi_{0}\cos\vartheta_{2}\sin\vartheta_{0}
\\&+\sqrt{2}\Gamma\cos\vartheta_{1}\sin\varphi_{0}\sin\vartheta_{0}			
+\sqrt{2}\Gamma\cos\vartheta_{2}\sin\varphi_{0}\sin\vartheta_{0}
\\&-\sqrt{3}\Gamma\cos\varphi_{0}\sin\vartheta_{0}\sin\vartheta_{1}			
-2\sqrt{2}\Gamma\cos\vartheta_{3}\sin\varphi_{0}\sin\vartheta_{0}
\\&+2\sqrt{3}\Gamma\cos\varphi_{0}\sin\vartheta_{0}\sin\vartheta_{2}			
+3\sqrt{3}\Gamma\cos\varphi_{0}\sin\vartheta_{0}\sin\vartheta_{3}
\\&-\sqrt{2}K\cos\vartheta_{1}\sin\varphi_{0}\sin\vartheta_{0}			
-\sqrt{2}K\cos\vartheta_{2}\sin\varphi_{0}\sin\vartheta_{0}
\\&+\sqrt{3}K\cos\varphi_{0}\sin\vartheta_{0}\sin\vartheta_{1}			
+2\sqrt{2}K\cos\vartheta_{3}\sin\varphi_{0}\sin\vartheta_{0}
\\&+\sqrt{3}K\cos\varphi_{0}\sin\vartheta_{0}\sin\vartheta_{2}
	\end{aligned}
	\label{VF}
\end{displaymath}
\begin{figure}[ht]
   \includegraphics[width=0.6\columnwidth]{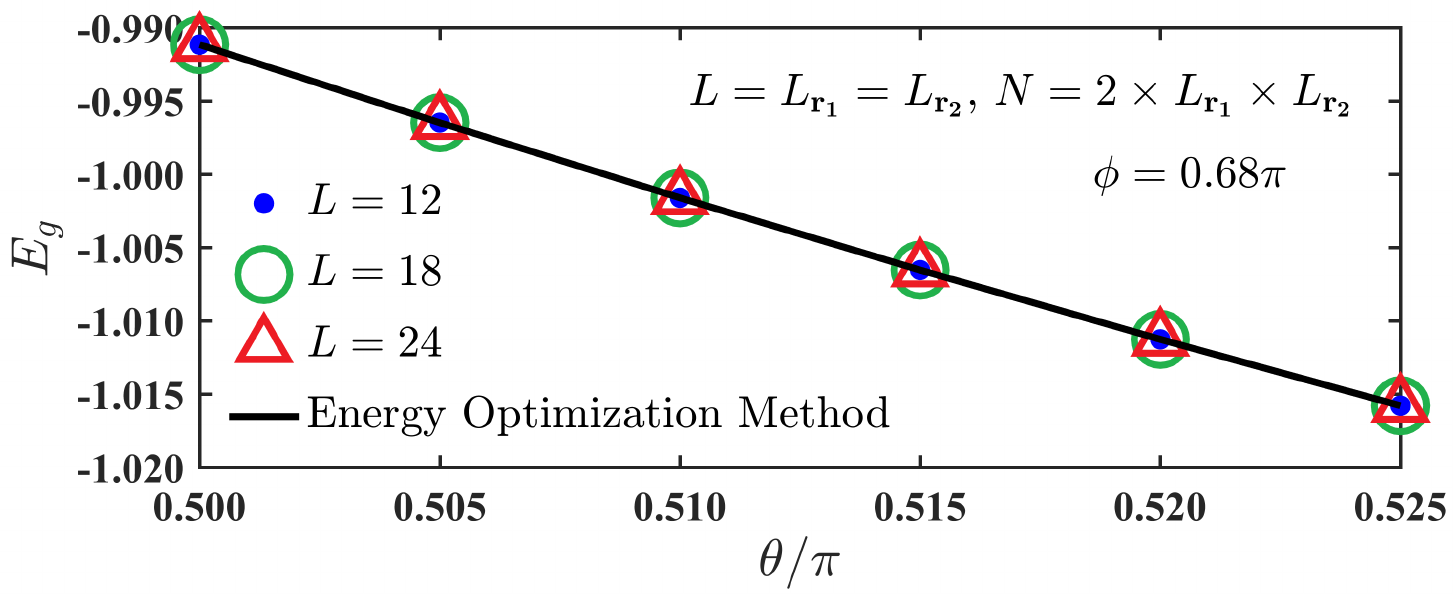}
\caption{The energy obtained by the energy optimization method and Monte Carlo is shown for comparison. $L=12$, 18 and 24 are the sizes in the Monte Carlo simulations.}
    \label{MCEO}
\end{figure}
The ground state energy $E_g\left(=\rm{min}(\mathcal{F}/18)\right)$ is readily available by numerically optimizing such a variational function. 
The equations $\partial{\mathcal{F}}/\partial{\xi}=0$($\xi=\vartheta_0, \varphi_0, \vartheta_1, \vartheta_2, \vartheta_3$) are safisfied at 
the minimum and they are used to check the results.
In Fig.~\ref{MCEO}, we compare the ground-state energy obtained by the energy optimization method and the Monte Carlo method.
We find that they are well consistent, demonstrating the validity of the function $\mathcal{F}(\vartheta_0,\varphi_0;\vartheta_1,\vartheta_2,\vartheta_3)$.

From the global phase diagram Fig.~\ref{PDM}, one can see that
ZZ, $18_\mathrm{B}$, TmX, 20, 8$_\mathrm{B}$, AFM$_\mathrm{V}$ and 120$^\circ_\mathrm{I}$ phases intersect at
the point $\Gamma=1$~(i.e., $A=0,K=0$). The ground state of such pure $\Gamma$ model is exactly known to be
classical spin liquid\cite{RousochatzakisPerkins2017} with its $E_g=-\Gamma$. Moreover, 
$E_g=-\Gamma$ for the $\Gamma$ model can also be obtained exactly from $\mathcal{F}(\vartheta_0,\varphi_0;\vartheta_1,\vartheta_2,\vartheta_3)$. 
This further suggests that the TmX state is rooted in the $\Gamma$ model and
it is one of the highly degenerate ground states of the $\Gamma$ model.
To obtain a stable TmX state, a reasonable strategy is to add proper interactions to lift the degeneracy
so that the TmX state becomes energetically favored. The Kitaev interaction
can partly lift the degeneracy and the degenerate TmX and 18$_\mathrm{B}$ become the ground states.
As we have shown in the main text, the degeneracy of the TmX and 18$_\mathrm{B}$ is further lifted by the single-ion anisotropy $A$.
For a negative $A$, the TmX state is energetically favored.

\begin{center}
\textsf{\textbf{Sec. IV: Topological Charge and Topological Hall Effect}}
\end{center}
\setcounter{subsection}{4}
\begin{figure}[hbt]
\centering
\includegraphics[width=0.6\columnwidth, clip]{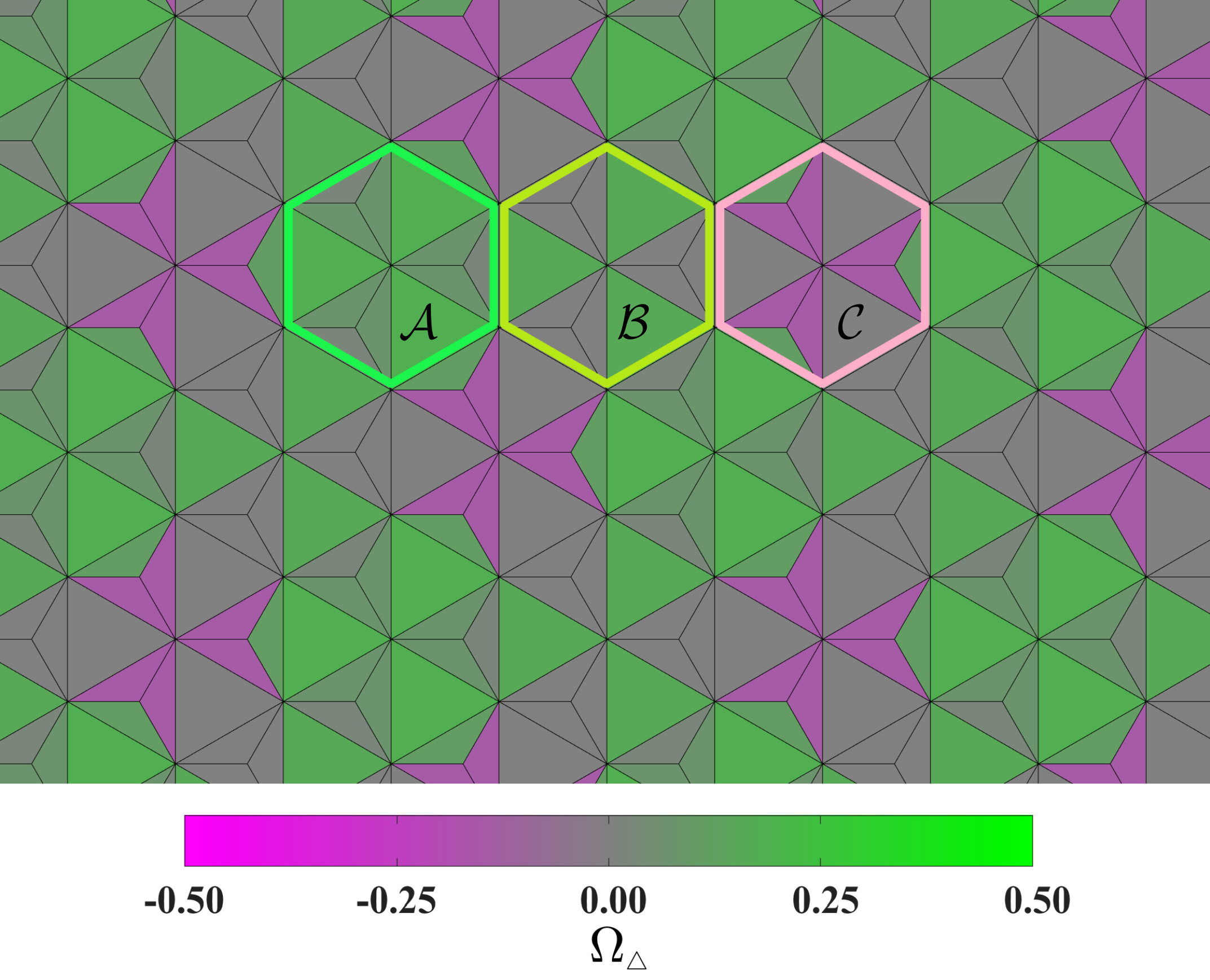}
\caption{The solid angle $\Omega_\triangle$~(in unit of $4\pi$) of each triangle is shown in the colormap. The spin configuration is same as that of
	Fig.~2(b) in the main text.}
\label{FigSM-3}
\end{figure}
\begin{figure}[hbt]
\centering
\includegraphics[width=0.6\columnwidth, clip]{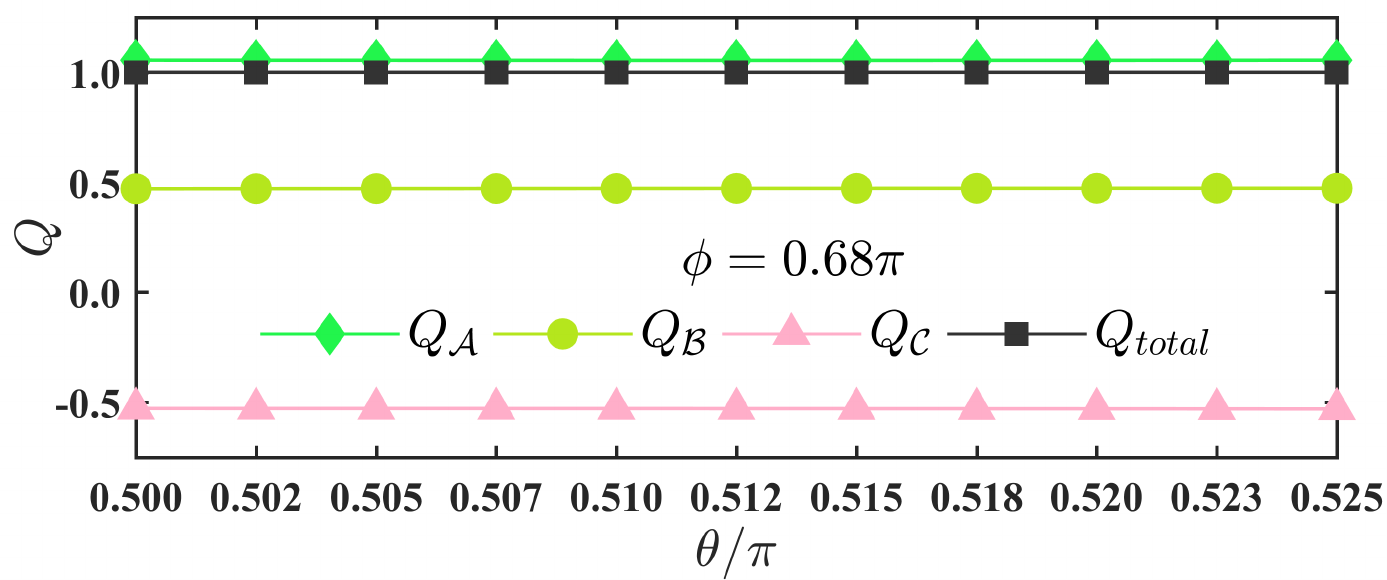}
	\caption{The topological charges $Q$ for the meron $\mathcal{A}$, $\mathcal{B}$ and $\mathcal{C}$ are shown as a function of $\theta$.
Their summation gives 1.0 with a numerical error of the order $10^{-14}$ or smaller at all the parameters.}
\label{FigSM-4}
\end{figure}
First, we discuss briefly how to calculate the topological charges $Q$ of the merons in our lattice model.
The definition was given by Berg and L\"uscher\cite{BergLuscher1981}, which is illustrated in Fig.~\ref{FigSM-3}.
Actually, in a lattice the topological charge can be obtained by summing up the solid angle on every elementary triangle.
For this purpose, the hexagon of each meron is splitted into six equilateral triangles, each is formed by a core spin
and its two next-nearest neighbors. Among these six triangles, three of them contain an additional
spin (the nearest neighbor of the core), which further divides the equilateral triangle into three isosceles triangles.
The solid angle for each triangle is thus calculated following the definition given by Berg and L\"uscher\cite{BergLuscher1981}.
In our work we present the values of the topological charges up to six digits for simplicity
although the numerical errors are of the order $10^{-14}$ or smaller.

In Fig.~\ref{FigSM-4} we plot the topological charges $Q$ of the three merons
and their summation as a function of $\theta$.
Although the outmost spins might be driven off $\mathbf{ab}$-plane, 
we can see that $Q_\mathcal{B}$ and $Q_\mathcal{C}$ are still close to $\pm1/2$ and $Q_\mathcal{A}$ is roughly 1.
Importantly, their summation is always 1.0. 
We want to mention that in the TmX phase we can find parameters with the topological charges of $Q_\mathcal{A}, Q_\mathcal{B}, Q_\mathcal{C}$ 
much closer to their theoretical values. For example, at $\theta=0.515\pi, \phi=0.656\pi$, $Q_\mathcal{A}=0.999881$, $Q_\mathcal{B}=0.500058$ 
and $Q_\mathcal{C}=-0.499939$, respectively. Moreover, due to the quadratic form of our model, a degenerate ground state is obtained
by flipping all the spins. Under this transformation, a straightforward conclusion is that the topological charges of
corresponding merons are sign-reversed.

Next, we discuss the details about calculating topological Hall effect~(THE). 
Since there is no THE in the meron-antimeron crystals\cite{HayamiOM2021}, 
the THE caused by the TmX is highly nontrivial for the merons crystals.  
The double-exchange model used in main text means the coupling between itinerant electrons and the local magnetization 
is much larger than the hoping amplitude, so the spin of itinerant electron should be parallel to the direction of local magnetization $\mathbf{S}_i$. Hence, the energy of Kondo coupling term turns to be a constant. By defining a rotation matrix $U_i$ on each site which rotates the eigenvector of the itinerant electron from $\sigma_z$ to $\bm{\sigma}\cdot \mathbf{S}_i$,  the above Hamiltonian can be transformed to a spinless free-fermion model\cite{Hamamoto2015}:
\begin{eqnarray}
  H_{eff} &=& \sum_{ij}t_{ij}^{eff} d_i^\dagger d_i,
\end{eqnarray}
where
\begin{eqnarray*}
	d_i & = U_i c_i=& \begin{pmatrix} \cos\bar{\theta}_i/2 & -\sin\bar{\theta}_i/2\\
	\sin\bar{\theta}_i/2 e^{i\bar{\phi}_i} &\cos\bar{\theta}_i/2 e^{i\bar{\phi}_i} \end{pmatrix}c_i,
\end{eqnarray*}
and
\begin{eqnarray*}
	t_{ij}^{eff} &= &\cos\frac{\bar{\theta}_i}{2}\cos\frac{\bar{\theta}_j}{2}+\sin\frac{\bar{\theta}_i}{2}\sin\frac{\bar{\theta}_j}{2}e^{-i(\bar{\phi}_i-\bar{\phi}_j)},
\end{eqnarray*}
with $\bar{\theta}_i$ and $\bar{\phi}_i$ the corresponding polar and azimuthal angle of the on-site magnetization $\mathbf{S}_i$.

The band structure $E_n\left({\mathbf{q}}\right)$ as well as wave functions $\phi_n\left(\mathbf{q}\right)$ can be directly obtained 
through the above effective Hamiltonian. The topological property of a band can be determined through its Chern number $\mathcal{C}_n$ ,
which by definition is the integration of the Berry curvature $\boldsymbol{\Sigma}$ over the first Brillouin zone:
\begin{eqnarray}
  \mathcal{C}_n=\frac{1}{2\pi} \int_{BZ} d\mathbf{s}\cdot \boldsymbol{\Sigma}= \frac{1}{2\pi} \int_{BZ} d\mathbf{s}\cdot
  \left[ \partial_{q_x}\mathbf{A}_y(q)-\partial_{q_y}\mathbf{A}_x(q)\right],
\end{eqnarray}
with $\mathbf{A}(\mathbf{q})=-i\langle \phi_n(\mathbf{q})|\nabla_\mathbf{q}|\phi_n(\mathbf{q})\rangle$ the Berry connection.
\begin{figure}[ht]
   \includegraphics[width=0.75\columnwidth]{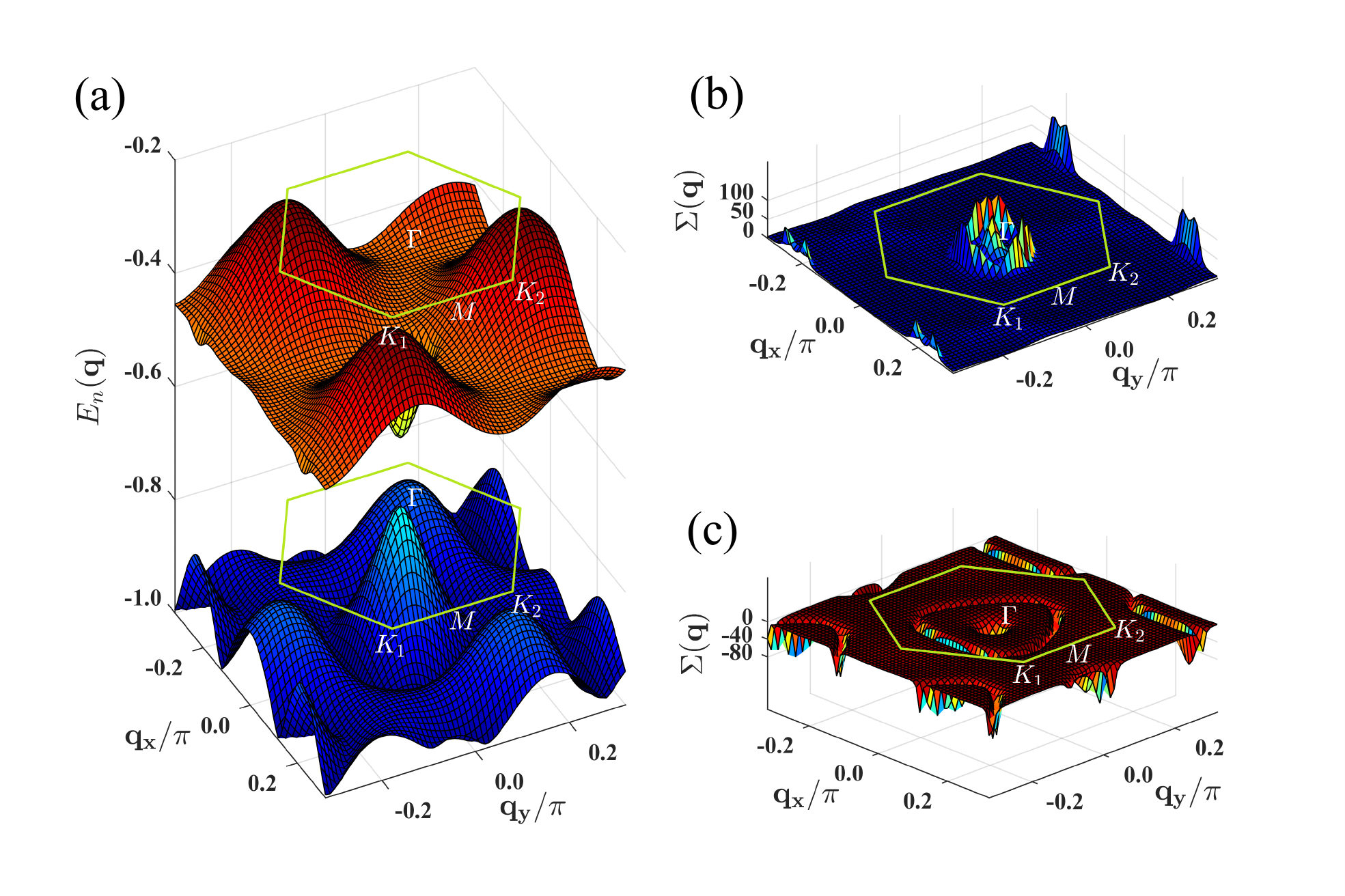}
\caption{(a) Band structures of band 6 and 7. (b) and (c) Berry curvatures of band 7 and 6, respectively.}
    \label{berry67}
\end{figure}

\begin{figure}[ht]
   \includegraphics[width=0.75\columnwidth]{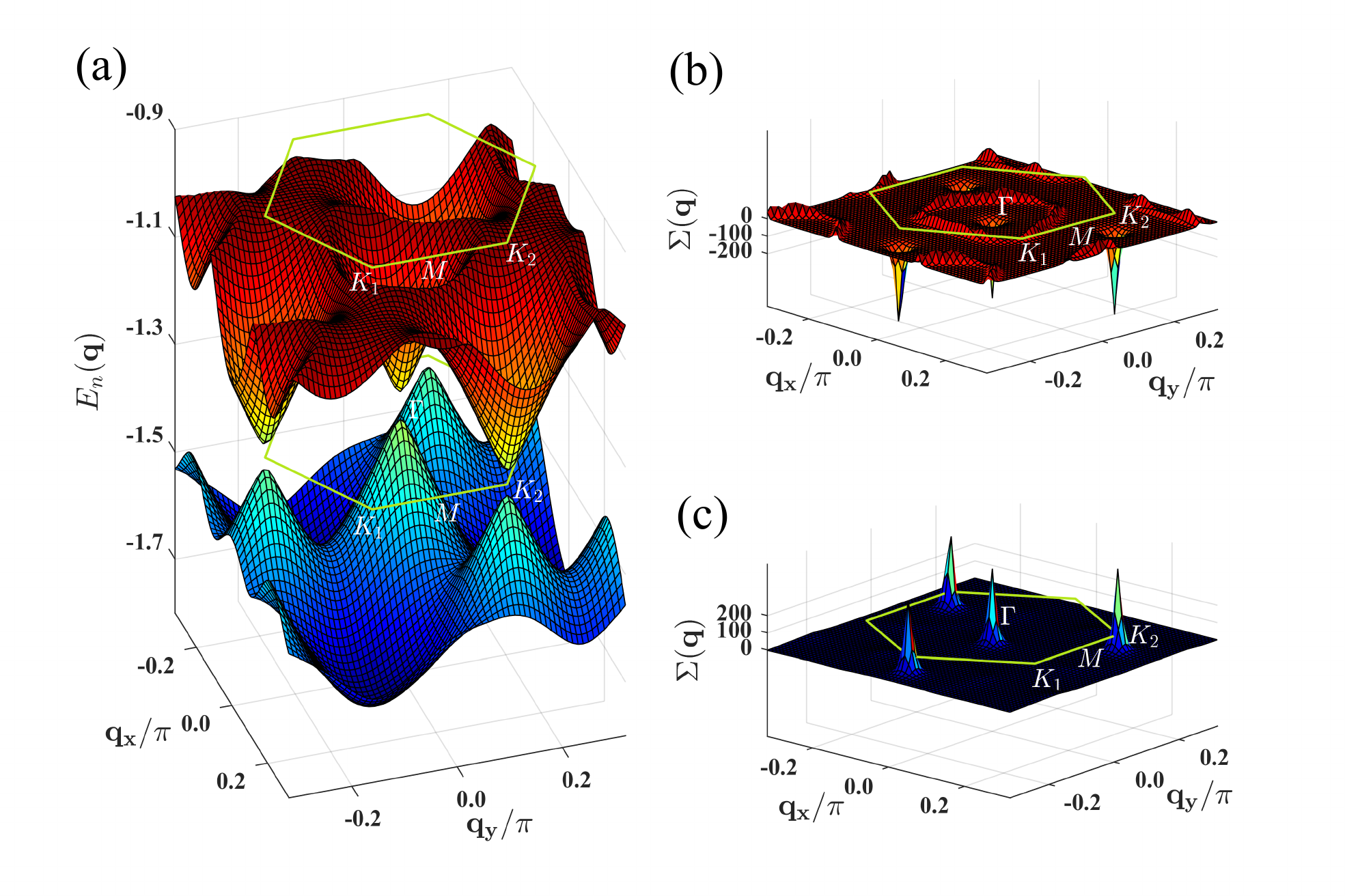}
\caption{(a) Band structures of band 4 and 5. (b) and (c) Berry curvatures of band 5 and 4, respectively.}
    \label{berry45}
\end{figure}
The Chern number of the first 9 bands presented in main text are 0, 0, 0, 1, 0, -2, 2, 1, 2.
Since there is no band gap between 7-8 band and between the 8-9 bands, we focus on the 4-5 and 6-7 bands.
Fig.~\ref{berry67} and Fig.~\ref{berry45} shows the full band structure and Berry curvature of selected bands.
There is a clear Dirac cone structure with certain large gap in the vicinity of $\Gamma$ point between 6 and 7 bands.
As a results, one can see that the Berry curvature mainly concentrates around $\Gamma$ point (Fig.~\ref{berry67}).
As a contrast, there are two Dirac cones at $\Gamma$ and $K_2$ points between 4 and 5 bands resulting in very narrow global gap,
and the Berry curvature dominates around $\Gamma$ and $K_2$ (Fig.~\ref{berry45}).

One may notice that the quantized Hall conductance shown in the main text is just the total Chern number when the chemical 
potential is inside the gap between the bands 6 and 7, i.e., $\sigma_{xy}=e^2/h\sum_{E_n< E_f}\mathcal{C}_n$, 
since the Berry curvature could be equivalently obtained through Kubo formula as well.
There is also a very narrow conductance platform between the bands 4 and 5, due to the tiny global gap there.

\begin{center}
\textsf{\textbf{Sec. V: Spin dynamics simulations}}
\end{center}
\setcounter{subsection}{5}
~Since the TmX state is found in the ground-state phase diagram, an important question to be answered is whether
we can obtain such state experimentally.
While Monte Carlo simulation is an elaborate and powerful metheod for equilibrium state problems,
the dynamics of magnetic properties like: domain-wall motion, Hysteresis loop, pining-depinning problem, etc.,
are usually treated with Landau-Lifshitz-Gilbert equation\cite{Landau1935, Gilbert2004}. To this end, we perform atomistic spin-dynamics 
simulations on field-cooling process of TmX state. Here, we choose again the representative point $\theta=0.515\pi, \phi=0.68\pi$.

The Landau-Lifshitz-Gilbert (LLG) equation is numerically solved in the \textbf{xyz} coordinate system:
\begin{eqnarray*}
	\frac{\partial \mathbf{S}_i}{\partial t} = -\frac{1}{\hbar}\left(\mathbf{S}_i \times \mathbf{h}^i_{eff} \right)
	-\frac{\alpha'}{\hbar}\mathbf{S}_i\times \left(\frac{\partial \mathbf{S}_i}{\partial t} \right),
\end{eqnarray*}
with
\begin{eqnarray*}
	\mathbf{h}^i_{eff} = -\sum_{\langle  i,j \rangle \in \pmb{\gamma}}\left [ KS_j^\gamma \pmb{\gamma}
	+\Gamma\left ( S_j^\beta \pmb{\alpha} + S_j^\alpha \pmb{\beta}\right) + 2A \left( \mathbf{S}_i \cdot \mathbf{c}\right)\mathbf{c}+ \mathbf{h}_{ext}^c  \right]
\end{eqnarray*}
the effective field felt by $\mathbf{S}_i$,  $\alpha'$ a dimensionless damping factor and $\mathbf{h}_{ext}^c$
an external magnetic field along $\mathbf{c}$ direction.
For finite temperatures, a stochastic fluctuation field $\mathbf{h}^{fl}$ is then added to the effective field to act as the thermal noise,
which should have a zero average value and be uncorrelated in spin component, space and time:
\begin{eqnarray*}
	\langle \mathbf{h}^{fl}_{i,\alpha}(0)\cdot \mathbf{h}^{fl}_{j,\beta}(t)\rangle = \epsilon^2 \delta_{ij} \delta_{\alpha\beta}\delta(t),
\end{eqnarray*}
with $\epsilon^2 = 2\alpha'k_B\mathcal{T}$ and $\mathcal{T}$ the temperature determining the strength of thermal fluctuations.
In our simulation, we set $k_B = \hbar =1$ (a dimensionless system),
$\alpha'=0.01$ (different values of $\alpha'$ have also been tested) and solve the LLG equation using the fourth-order Runge–Kutta method on a $288\times288\times2$ honeycomb lattice. $6\times10^6$ iterations are performed between every two successive temperatures.

\begin{figure*}[!tb]
	\centering
	\begin{tabular}{c}
		\begin{minipage}[c]{0.92\textwidth}
			\includegraphics[width=\textwidth, clip]{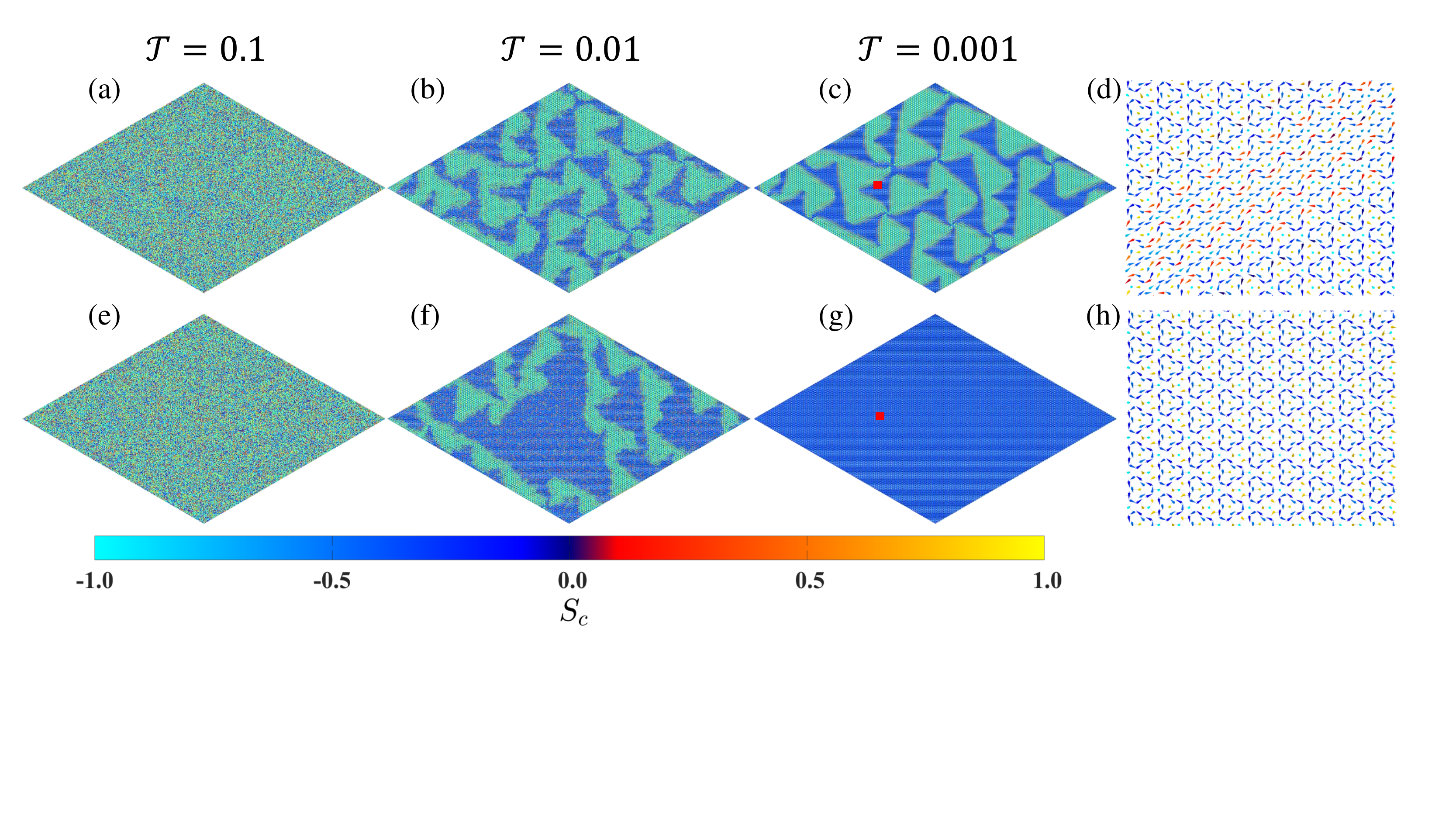}
		\end{minipage}
	\end{tabular}
	\vspace{-1.4cm}
	\caption{Snapshots of spin configurations obtained in field-cooling simulations with $H_{ext}^c = -0.25$ through atomistic spin dynamics.
		The parameters are $\theta=0.515\pi,\ \phi=0.68\pi$.
		{(a)}-{(c)}: Snapshots at the temperature $\mathcal{T} = 0.1,\ 0.01$ and $0.001$, respectively.
		The islands and the sea in both {(b)} and {(c)} are occupied by the TmX states but their cores are on the different sublattice.
		{(d)}: The zoom-in of the spin configuration of the red square in the panel~{(c)}.
		{(e)}-{(h)}: Same parameters to the panels~{(a)}-{(d)} except that the coefficient of the single-ion anisotropy on one
		sublattice is changed from $A$ to $0.9A$ to lift the degeneracy.}
	\label{FigSM-5a}
\end{figure*}
\begin{figure*}[!tb]
\centering
\begin{tabular}{c}
   \begin{minipage}[c]{0.92\textwidth}
      \includegraphics[width=\textwidth, clip]{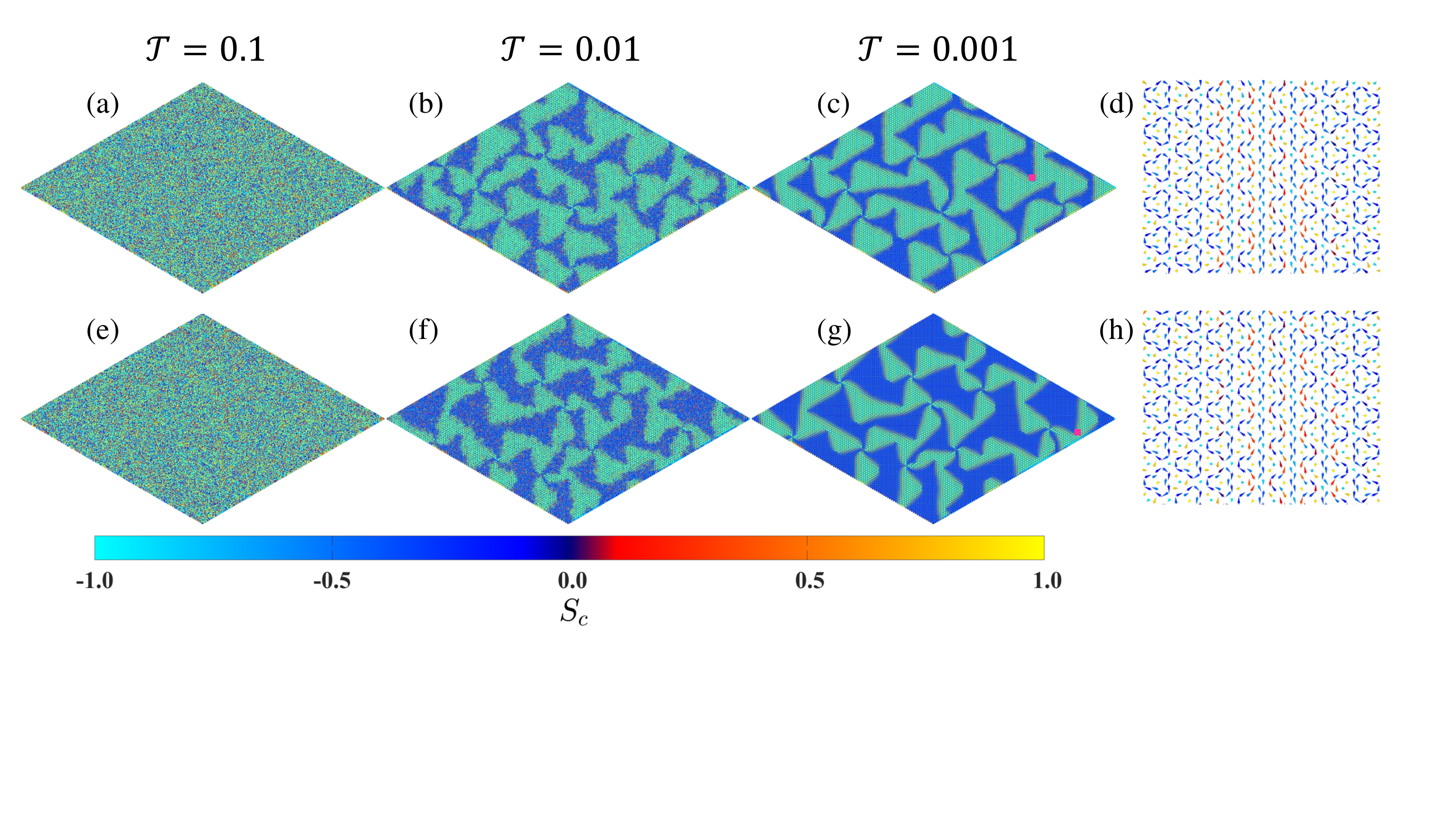}
   \end{minipage}
\end{tabular}
\vspace{-1.4cm}
\caption{Same as those in Fig.~\ref{FigSM-5a} but with a different initial random configuration.}
\label{FigSM-5}
\end{figure*}
As we have already noticed, the TmX state is degenerate but with an opposite $Q$ under the transformation $\bf{S}\rightarrow -\bf{S}$.
These two states might appear simultaneously and annihilate each other when starting from a random initial state.
Such a degeneracy can be easily lifted by applying an external magnetic field $h_{ext}^{c}$ along the $c$-axis.
As shown in Fig.~\ref{FigSM-5a}{(a)-(c)}, one can already find the TmX state when the temperature $\mathcal{T}=0.01$.
As $\mathcal{T}$ decreases, the TmX state becomes more clear, suggesting the field cooling is indeed
effective to obtain a stable TmX state~(Supplementary Movie Sabc). However, one may notice that there are two kinds of TmX states with their
core spins on different sublattices, say, a and b, which are separated by narrow domains~(Fig.~\ref{FigSM-5a}{(c)} and Fig.~\ref{FigSM-5a}{(d)}).
Such a phenomenon arises from the bipartite nature of the honeycomb lattice.
To eliminate such domains, it is necessary to lift the degeneracy further. In our numerical simulations,
a simple strategy is to change $A$ on one sublattice, say a, to $\alpha A$ with $\alpha$ equal to, for example, $0.9$.
We want to mention that in experiments one can break the symmetry of the lattice by applying strain.
The corresponding results are shown in Fig.~\ref{FigSM-5a}{(e)-(h)}. Though there are still domains at $\mathcal{T}=0.01$, one can find a
nearly perfect TmX state at $\mathcal{T}=0.001$~(Supplementary Movie Sefg).
We want to stress that lifting such degeneracy
is definitely helpful to eliminate domains but the effect depends on $\alpha$ as well as the initial configuration.
As shown in Fig.~\ref{FigSM-5}, the parameters and simulation details in Fig.~\ref{FigSM-5} and  Fig.~\ref{FigSM-5a} are 
the same except that the random initial configuration is different. However, in Fig.~\ref{FigSM-5}(g) there are still domains.
Even though, the effect of lifting the degeneracy can be demonstrated by counting the number of merons with their cores on the different sublattice, say $N_a$ or $N_b$. In Fig.~\ref{FigSM-5}{(c)} where $\mathcal{T}=0.001$, $N_a:N_b = 7524:7328$, which is very close to $1:1$. 
This ratio again evidences the degeneracy of the TmX state.
As we change the coefficient of the single-ion anisotropy $A$ on the sublattice a to $0.9A$ in Fig.~\ref{FigSM-5}{(e)-(h)},
at $\mathcal{T}=0.001$ (Fig.~\ref{FigSM-5}(g)), the region of the energetically favored Tmx state is obviously increased, i.e., $N_a:N_b = 12337 : 3937$. One may naively expect that core spins on the $b$ sublattice is energetically favored because $\alpha<1.0$ and $A<0$. Actually, this is not true. Changing $\alpha$ simultaneously changes the spin directions of NNs and NNNs and one can not just compare the energy of the core spins. We confirm the core spins prefer staying on the $a$ sublattice by the Monte Carlo simulations.

\begin{center}
\textsf{\textbf{Sec. VI: Complete phase diagram, Spin configurations and static spin structure factors}}
\end{center}
\setcounter{subsection}{6}

In this section, we show the complete phase diagram, the spin configurations and the corresponding static spin structure factors
${{\mathcal{F}}_{\bf{q}}^{\xi\xi}} = \frac{1}{N}\sum_{ij}e^{i{\bf{q}}\cdot(\bf{R}_i-\bf{R}_j)}\langle{S_i^{{\xi}} S_j^{\xi}}\rangle$~($\xi = {\bf{a, b, c}}$).
\begin{figure}
    \includegraphics[width=\textwidth]{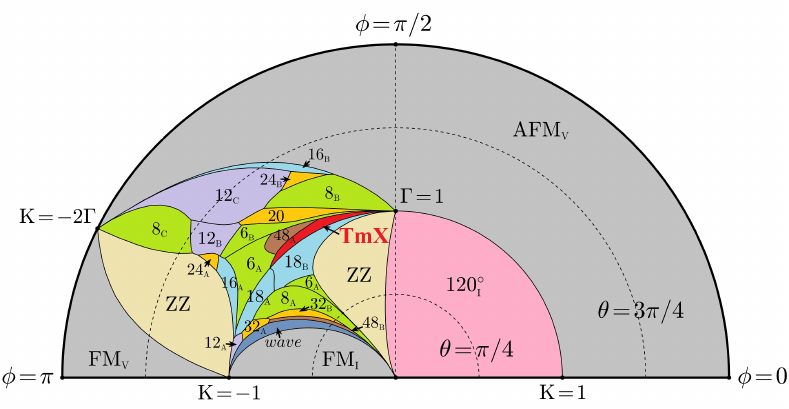}
	\caption{The complete phase diagram of the classical $\rm K\Gamma{A}$ model. The parameters $K, \Gamma$ and $A$ are parameterized
by $\left(\theta, \phi\right)$, where $\theta\in(0,\pi), \phi\in[0,\pi)$ are plotted as the radial distance and polar angle,
respectively. FM, AFM, ZZ, $\rm 120^\circ$ and $wave$ mean ferromagnetism, antiferromagnetism,
zigzag, $\rm 120^\circ$ and wave-like magnetic orders, respectively.
The subscript $\mathbf{I}$ and $\mathbf{V}$ indicate the spins lie in the plane and point out of the plane, respectively.
The numbers 6, 8, 12, 16, 18, 20, 24, 32 and 48 mark the number of spins in one magnetic unit cell and the subscripts A, B, C and D
are used to distinguish those phases marked by the same number. The TmX phase is discussed in the main text.
}
\label{PDM}	
\end{figure}

The complete phase diagram of the $\rm K\Gamma{A}$ model is shown in Fig.~\ref{PDM} as a semicircle plot.
The radial distance represents $\theta$ and the polar angle represents $\phi$.
In the region $\pi/2<\phi<\pi$, the model is highly frustrated and a large number of phases are found.
The number in the phase diagram marks the number of spins in one magnetic unit cell and the
subscripts A, B, C and D are used to distinguish different phases marked by the same number.
The phase transitions of all the ordered phases are of the first order except that the transitions to the \textit{wave} phase is unclear. 
Note that on the line $\theta=\pi$, where $K=0$ and $\Gamma=0$, the Hamiltonian is decoupled and then become trivial. From the symmetry, it is ready to know 
that the line $\phi=\pi$, or equivalently $\phi=0$, is the phase transition line.

\begin{figure*}[!h]
	\begin{tabular}{cc}
		\hspace{-.25in}
		\begin{minipage}[t]{0.45\textwidth}
			%\vspace{-3.25cm}
			\mbox{
				     \includegraphics[width=0.5\textwidth]{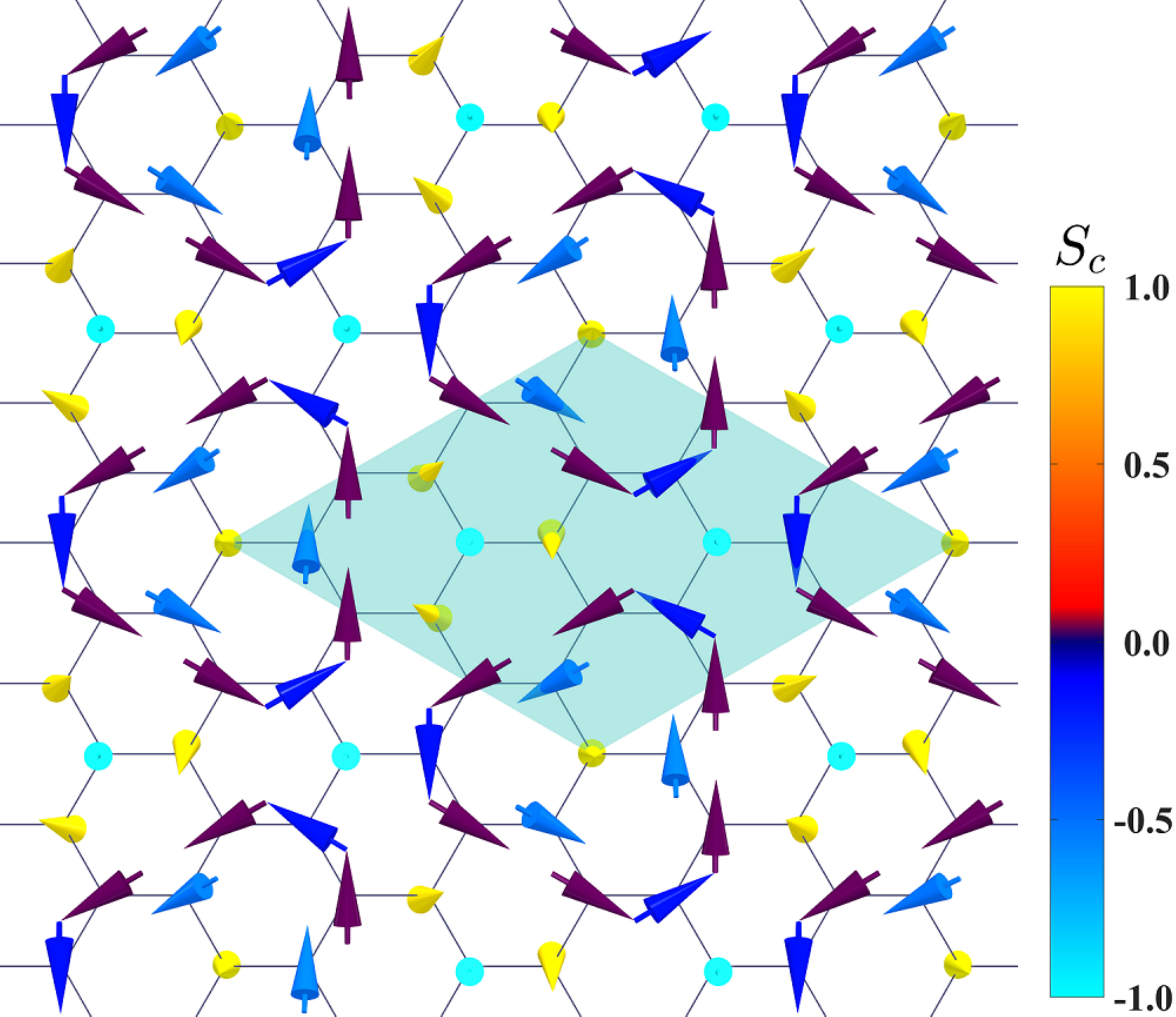}
				     \includegraphics[width=0.5\textwidth]{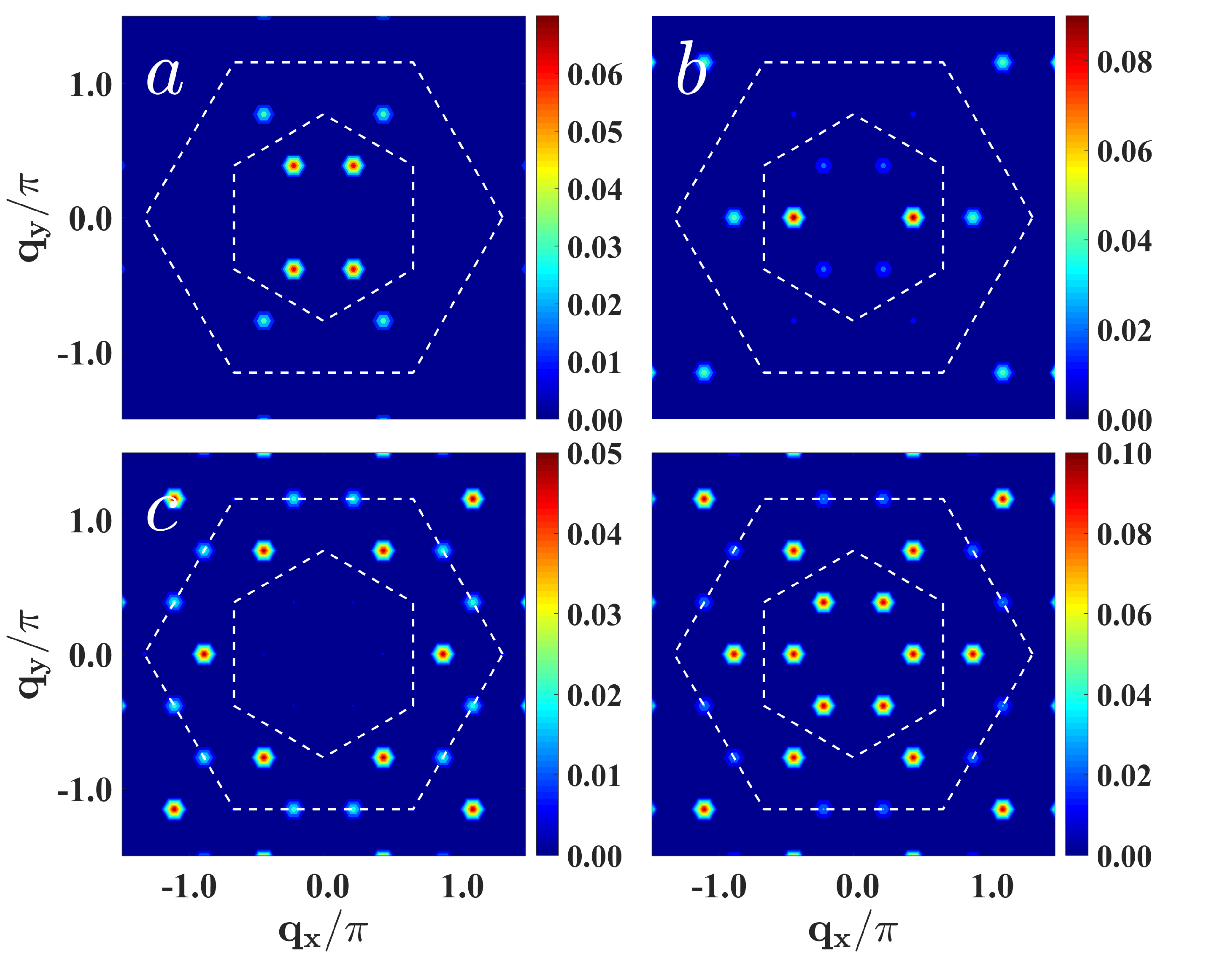}
			     }
			\caption{\textcolor{red}{TmX} phase ($\theta=0.515\pi$, $\phi=0.68\pi$)}
			\label{FigSMP-1}
		\end{minipage}
		
		\hspace{.15in}
		\begin{minipage}[t]{0.45\textwidth}
			%   \vspace{-3.25cm}
		    \mbox{
				      \includegraphics[width=0.5\textwidth]{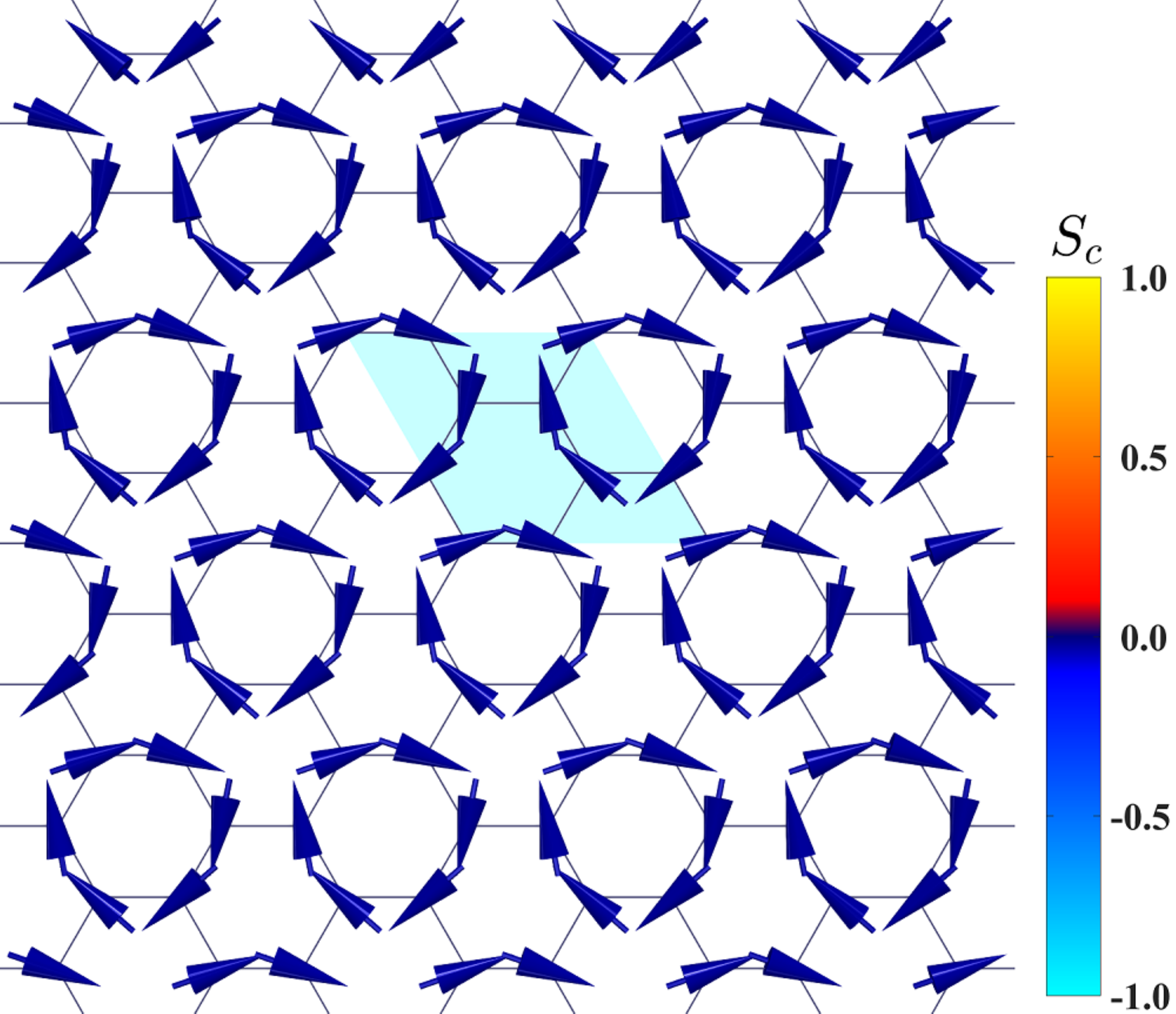}
				      \includegraphics[width=0.5\textwidth]{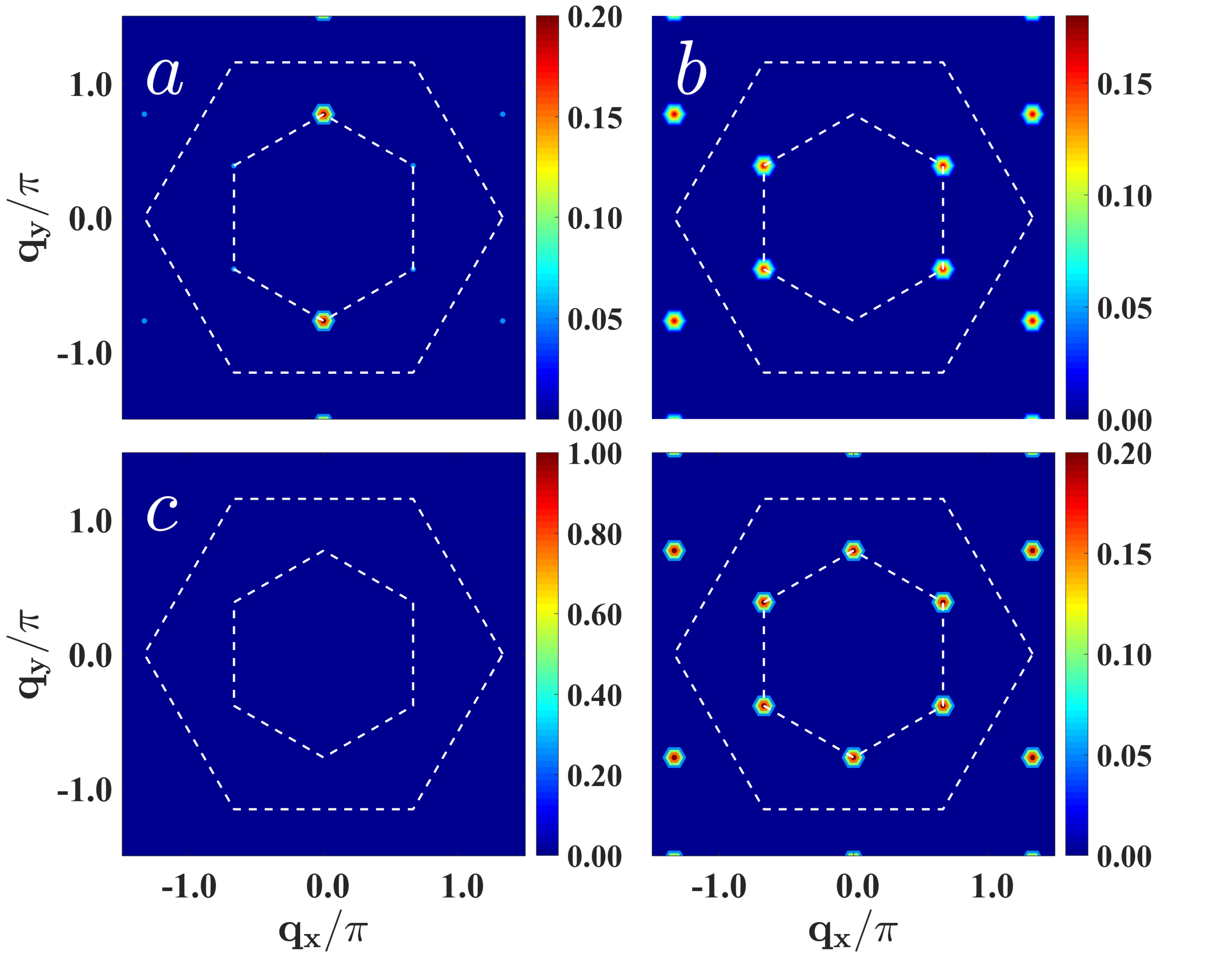}
			     }
			\caption{\textcolor{red}{$\rm 120^\circ_I$} phase ($\theta=0.49\pi$, $\phi=0.25\pi$)}
			\label{FigSMP-2}
		\end{minipage}
	\end{tabular}
\end{figure*}

\begin{figure*}[!h]
	\begin{tabular}{cc}
		\hspace{-.25in}
		\begin{minipage}[t]{0.45\textwidth}
			%\vspace{-3.25cm}
		     \mbox{
		            	\includegraphics[width=0.5\textwidth]{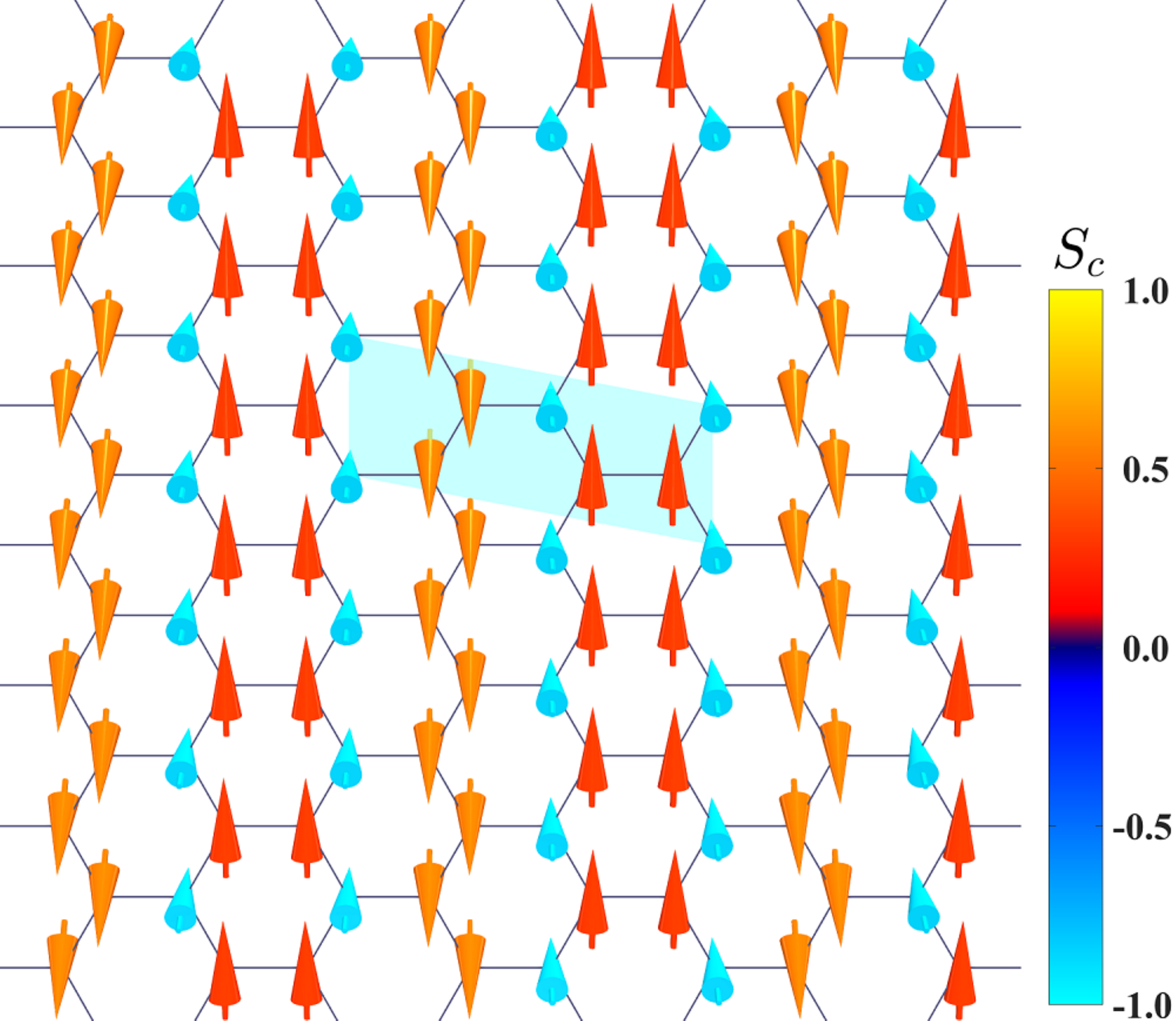}
		            	\includegraphics[width=0.5\textwidth]{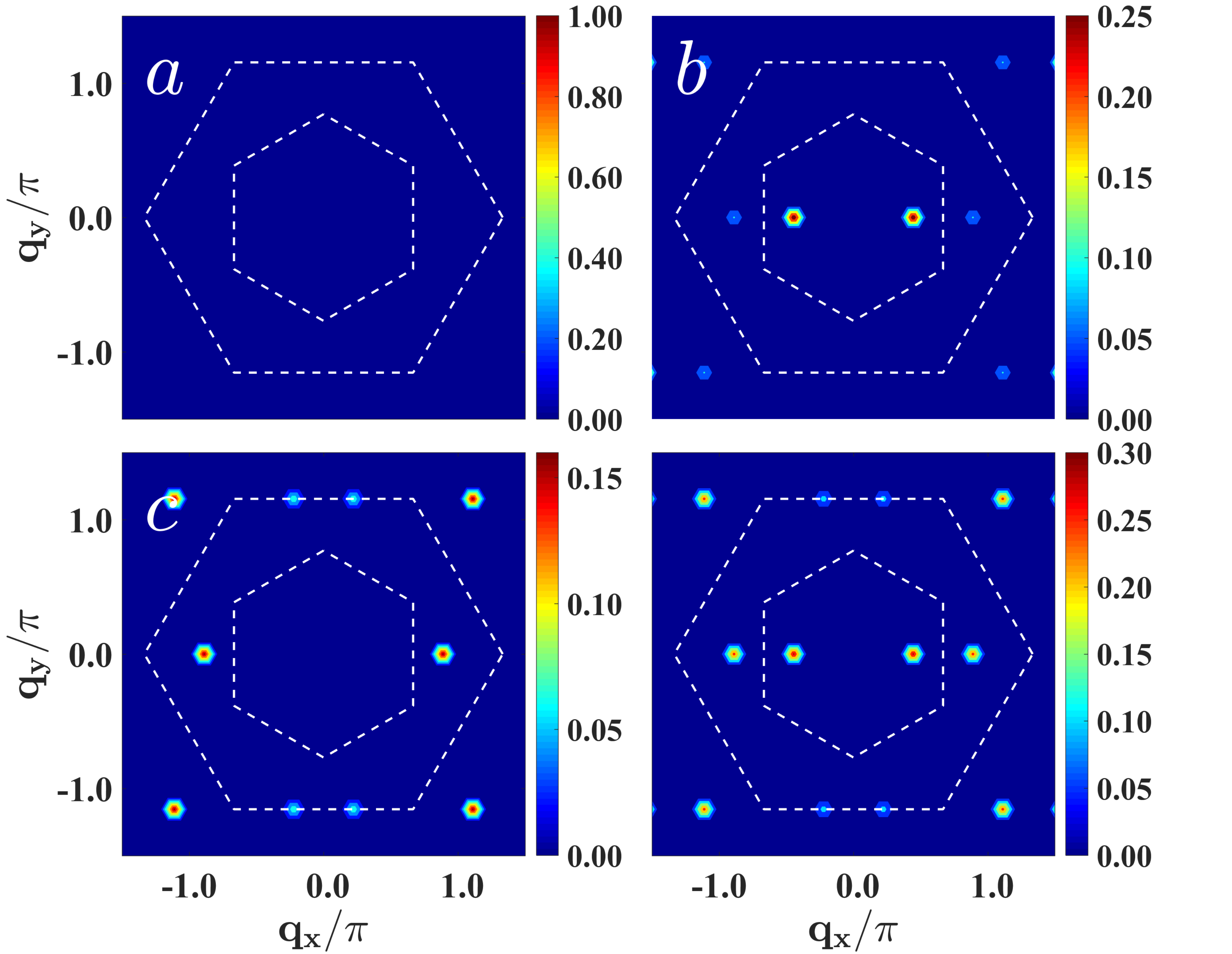}
		          }
			\caption{\textcolor{red}{$\rm 6_A$} phase ($\theta=0.515\pi$, $\phi=0.8\pi$)}
		    \label{FigSMP-3}
		\end{minipage}
		
		\hspace{.15in}
		\begin{minipage}[t]{0.45\textwidth}
			%\vspace{-3.25cm}
		       	\mbox{
		       	\includegraphics[width=0.5\textwidth]{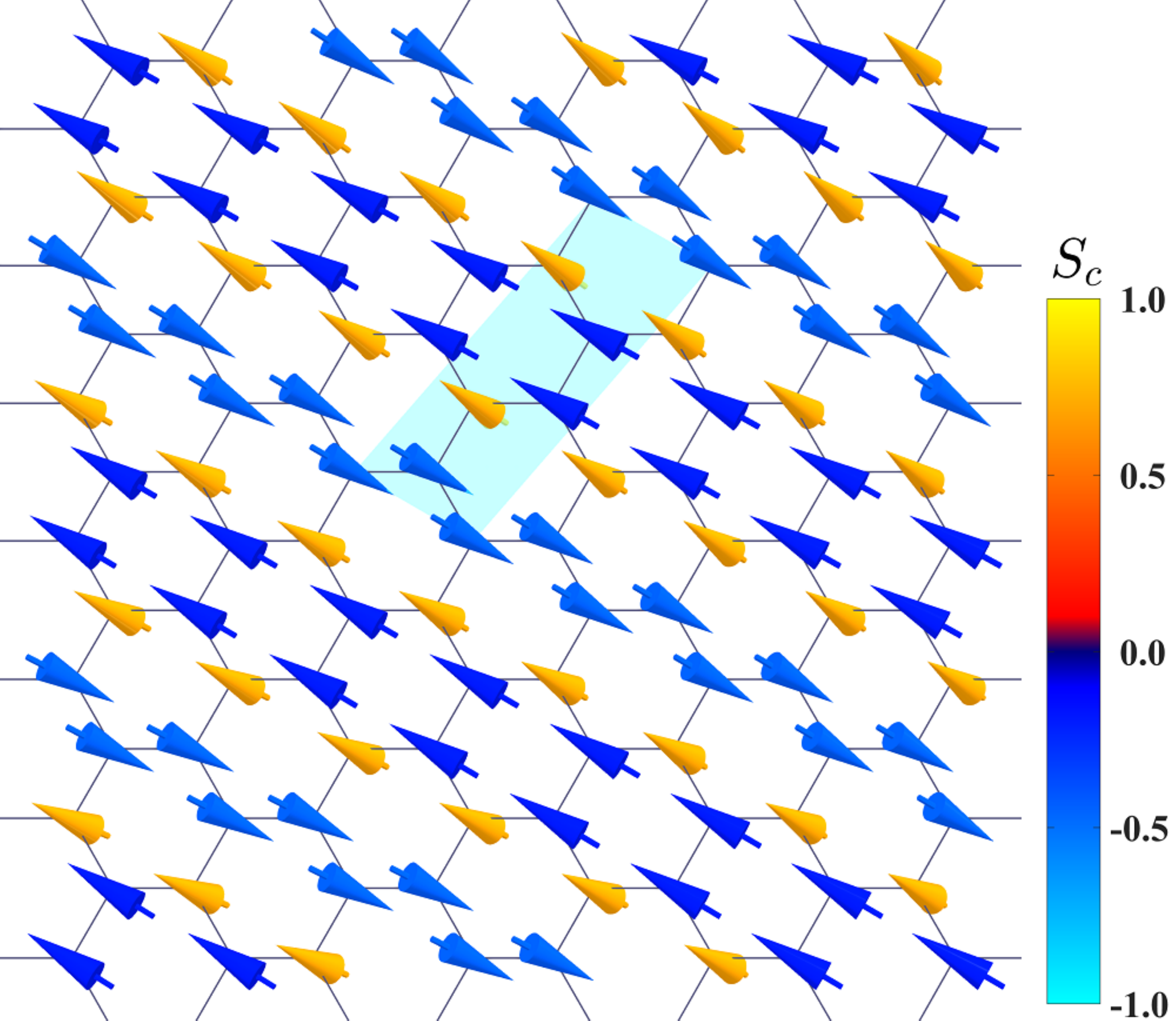}
		       	\includegraphics[width=0.5\textwidth]{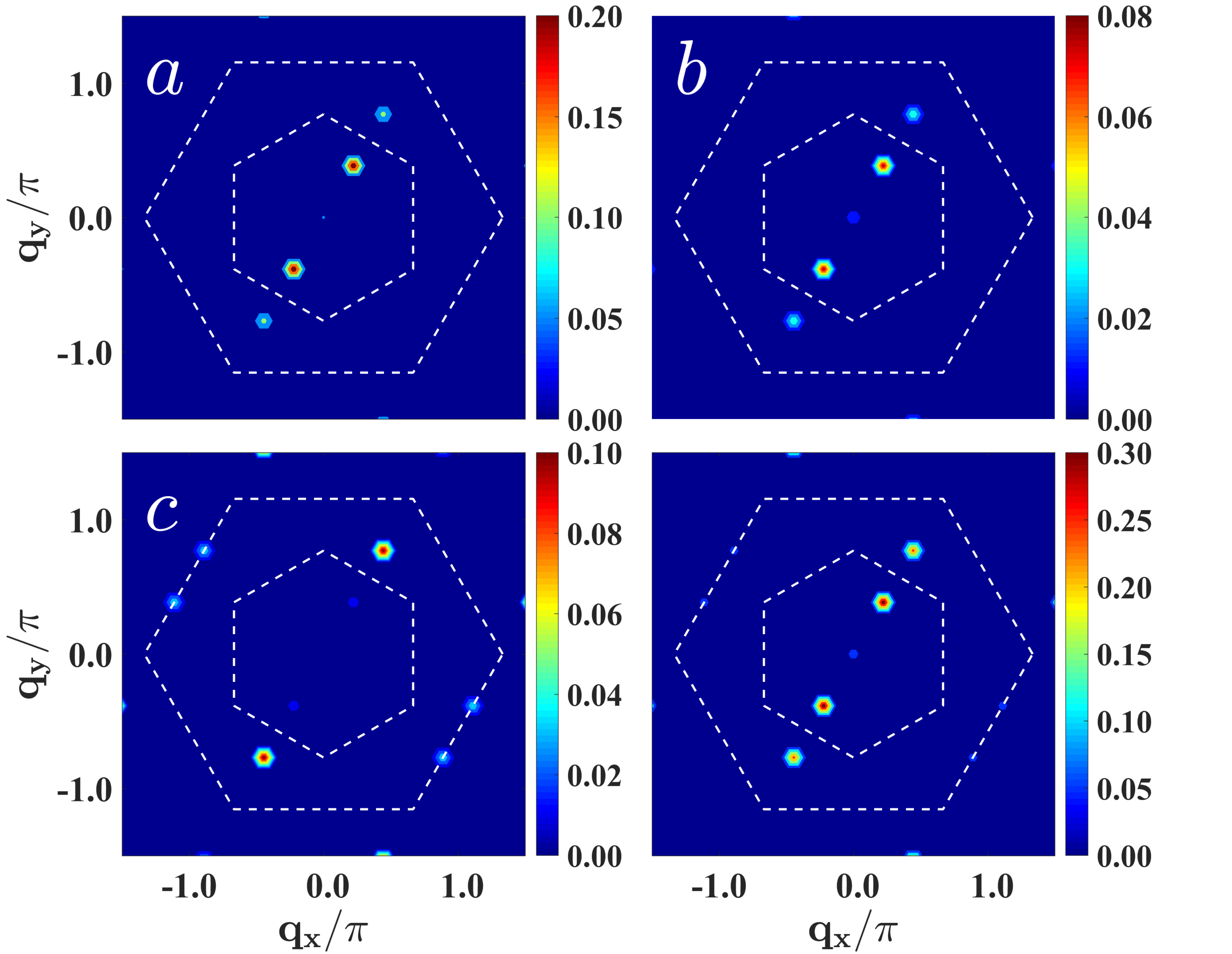}
		       }
			\caption{\textcolor{red}{$\rm 6_A$} phase ($\theta=0.4\pi$, $\phi=0.75\pi$)}
		\label{FigSMP-4}
		\end{minipage}
	\end{tabular}
\end{figure*}

\begin{figure*}[!h]
	\begin{tabular}{cc}
		\hspace{-.25in}
		\begin{minipage}[t]{0.45\textwidth}
			%\vspace{-3.25cm}
		   \mbox{
		         	 \includegraphics[width=0.5\textwidth]{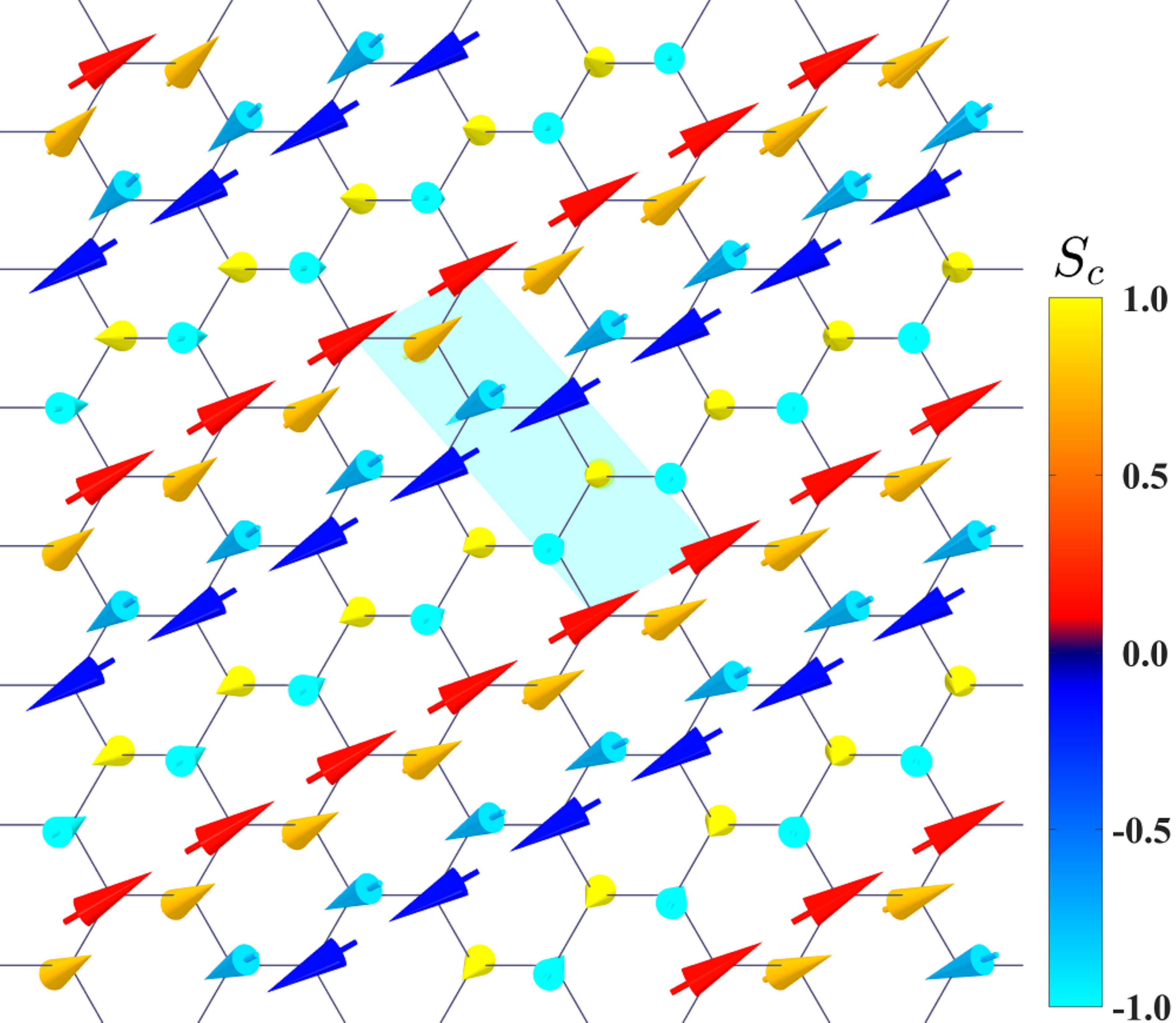}
		         	 \includegraphics[width=0.5\textwidth]{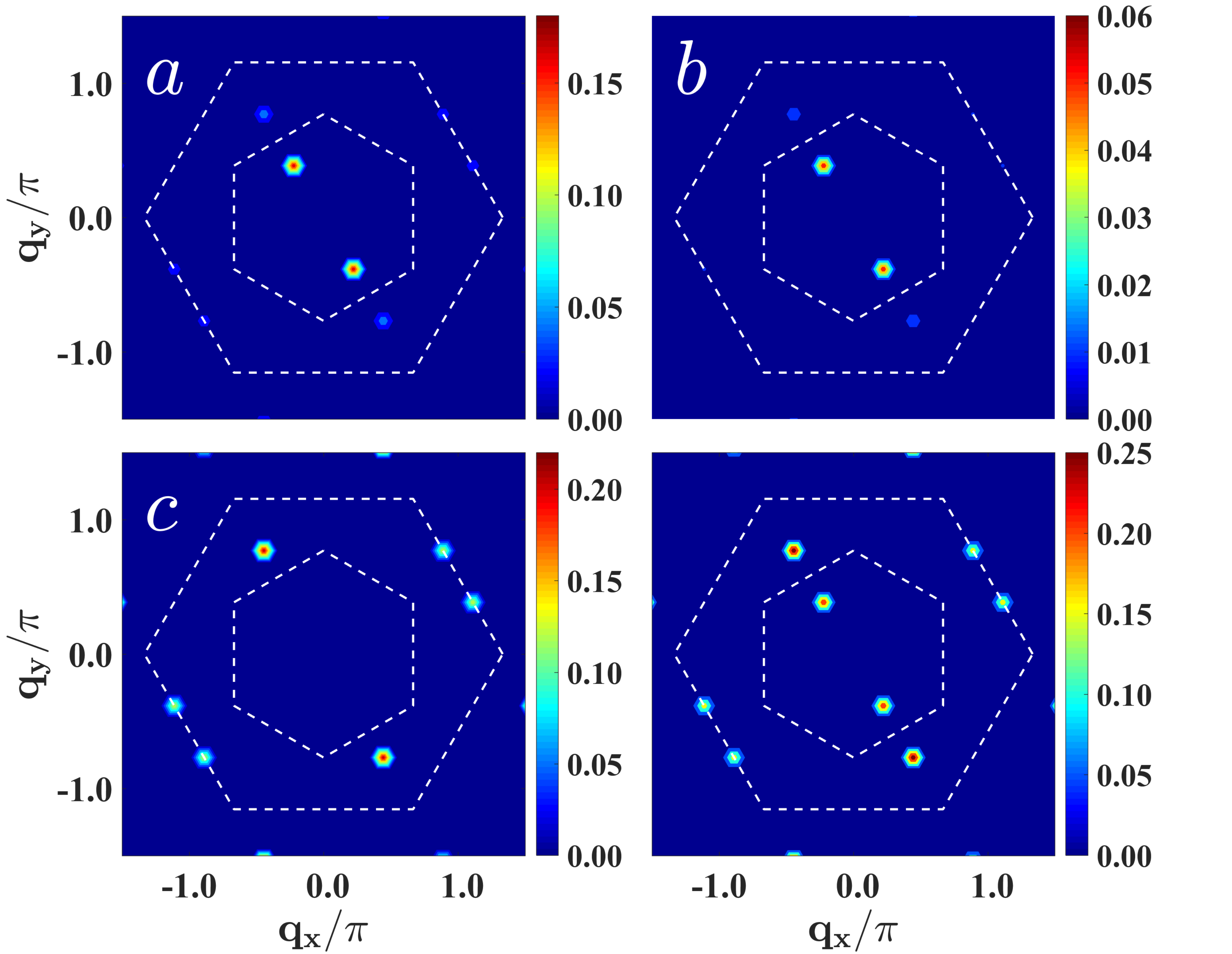}
		        }
			\caption{\textcolor{red}{$\rm 6_B$} phase ($\theta=0.625\pi$, $\phi=0.75\pi$)}
		    \label{FigSMP-5}
		\end{minipage}
		
		\hspace{.15in}
		\begin{minipage}[t]{0.45\textwidth}
			%   \vspace{-3.25cm}
		    \mbox{
		              \includegraphics[width=0.5\textwidth]{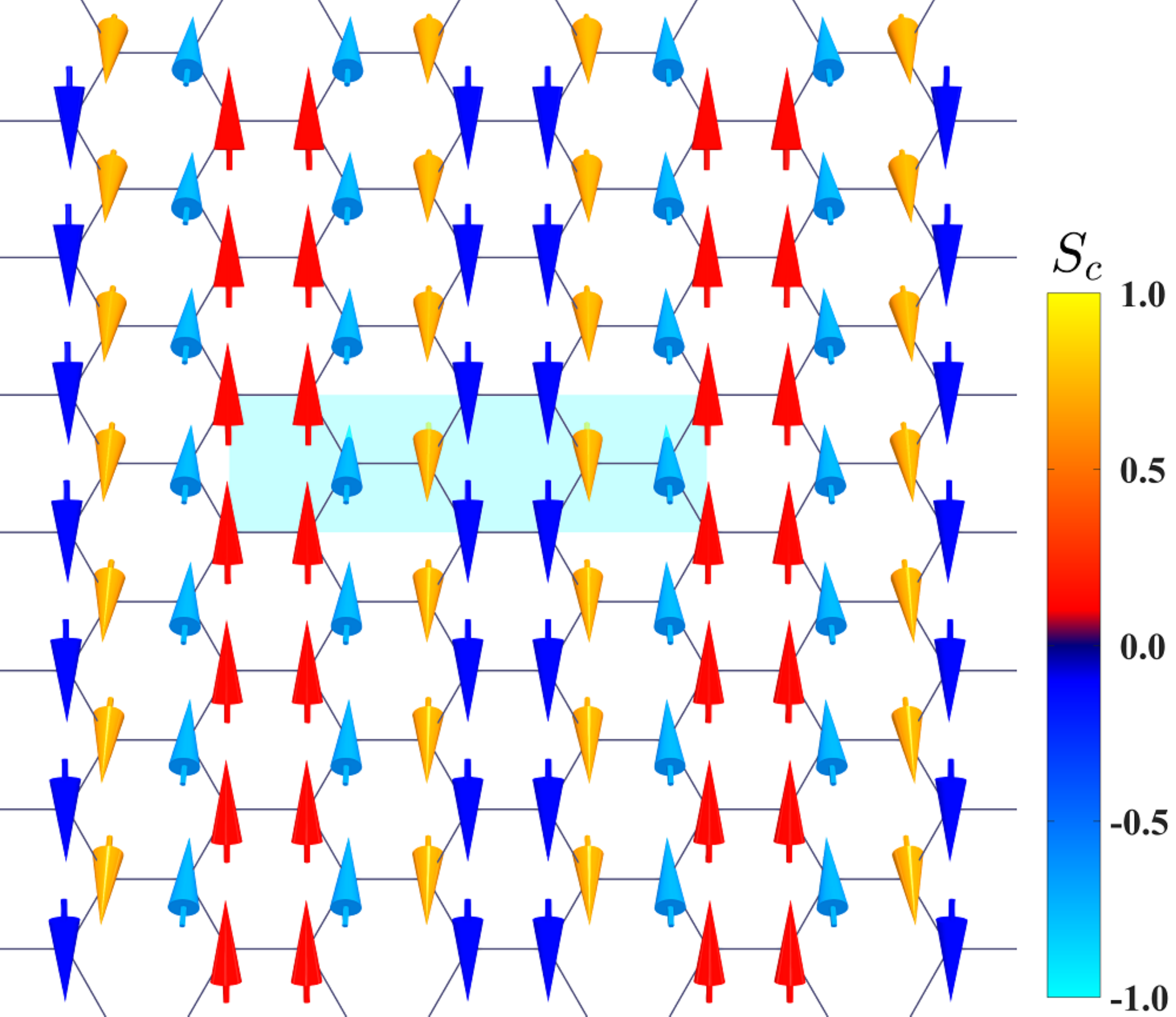}
		              \includegraphics[width=0.5\textwidth]{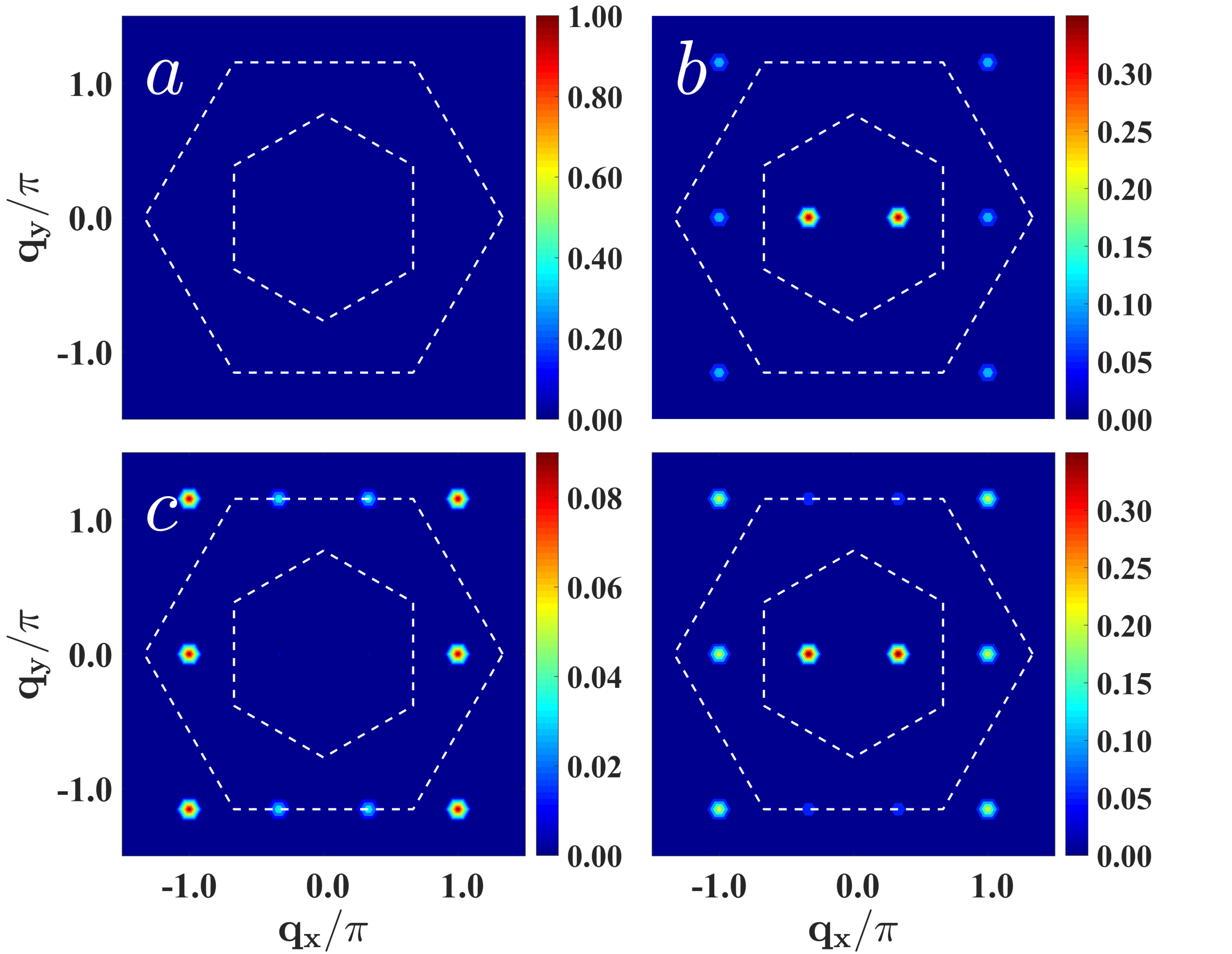}
		         }
			\caption{\textcolor{red}{$\rm 8_A$} phase ($\theta=0.4\pi$, $\phi=0.8\pi$)}
		    \label{FigSMP-6}
		\end{minipage}
	\end{tabular}
\end{figure*}

\begin{figure*}[!h]
	\begin{tabular}{cc}
		\hspace{-.25in}
		\begin{minipage}[t]{0.45\textwidth}
			%\vspace{-3.25cm}
		 		  \mbox{
		 	\includegraphics[width=0.5\textwidth]{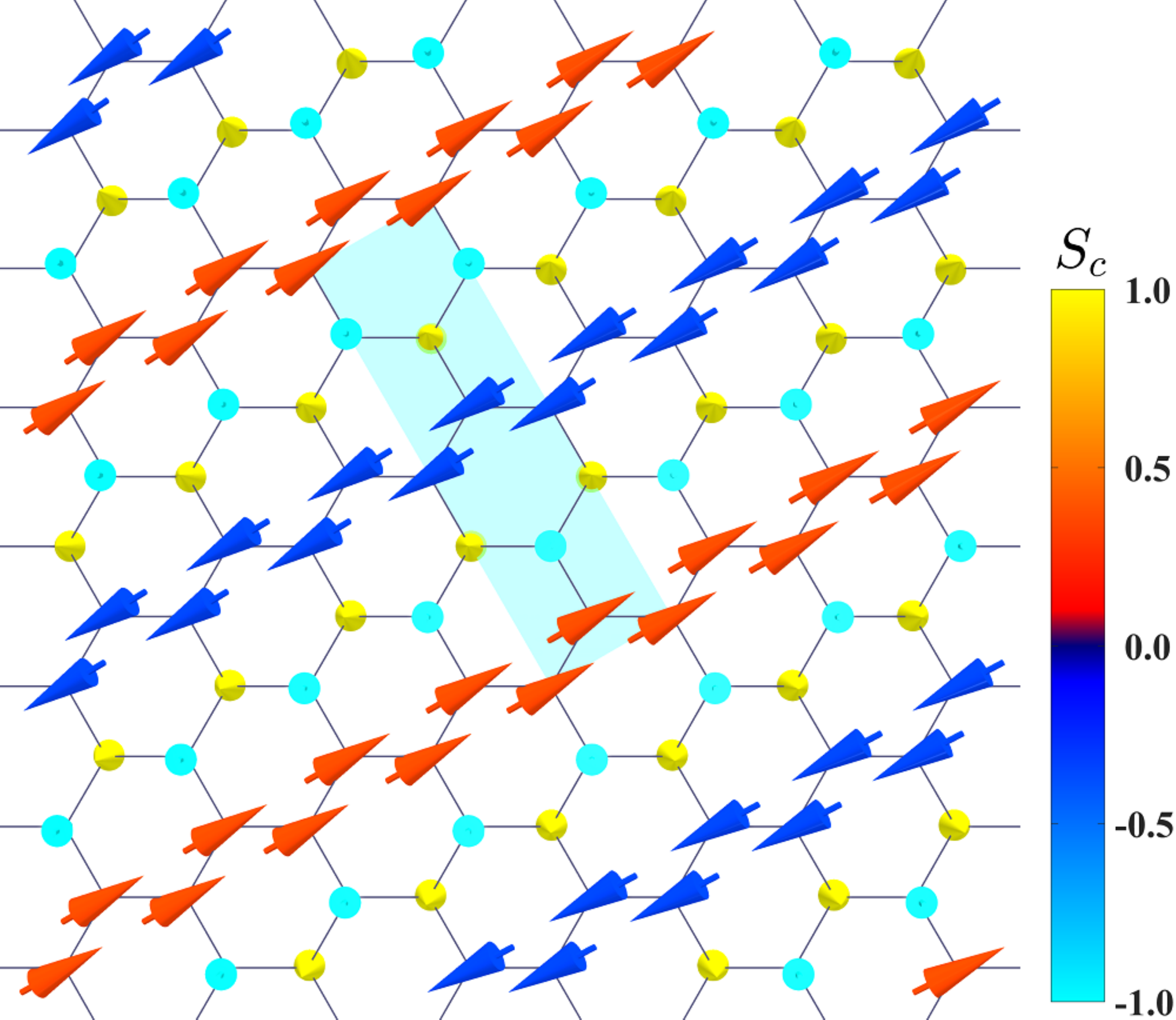}
		 	\includegraphics[width=0.5\textwidth]{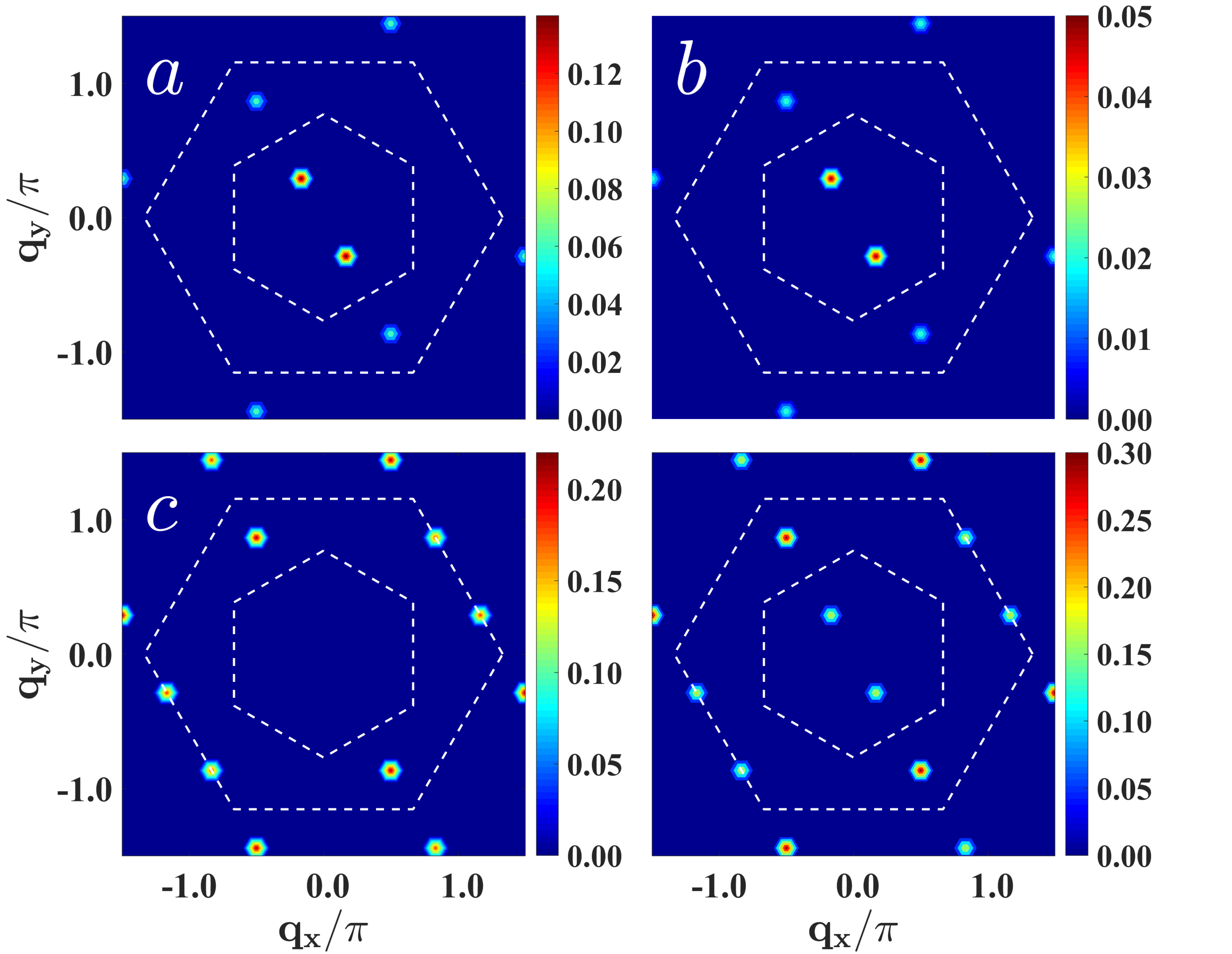}
		 }
		 \caption{\textcolor{red}{$\rm 8_B$} phase ($\theta=0.515\pi$, $\phi=0.54\pi$)}
		       \label{FigSMP-7}
		\end{minipage}
						
		\hspace{.15in}
		\begin{minipage}[t]{0.45\textwidth}
			%\vspace{-3.25cm}
			 \mbox{
				\includegraphics[width=0.5\textwidth]{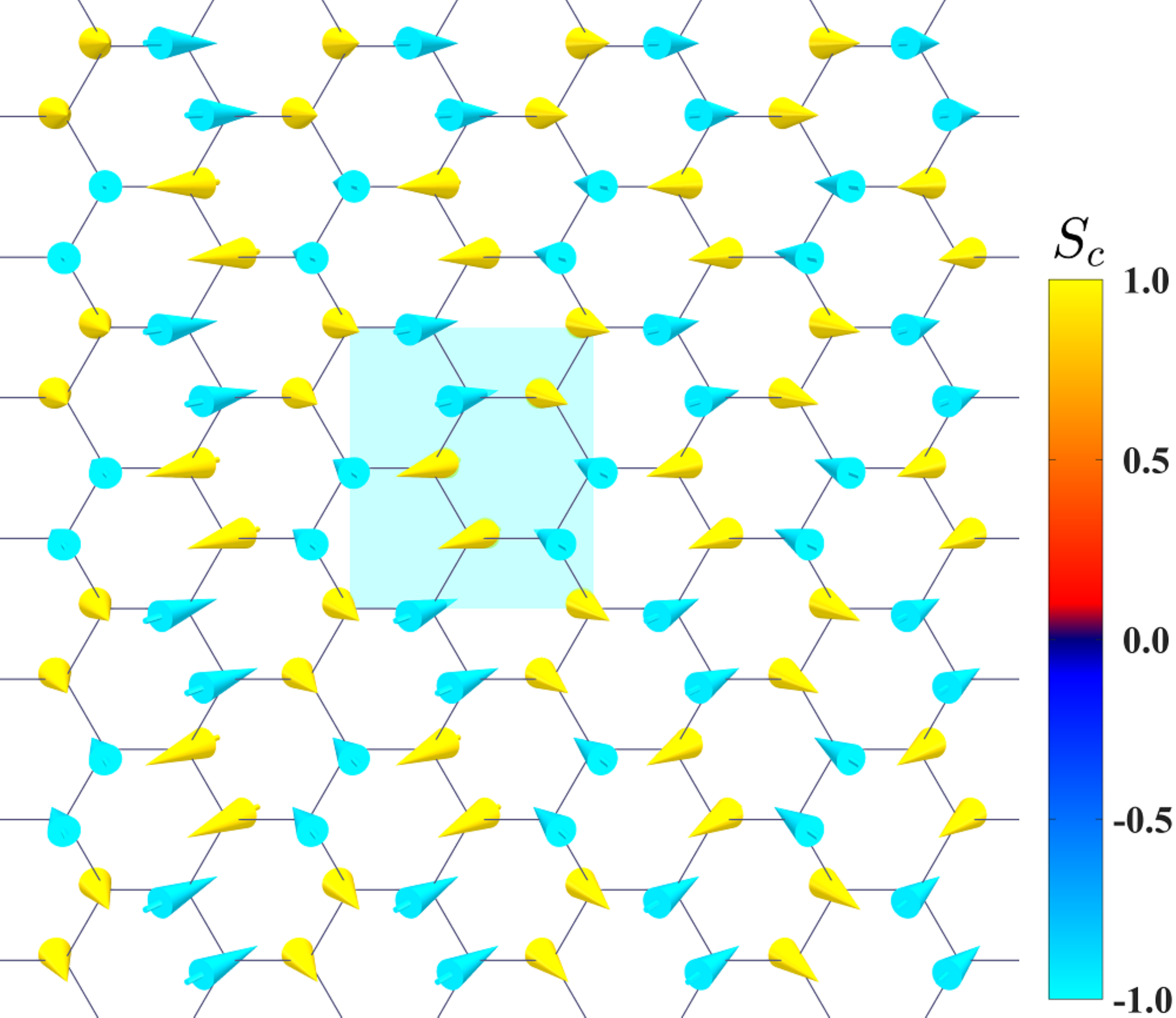}
				\includegraphics[width=0.5\textwidth]{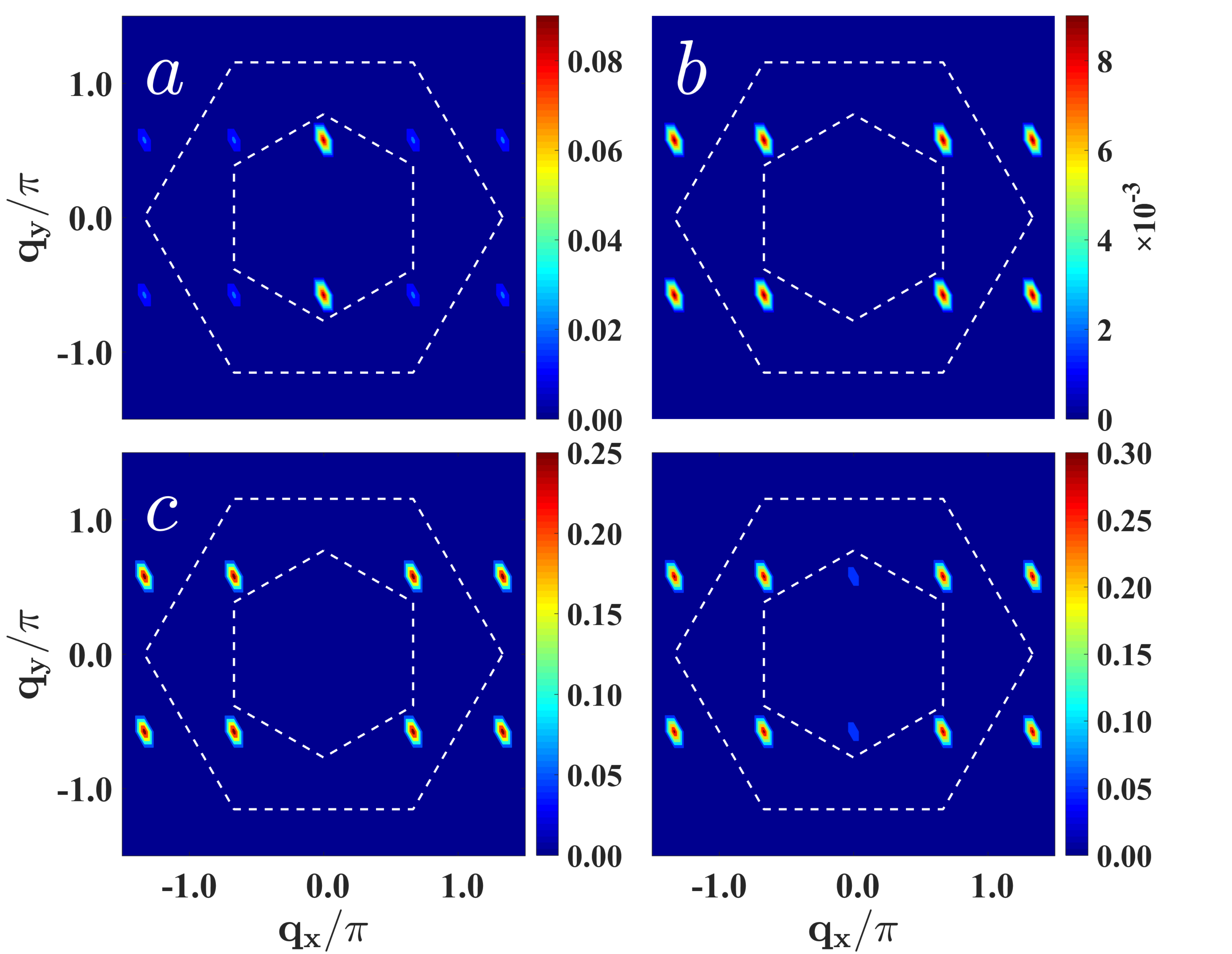}
			}
			\caption{\textcolor{red}{$\rm 8_C$} phase ($\theta=0.75\pi$, $\phi=0.815\pi$)}
		       \label{FigSMP-8}
		\end{minipage} 		
	\end{tabular}
\end{figure*}

\begin{figure*}[!h]
	\begin{tabular}{cc}
		\hspace{-.25in}
		\begin{minipage}[t]{0.45\textwidth}
			%\vspace{-3.25cm}
	      \mbox{
       	           \includegraphics[width=0.5\textwidth]{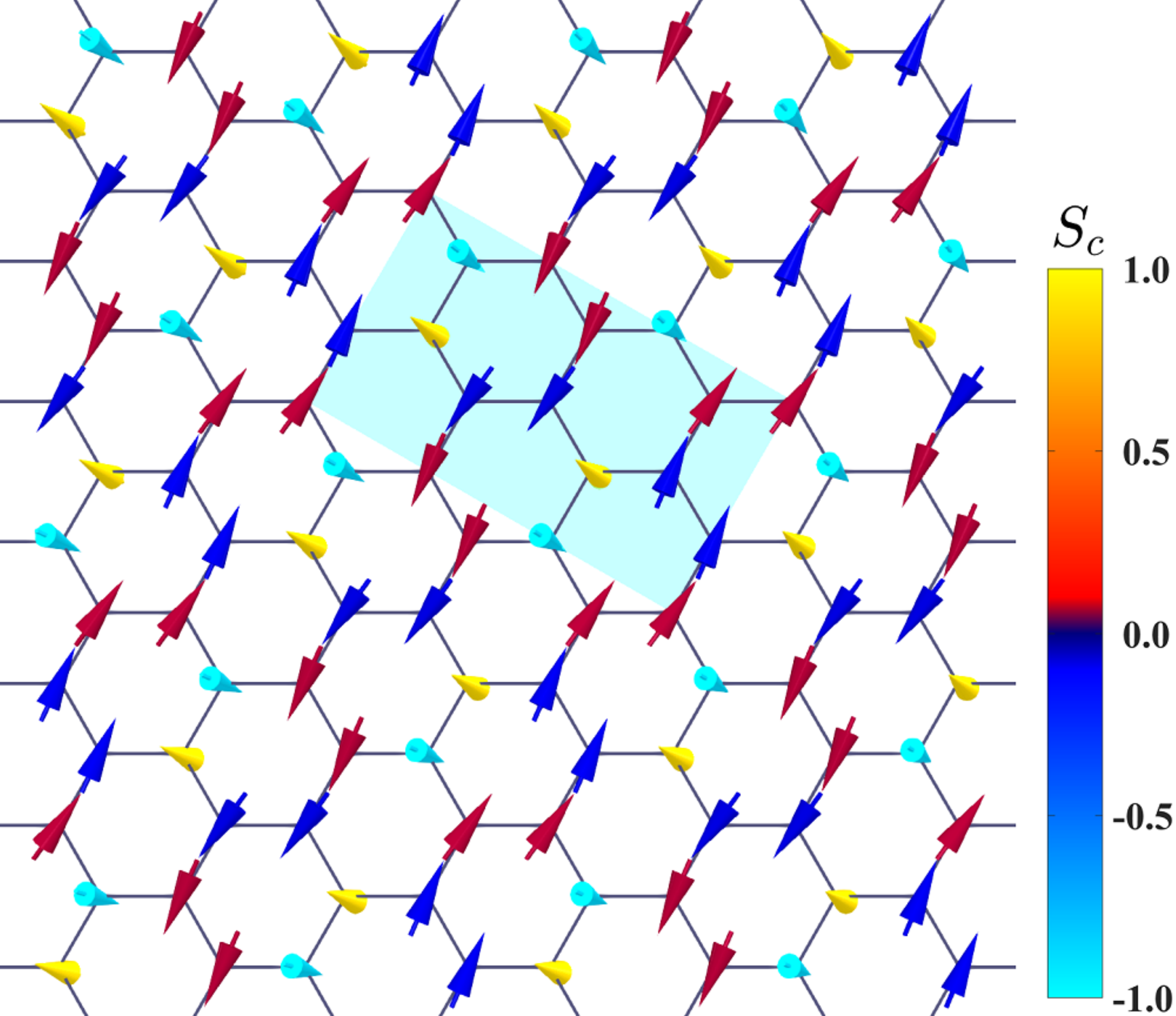}
       	           \includegraphics[width=0.5\textwidth]{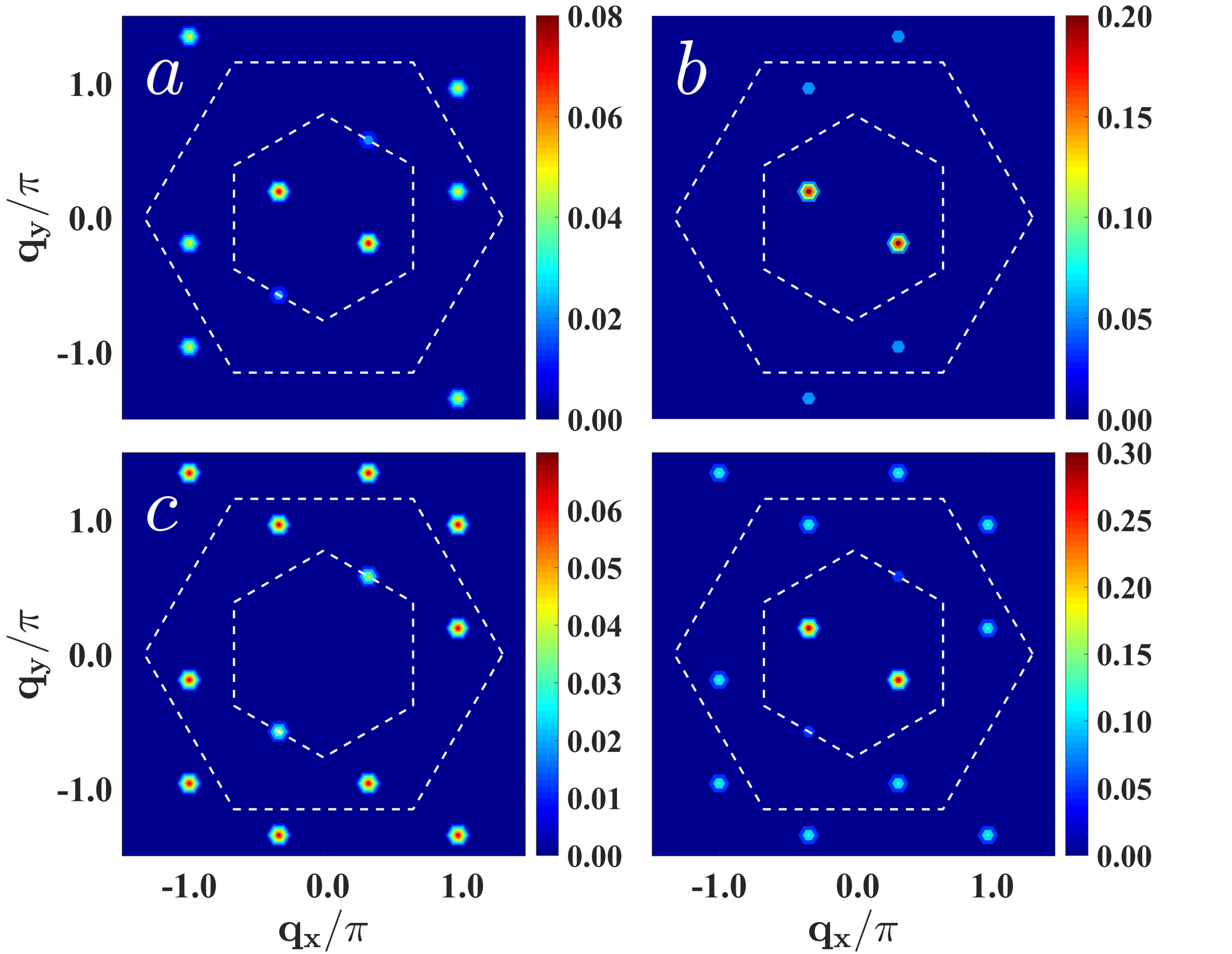}
               }
		\caption{\textcolor{red}{$\rm 12_A$} phase ($\theta=0.49\pi$, $\phi=0.94\pi$)}
		\label{FigSMP-9}
		\end{minipage}
		
		\hspace{.15in}
		\begin{minipage}[t]{0.45\textwidth}
			%   \vspace{-3.25cm}
		   \mbox{
		         	   \includegraphics[width=0.5\textwidth]{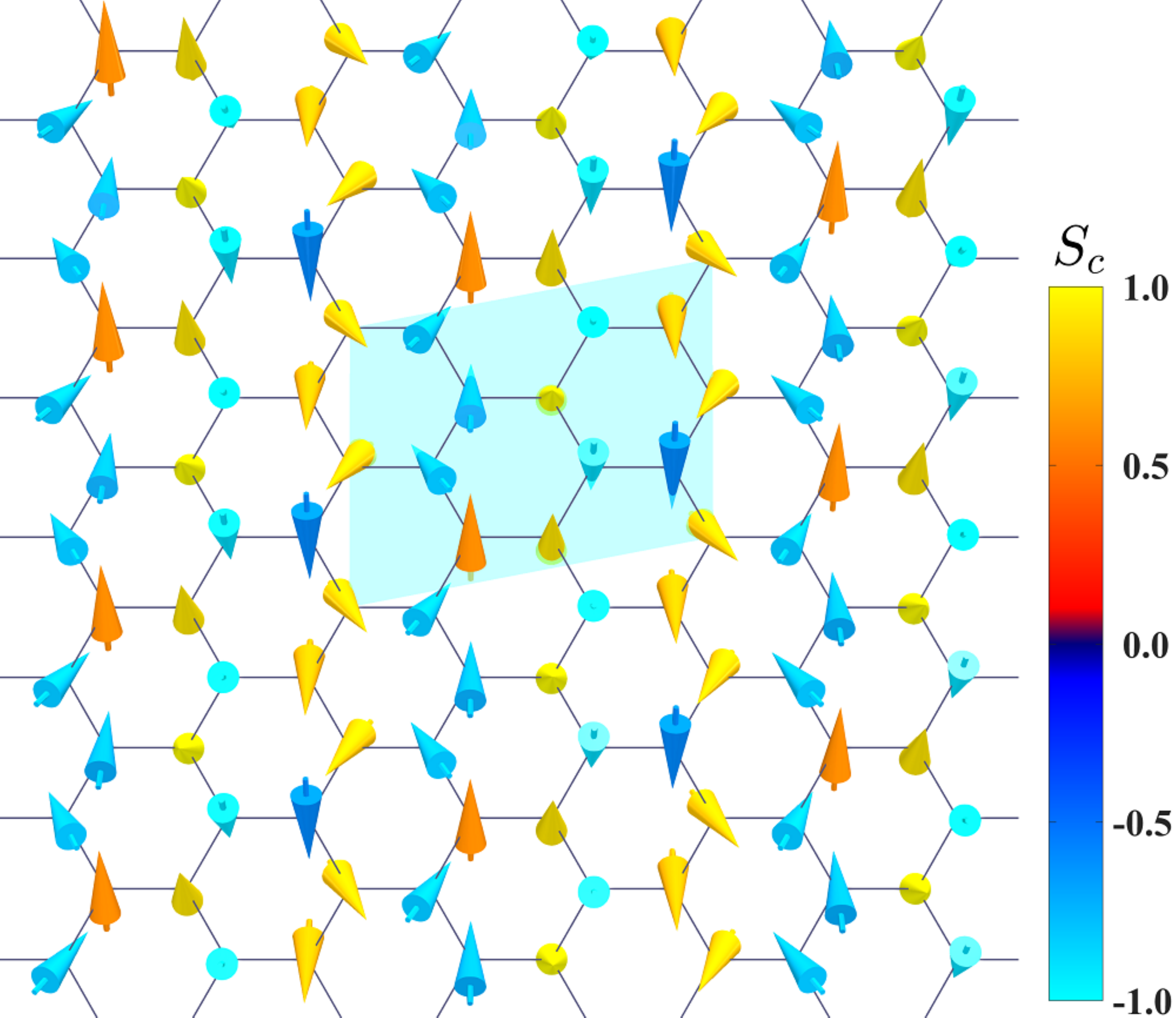}
		         	   \includegraphics[width=0.5\textwidth]{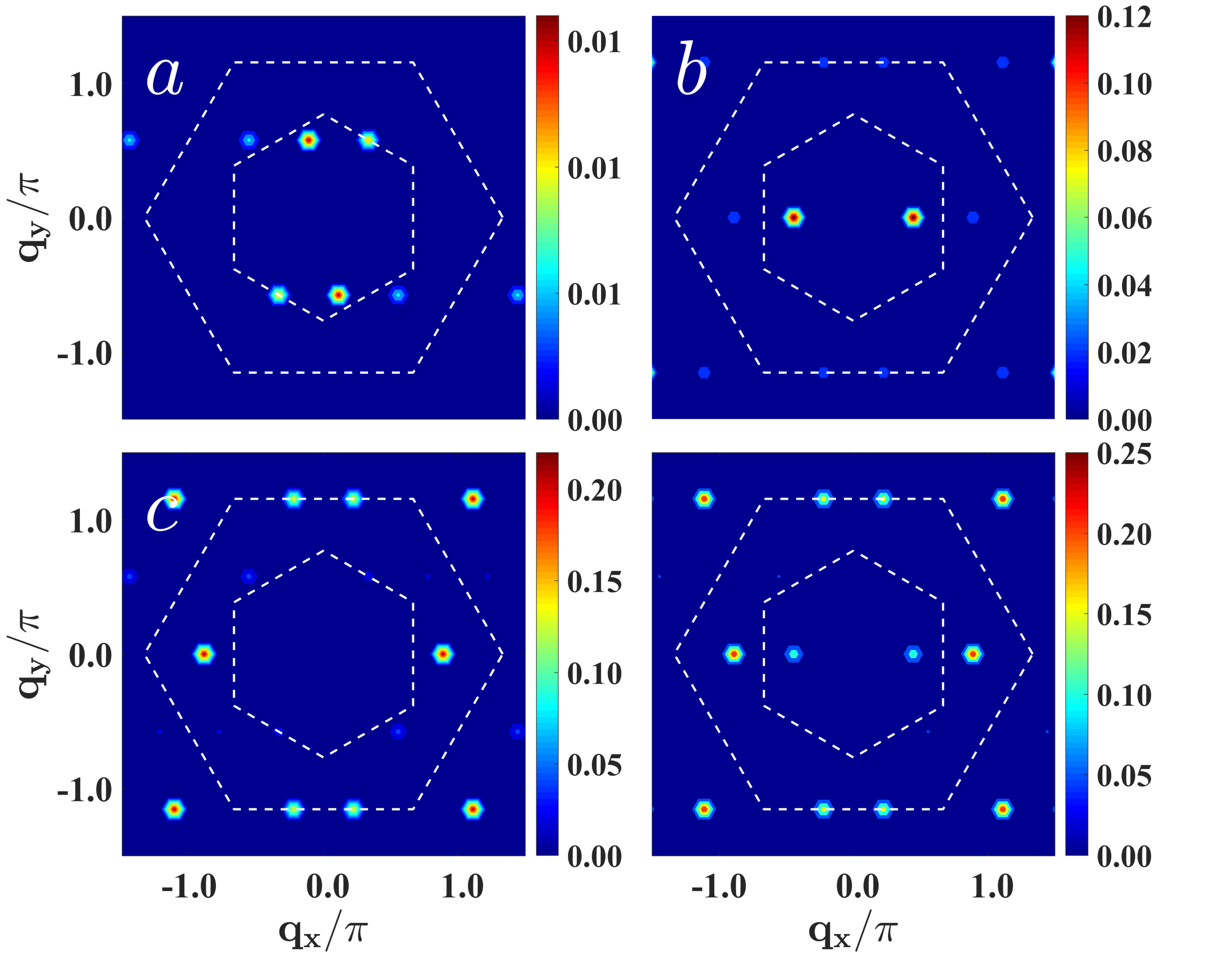}
		        }
			\caption{\textcolor{red}{$\rm 12_B$} phase ($\theta=0.7\pi$, $\phi=0.79\pi$)}
			\label{FigSMP-10}
		\end{minipage}   		
	\end{tabular}
\end{figure*}

\begin{figure*}[!h]
	\begin{tabular}{cc}
		\hspace{-.25in}
		\begin{minipage}[t]{0.45\textwidth}
			%   \vspace{-3.25cm}
		    \mbox{
		         	  \includegraphics[width=0.5\textwidth]{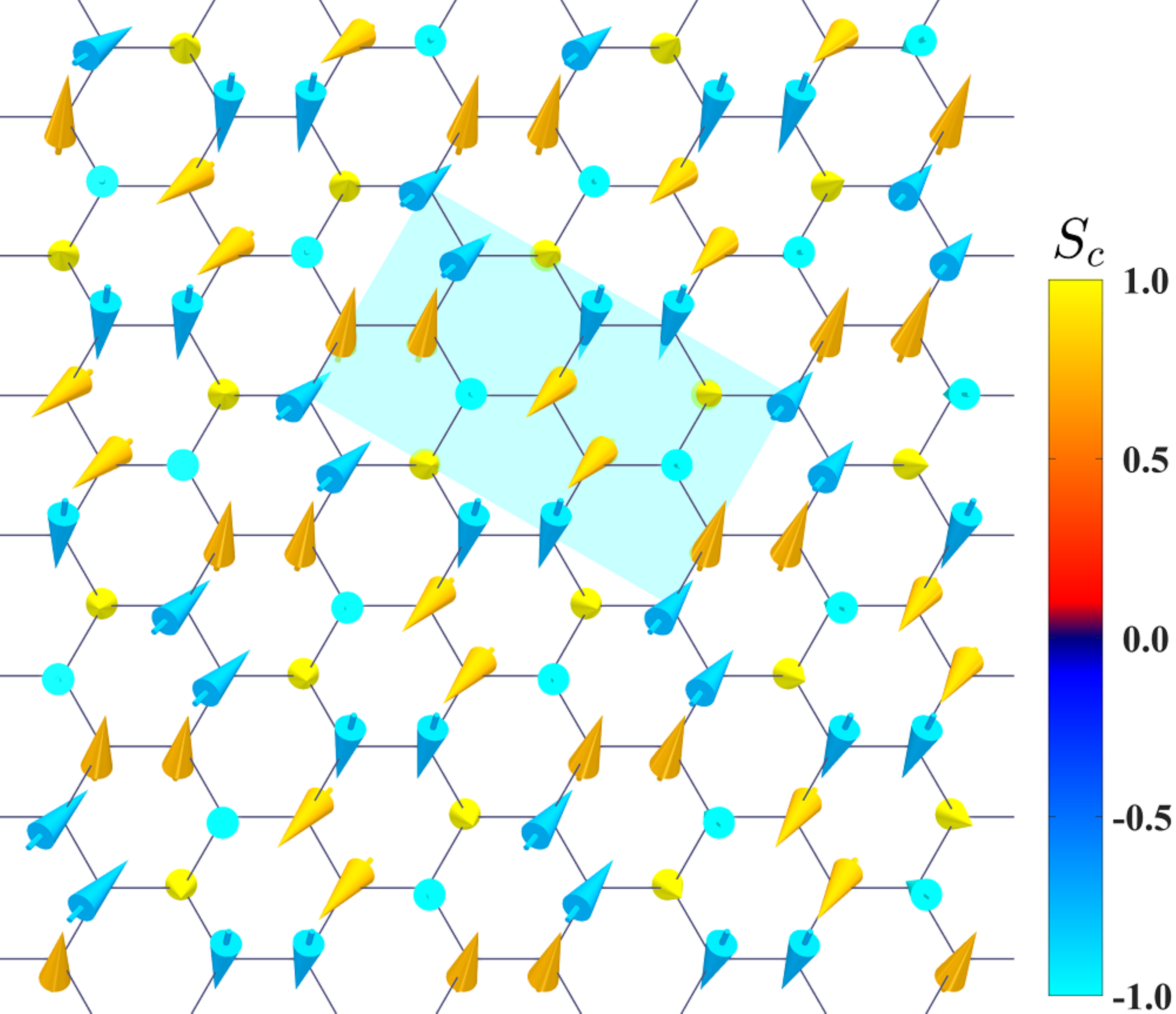}
		         	  \includegraphics[width=0.5\textwidth]{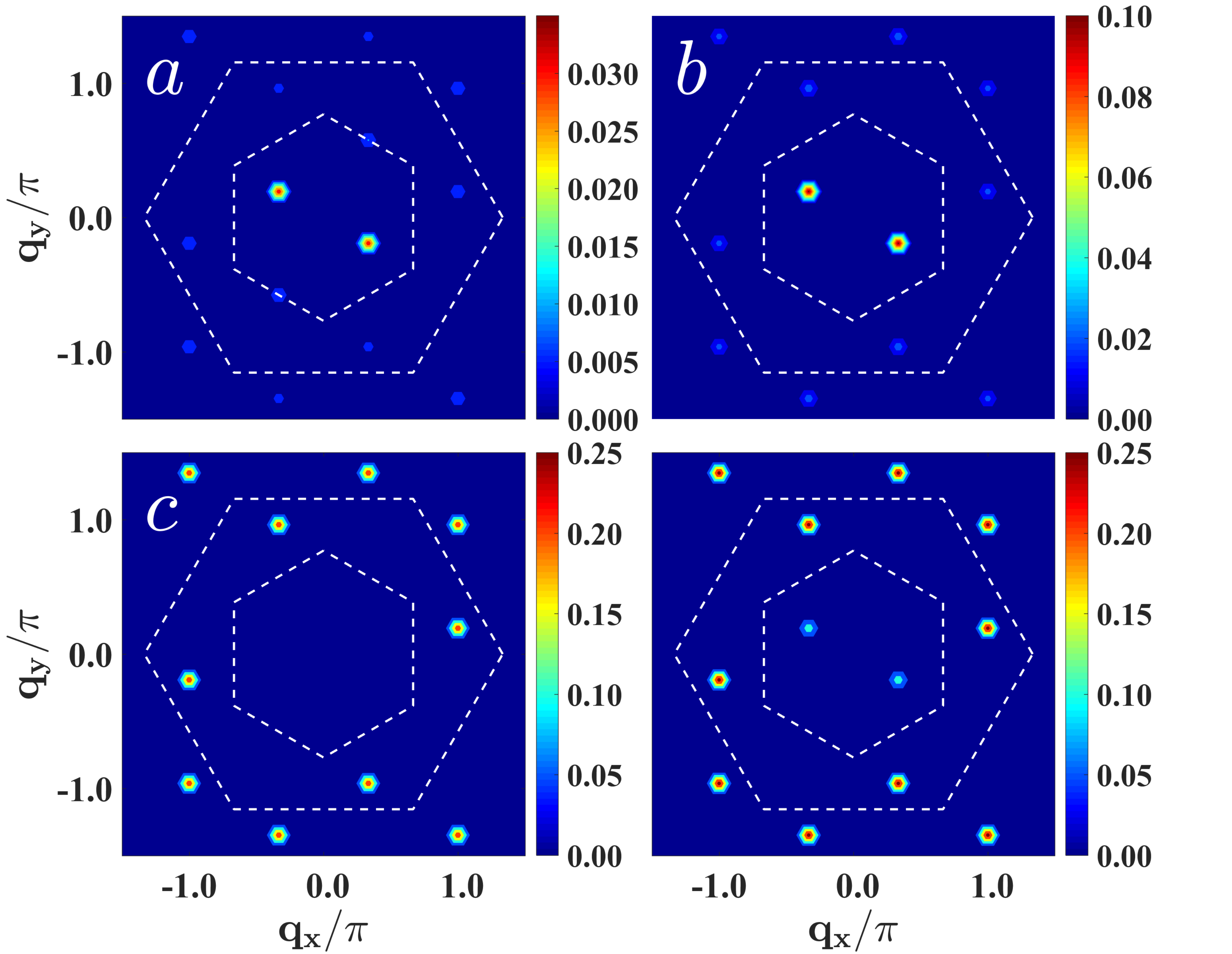}
		         }
			 \caption{\textcolor{red}{$\rm 12_C$} phase ($\theta=0.7\pi$, $\phi=0.72\pi$)}
			 \label{FigSMP-11}
		\end{minipage}
	
		\hspace{.15in}
		\begin{minipage}[t]{0.45\textwidth}
			%\vspace{-3.25cm}
		   \mbox{
		        	 \includegraphics[width=0.5\textwidth]{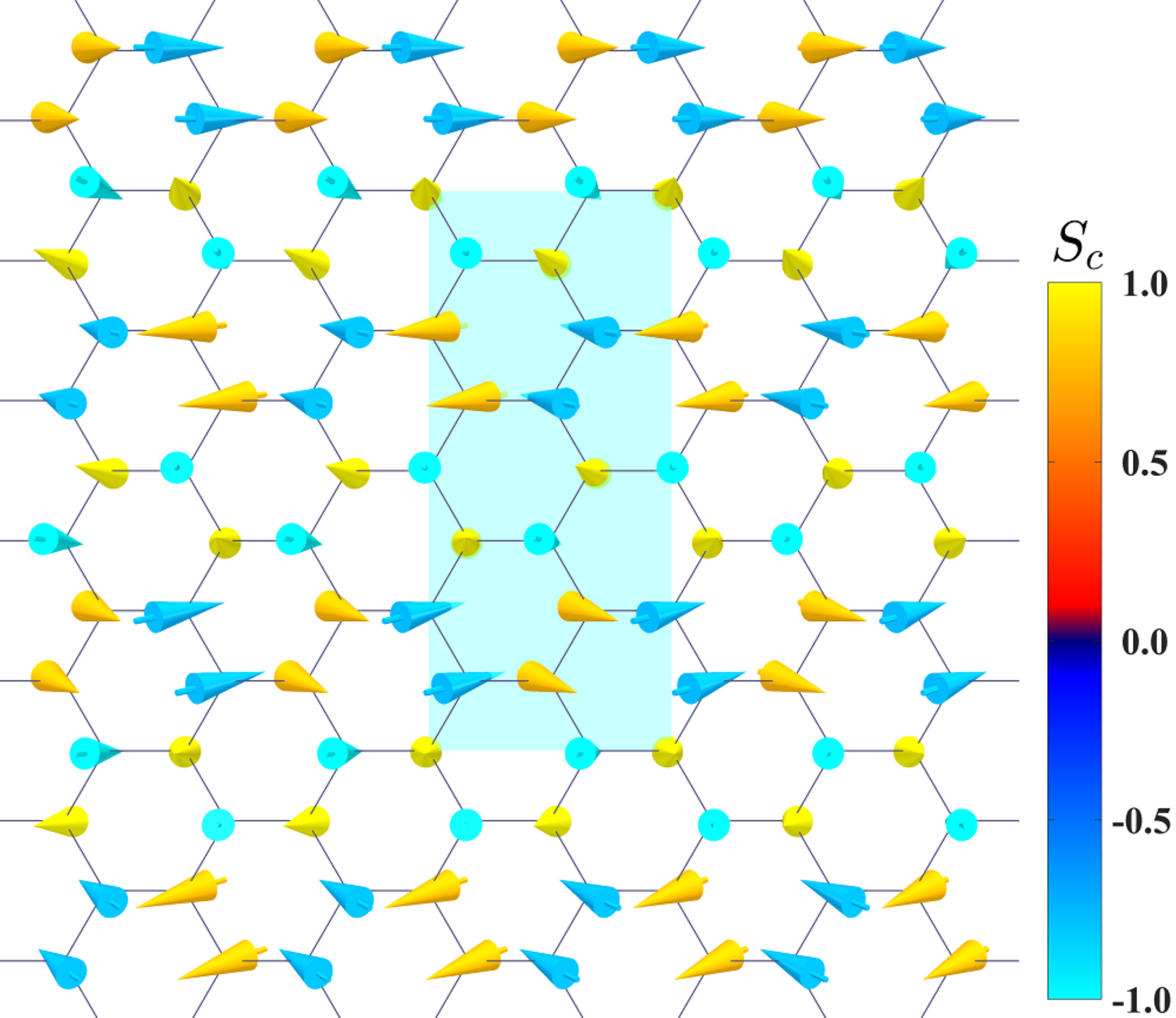}
		        	 \includegraphics[width=0.5\textwidth]{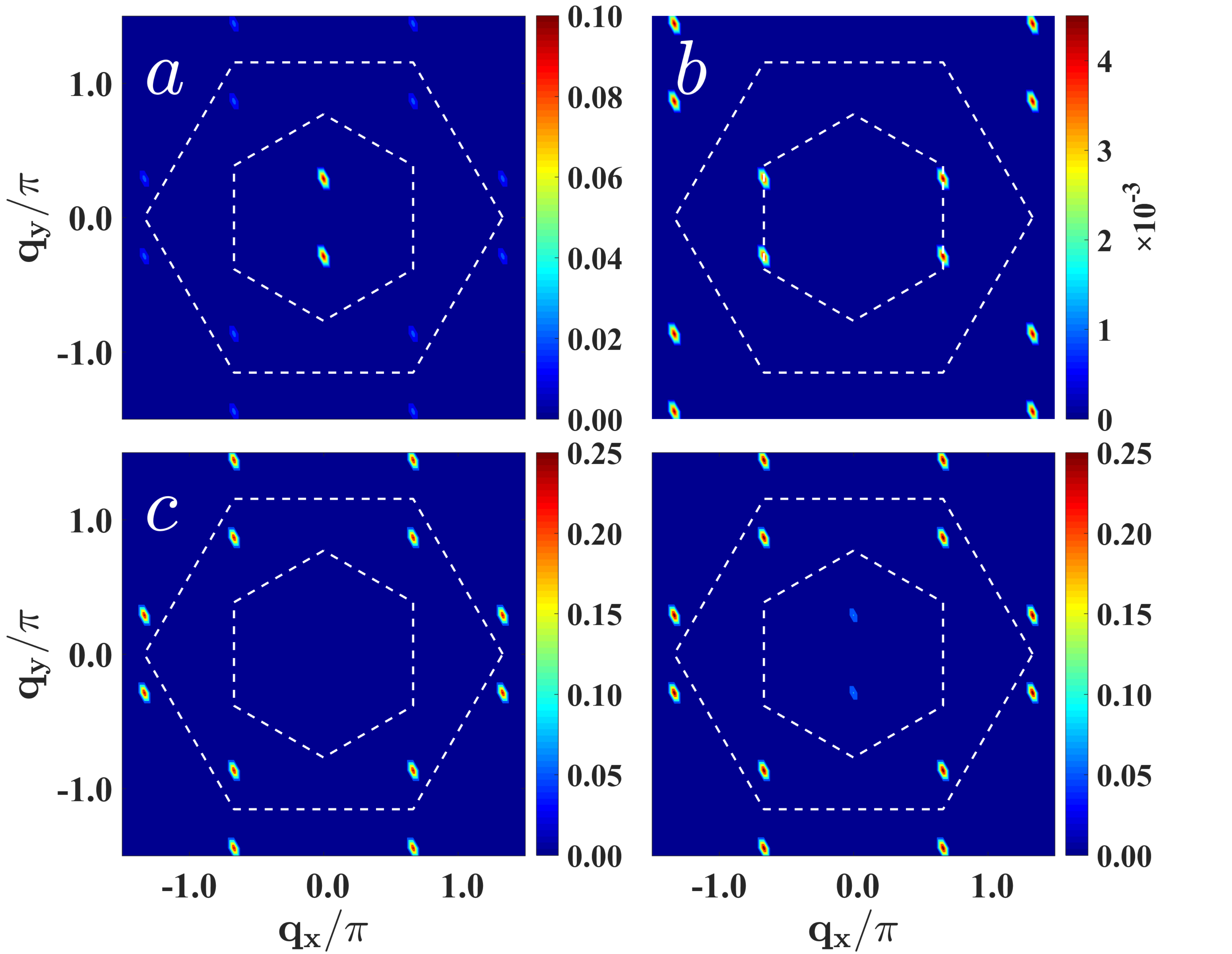}
		        }
			\caption{\textcolor{red}{$\rm 16_B$} phase ($\theta=0.7\pi$, $\phi=0.645\pi$)}
			\label{FigSMP-12}
		\end{minipage} 		
	\end{tabular}
\end{figure*}

\begin{figure*}[!h]
	\begin{tabular}{cc}
		\hspace{-.25in}
	   \begin{minipage}[t]{0.45\textwidth}
				%\vspace{-3.25cm}
            \mbox{
           	         \includegraphics[width=0.5\textwidth]{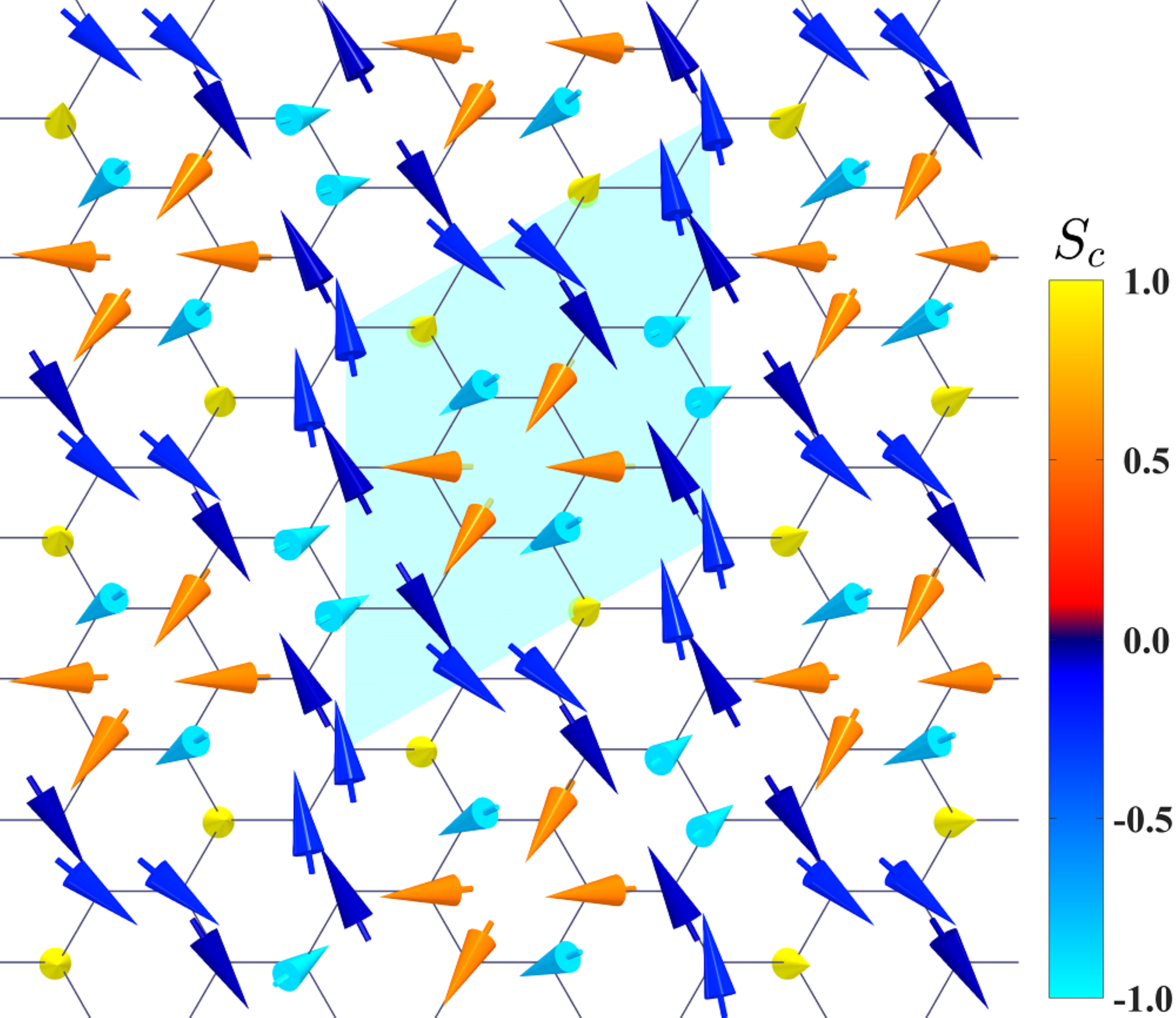}
           	         \includegraphics[width=0.5\textwidth]{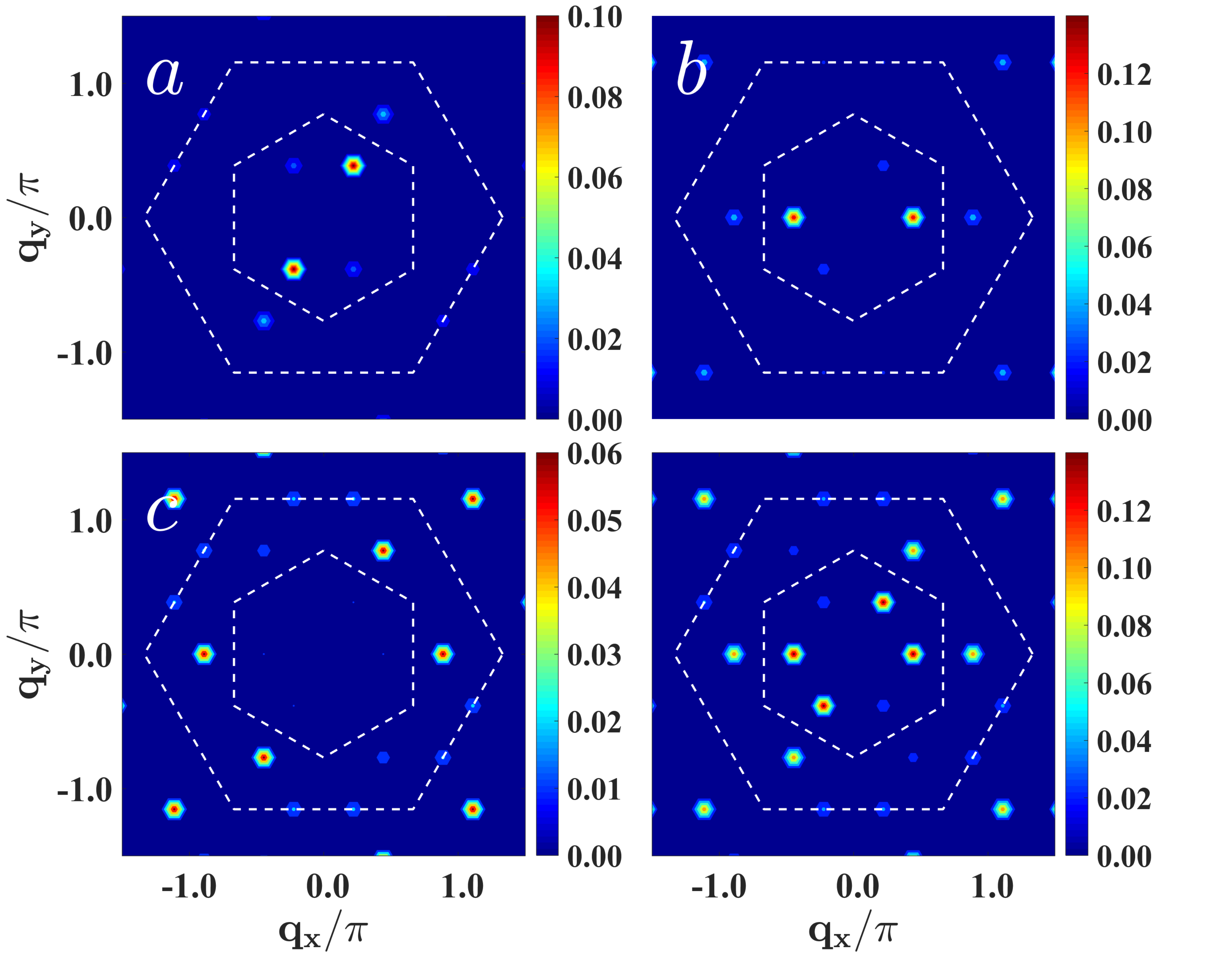}
                 }
		   \caption{\textcolor{red}{$\rm 18_A$} phase ($\theta=0.49\pi$, $\phi=0.845\pi$)}
		 \label{FigSMP-13}
	    \end{minipage}
				
				\hspace{.15in}				
		\begin{minipage}[t]{0.45\textwidth}
			%   \vspace{-3.25cm}
		    \mbox{
		         	  \includegraphics[width=0.5\textwidth]{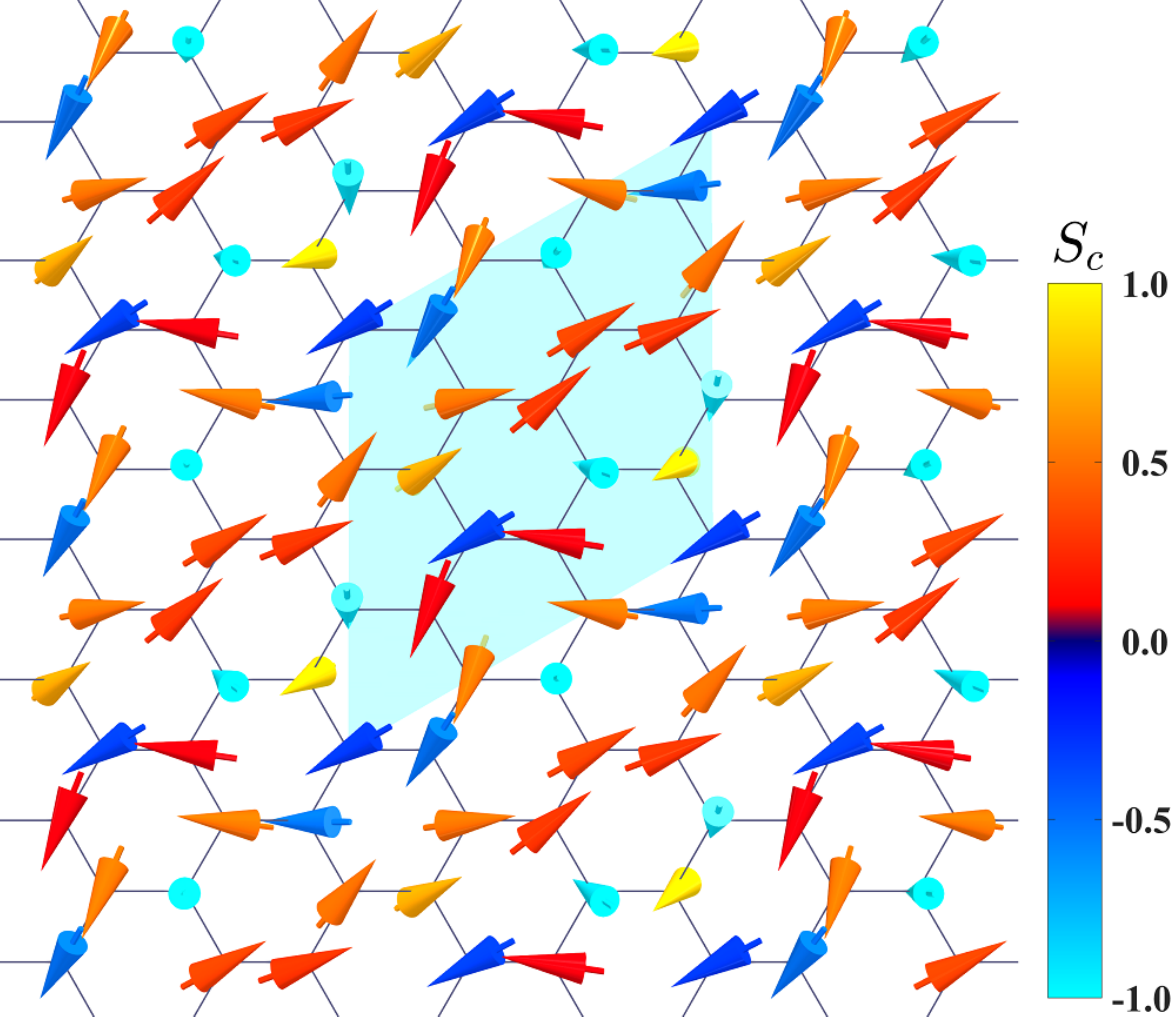}
		         	  \includegraphics[width=0.5\textwidth]{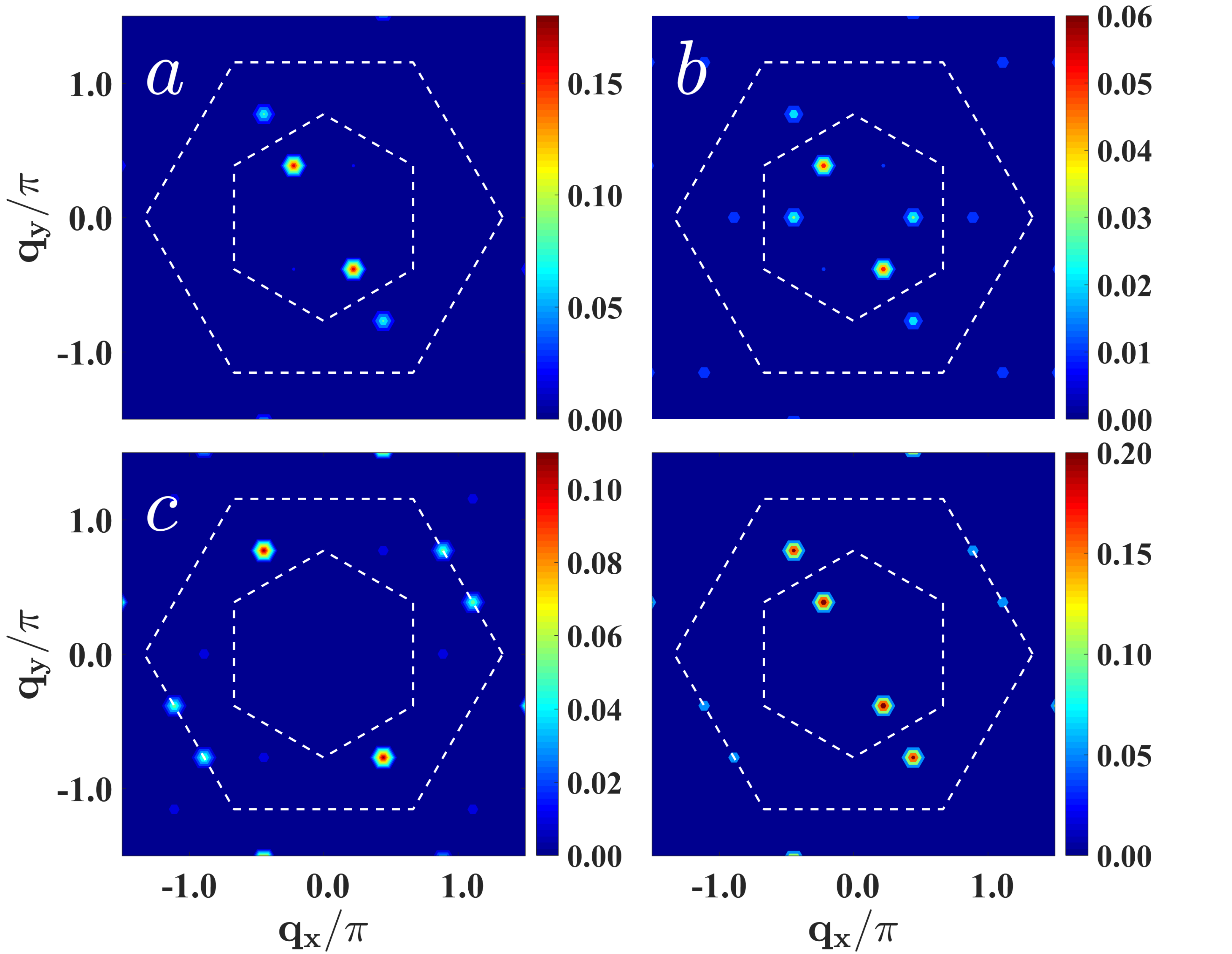}
		         }
			 \caption{\textcolor{red}{$\rm 18_B$} phase ($\theta=0.49\pi$, $\phi=0.68\pi$)}
			 \label{FigSMP-14}
		\end{minipage}   		
	\end{tabular}
\end{figure*}

In the following, we will show the spin configuration and
the static spin structure factors of each phase. We will not show the five phases discussed
in Sec. I since they are common and their properties are well known in magnets.
For simplicity, we choose one representive point in each phase to illustrate the spin structure,
which are shown in Fig.~\ref{FigSMP-1}-Fig.~\ref{FigSMP-14} with the phase name and corresponding parameters given in the caption.
In each figure, the spin configurations are shown on the left panels and the corresponding static spin structure factors are on the right
panels. On the right panel, the $a, b$ or $c$ in the subfigures marks the spin component and the right bottom subfigures show the summation
of all three components. Since there two regions belonging to the 6$\rm_A$ phase,
we choose one point in each region and plot them in Fig.~\ref{FigSMP-3} and Fig.~\ref{FigSMP-4}. The magnetic unit cell of the $\it{wave}$ phase
is rather complex. It may vary as a function of $\theta$ and $\phi$. At present we can not draw a reliable conclusion.
The two figures~\ref{FigSMP-23} and \ref{FigSMP-24} are plotted just for reference.

\vspace{3cm}

\begin{figure*}[!h]
	\begin{tabular}{cc}
		\hspace{-.25in}
		\begin{minipage}[t]{0.45\textwidth}
			%\vspace{-3.25cm}
		  \mbox{
		       	    \includegraphics[width=0.5\textwidth]{figSP/24a.pdf}
		       	    \includegraphics[width=0.5\textwidth]{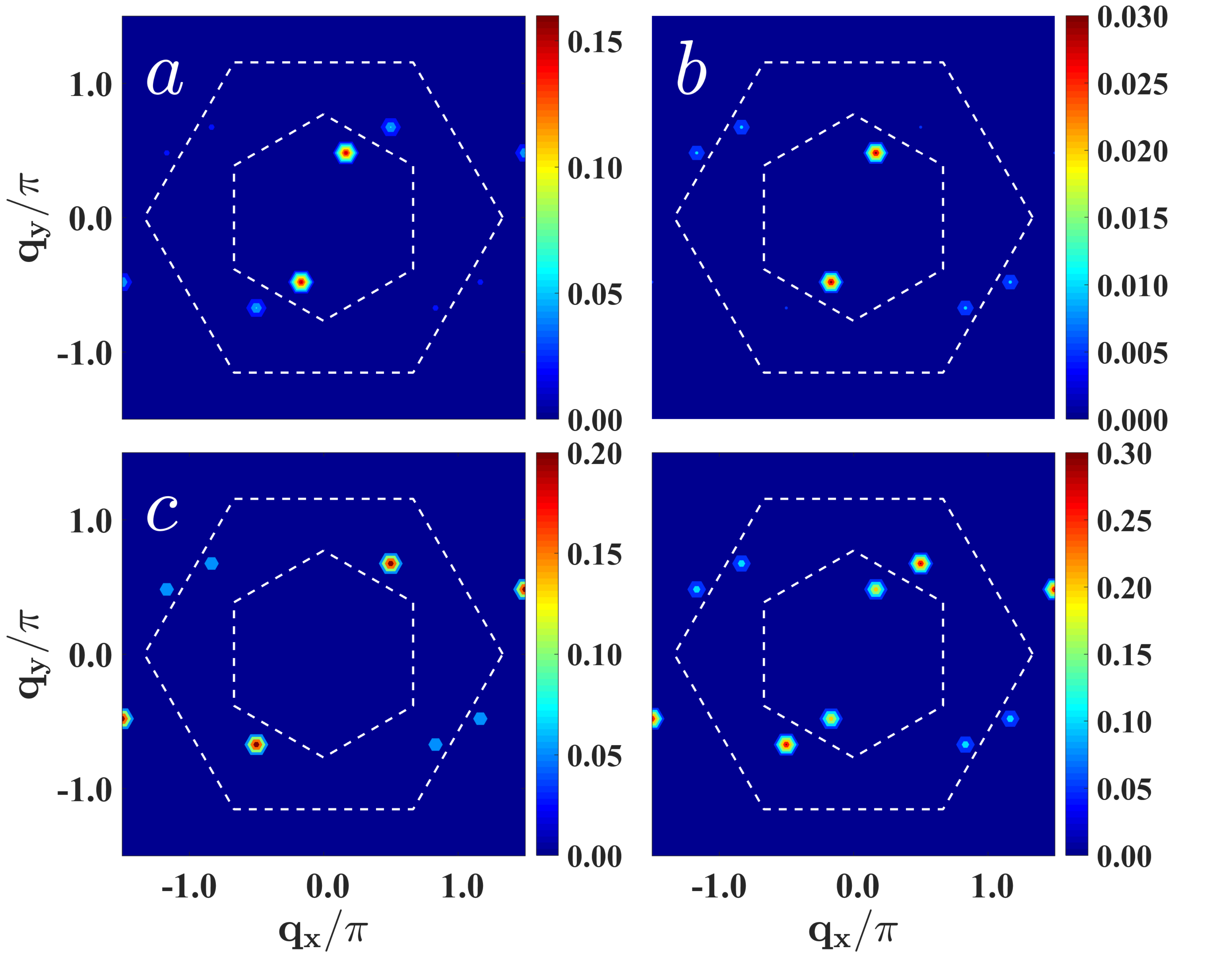}
		       }
			\caption{\textcolor{red}{$\rm 24_A$} phase ($\theta=0.65\pi$, $\phi=0.82\pi$)}
			\label{FigSMP-15}
		\end{minipage}
		
		\hspace{.15in}
		\begin{minipage}[t]{0.45\textwidth}
			%\vspace{-3.25cm}
		     \mbox{
		          	   \includegraphics[width=0.5\textwidth]{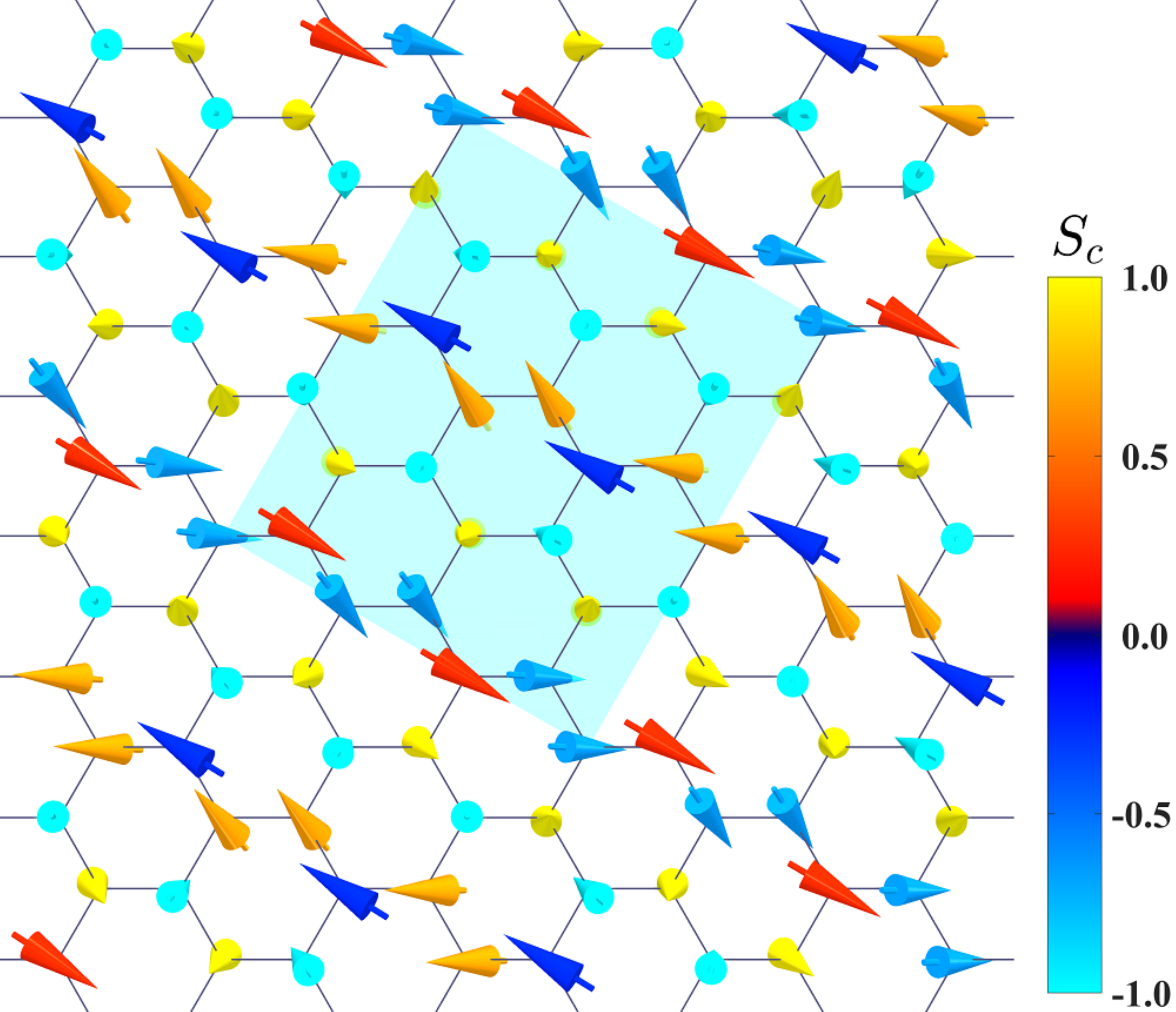}
		          	   \includegraphics[width=0.5\textwidth]{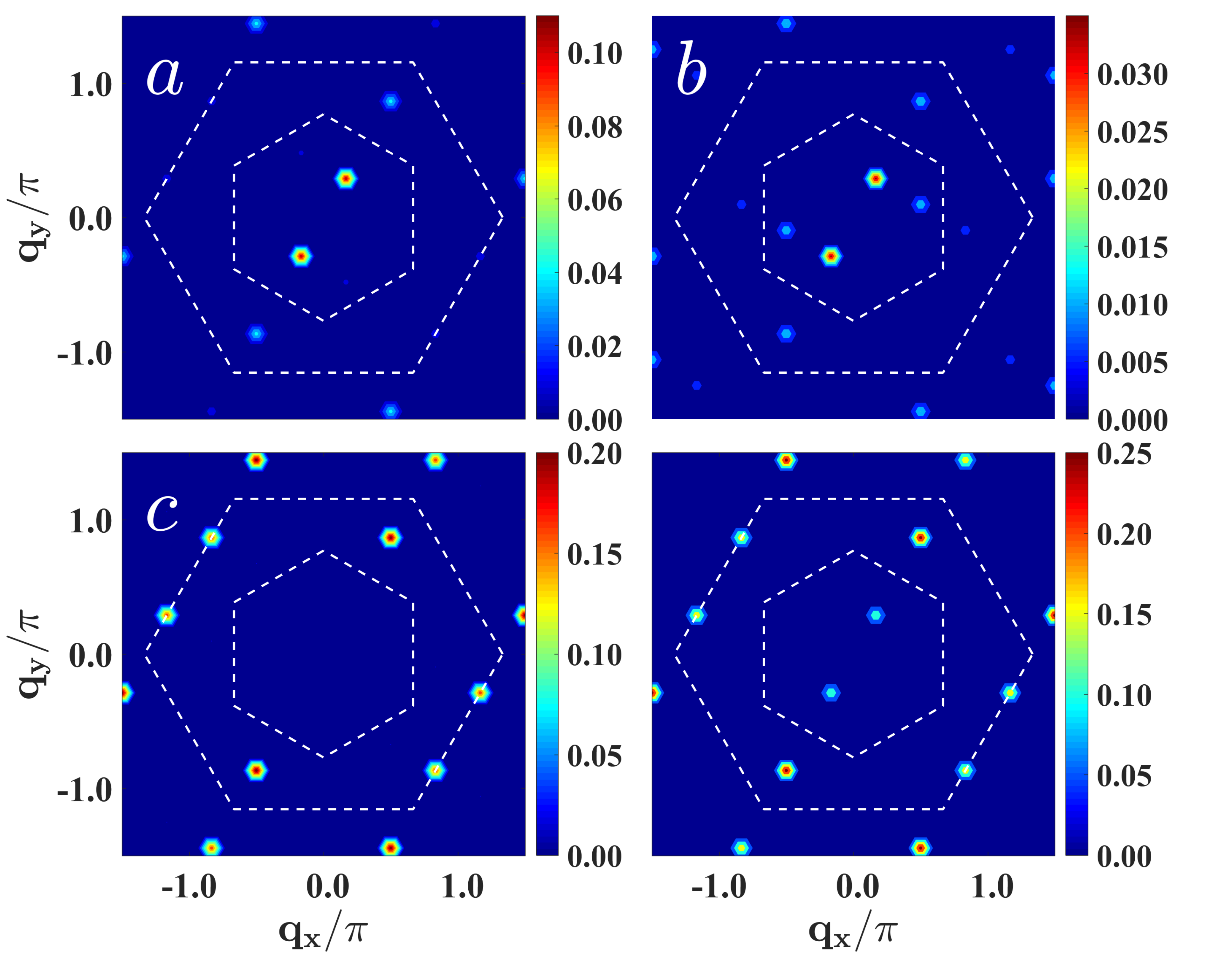}
		          }
			  \caption{\textcolor{red}{$\rm 24_B$} phase ($\theta=0.65\pi$, $\phi=0.615\pi$)}
			  \label{FigSMP-16}
		\end{minipage}  		
	\end{tabular}
\end{figure*}

\begin{figure*}[!h]
	\begin{tabular}{cc}
		\hspace{-.25in}
		\begin{minipage}[t]{0.45\textwidth}
			%   \vspace{-3.25cm}
		 \mbox{
		      	\includegraphics[width=0.5\textwidth]{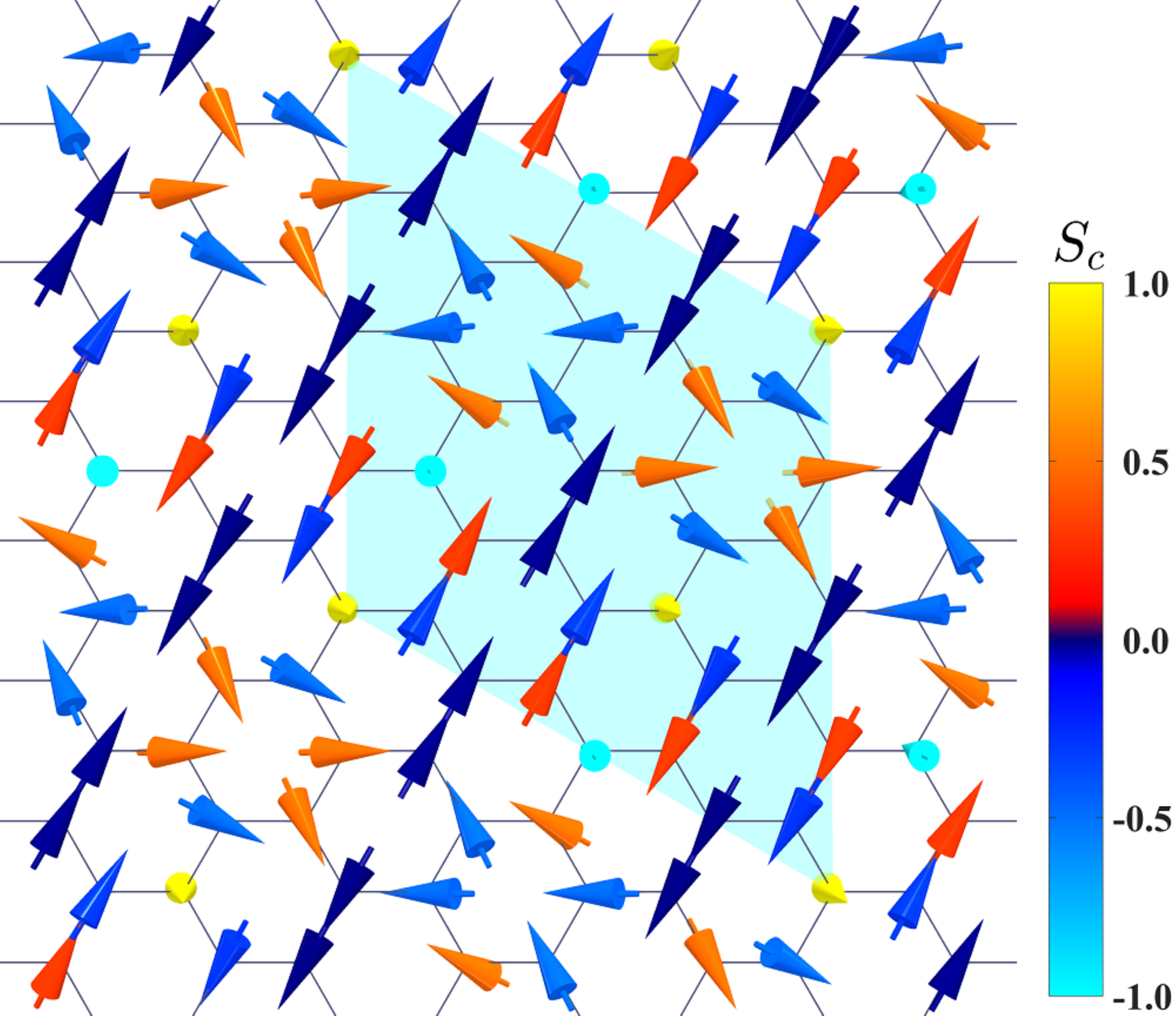}
		      	\includegraphics[width=0.5\textwidth]{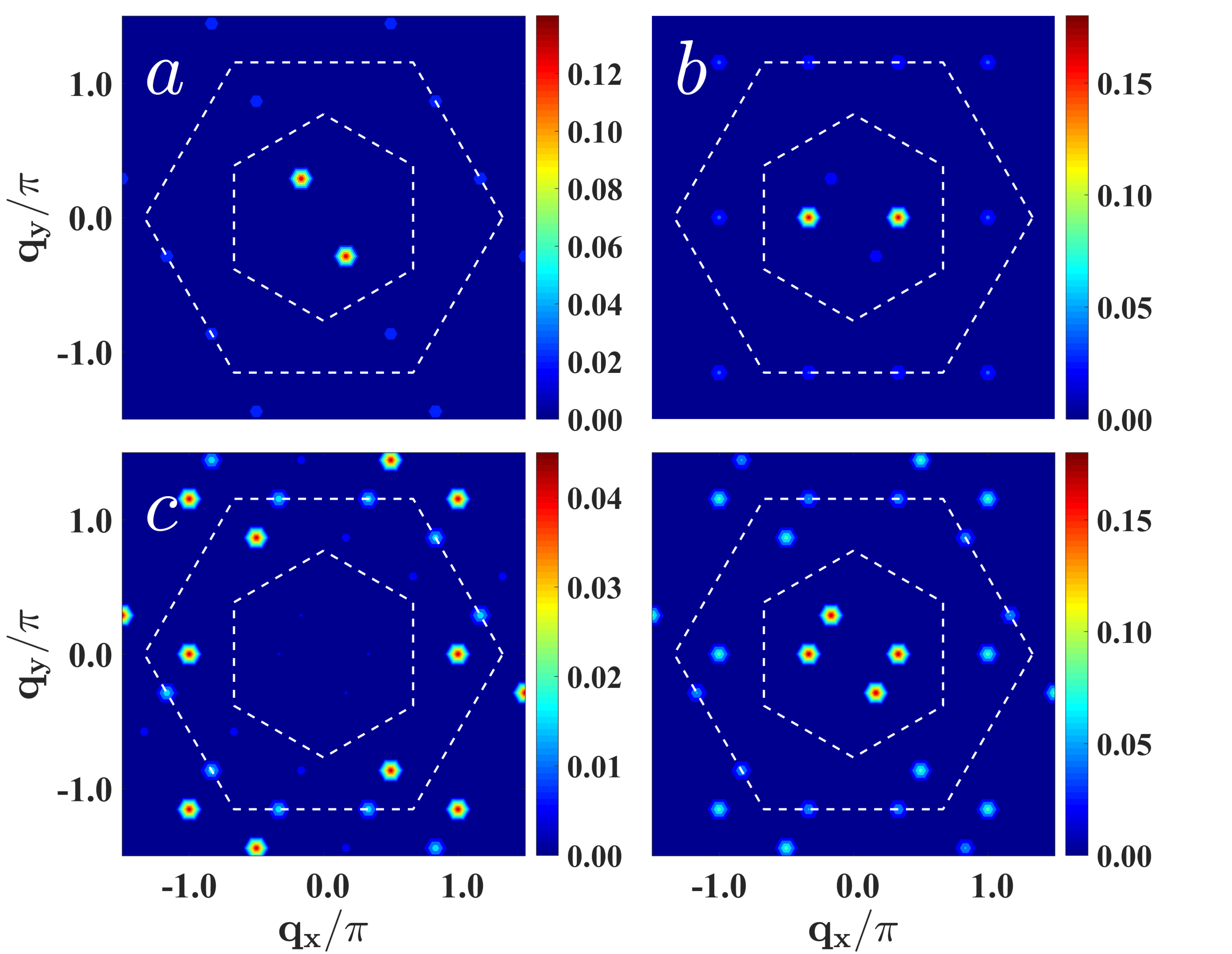}
		      }
		      \caption{\textcolor{red}{$\rm 32_A$} phase ($\theta=0.47\pi$, $\phi=0.9\pi$)}
			  \label{FigSMP-17}
		\end{minipage}
		
		\hspace{.15in}
		\begin{minipage}[t]{0.45\textwidth}
			%\vspace{-3.25cm}		
			    \mbox{
			   	\includegraphics[width=0.5\textwidth]{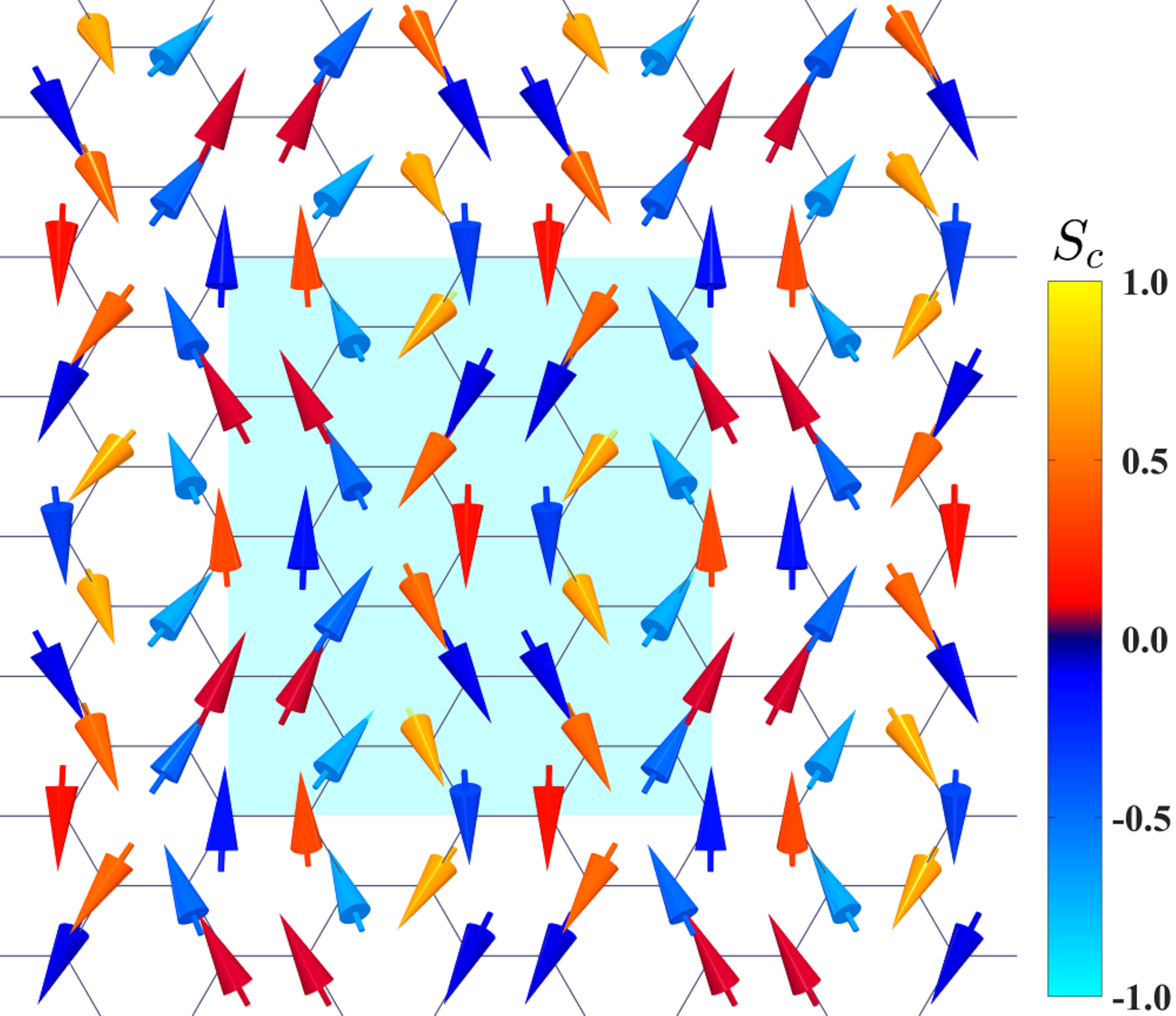}
			   	\includegraphics[width=0.5\textwidth]{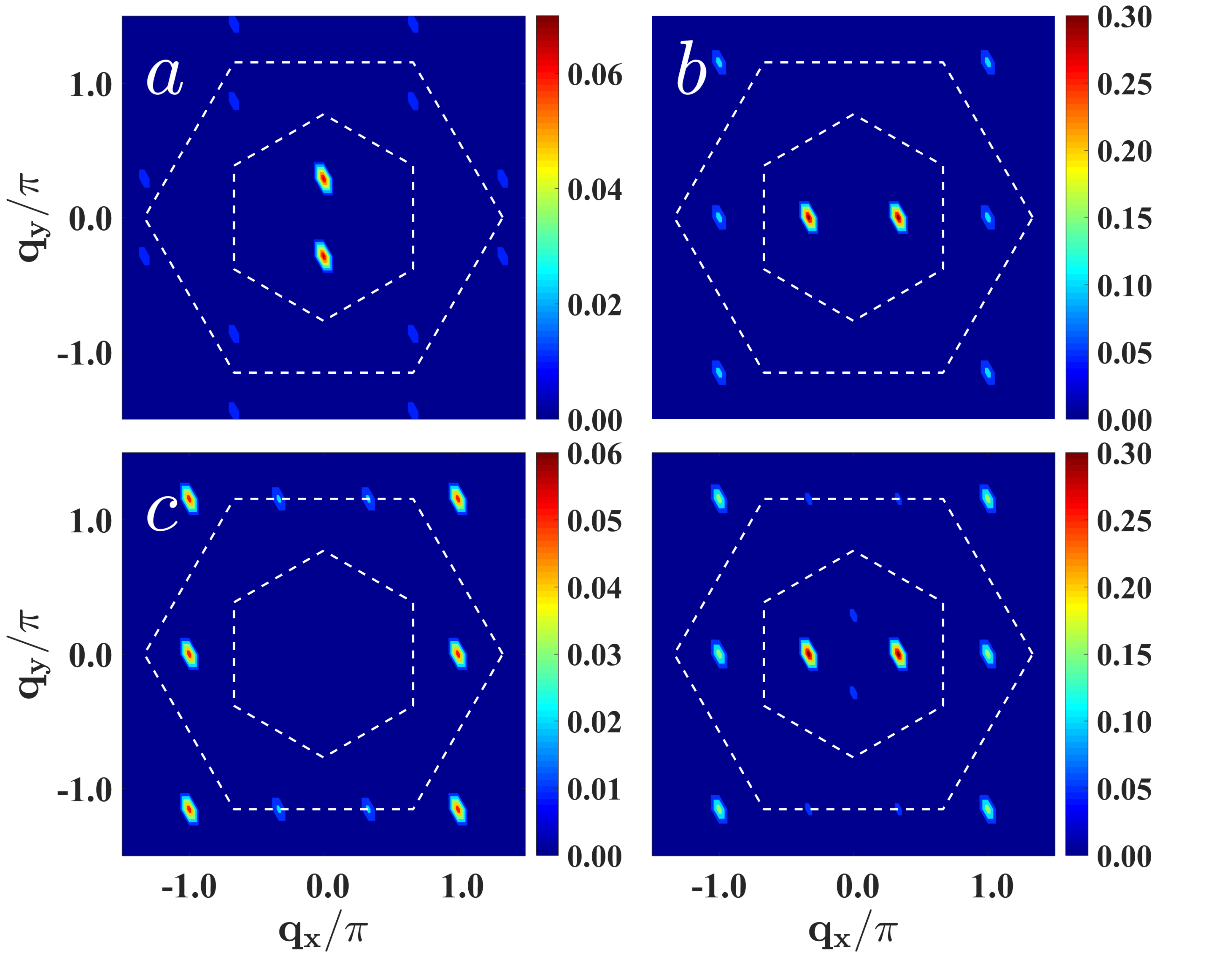}
			   }
			   \caption{\textcolor{red}{$\rm 32_B$} phase ($\theta=0.35\pi$, $\phi=0.82\pi$)}
			   \label{FigSMP-18}
		\end{minipage}
		
	\end{tabular}
\end{figure*}

\begin{figure*}[!h]
	\begin{tabular}{cc}
		\hspace{-.25in}
		\begin{minipage}[t]{0.45\textwidth}
			%\vspace{-3.25cm}
           \mbox{
               	\includegraphics[width=0.5\textwidth]{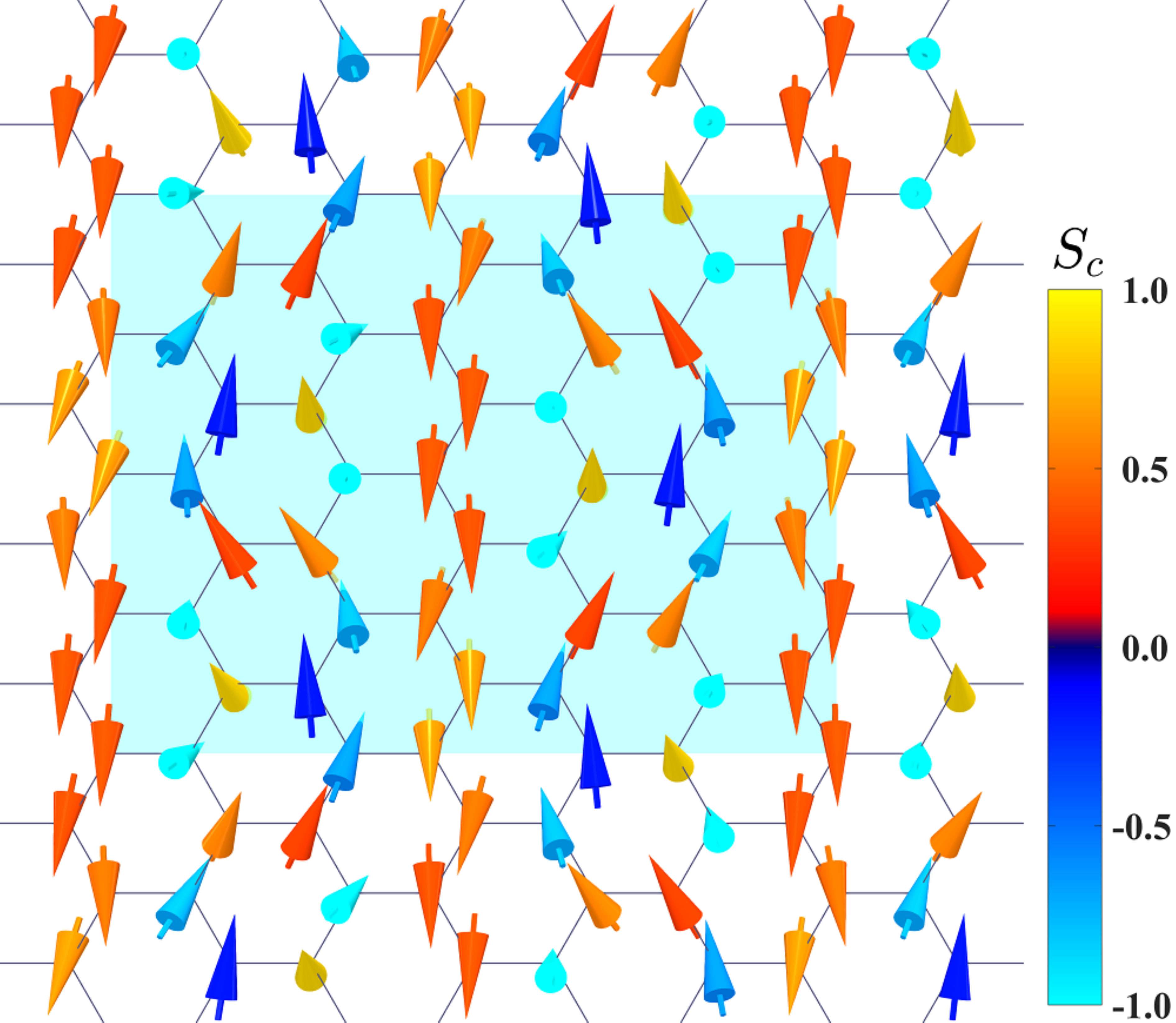}
               	\includegraphics[width=0.5\textwidth]{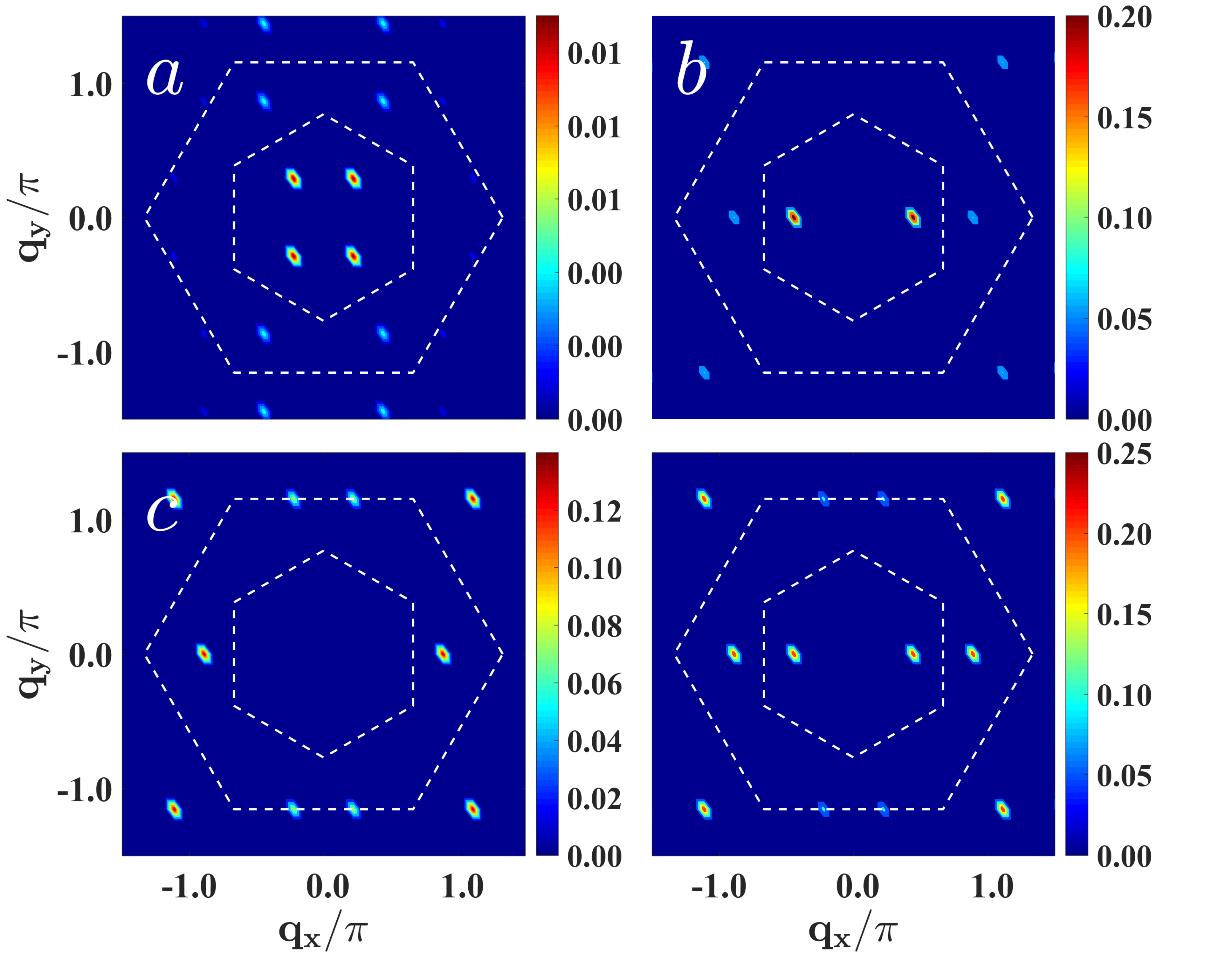}
               }
               \caption{\textcolor{red}{$\rm 48_A$} phase ($\theta=0.535\pi$, $\phi=0.68\pi$)}		
			 \label{FigSMP-19}
		\end{minipage}
		
		\hspace{.15in}
		\begin{minipage}[t]{0.45\textwidth}
			%   \vspace{-3.25cm}
			  \mbox{
			 	\includegraphics[width=0.5\textwidth]{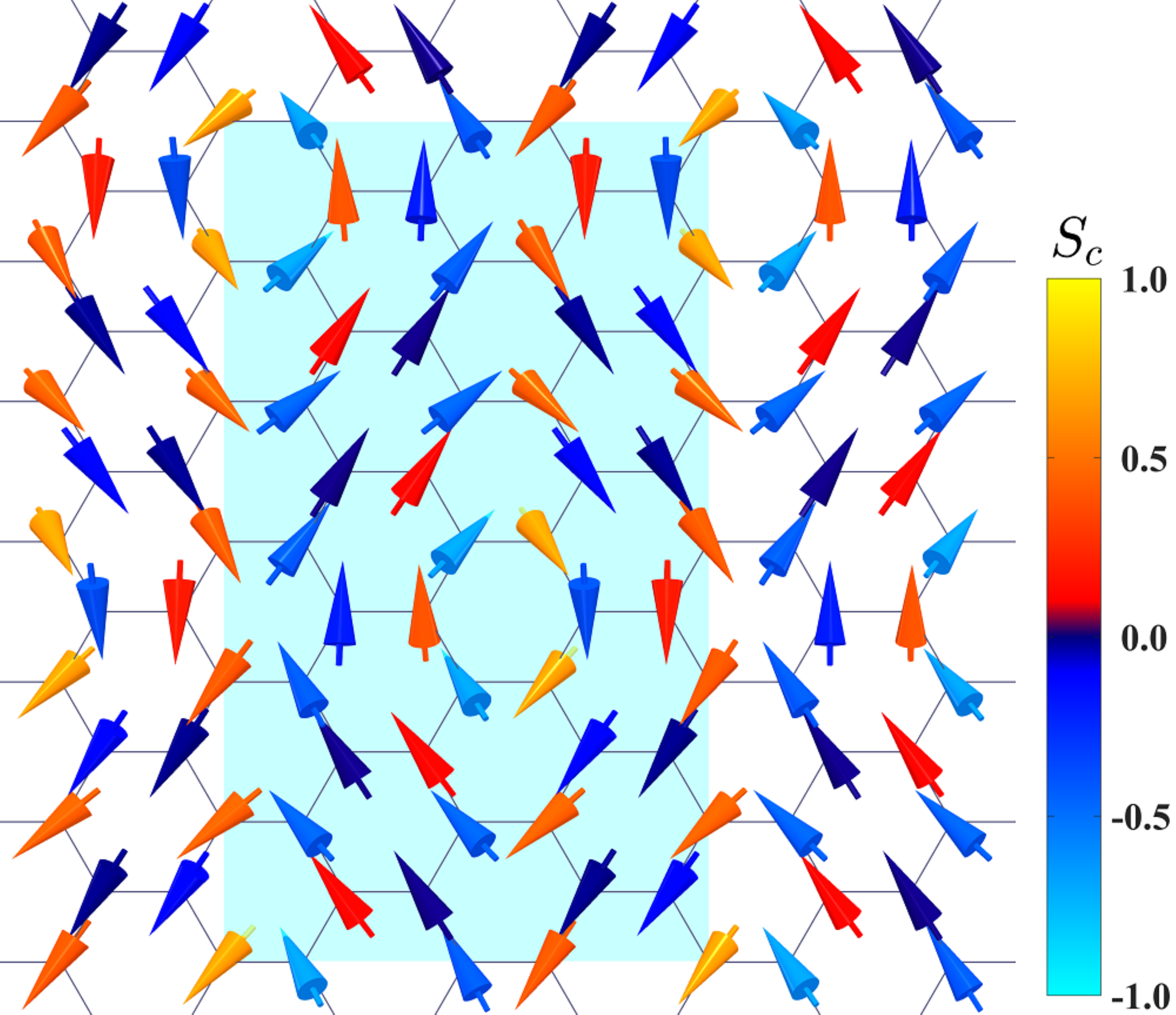}
			 	\includegraphics[width=0.5\textwidth]{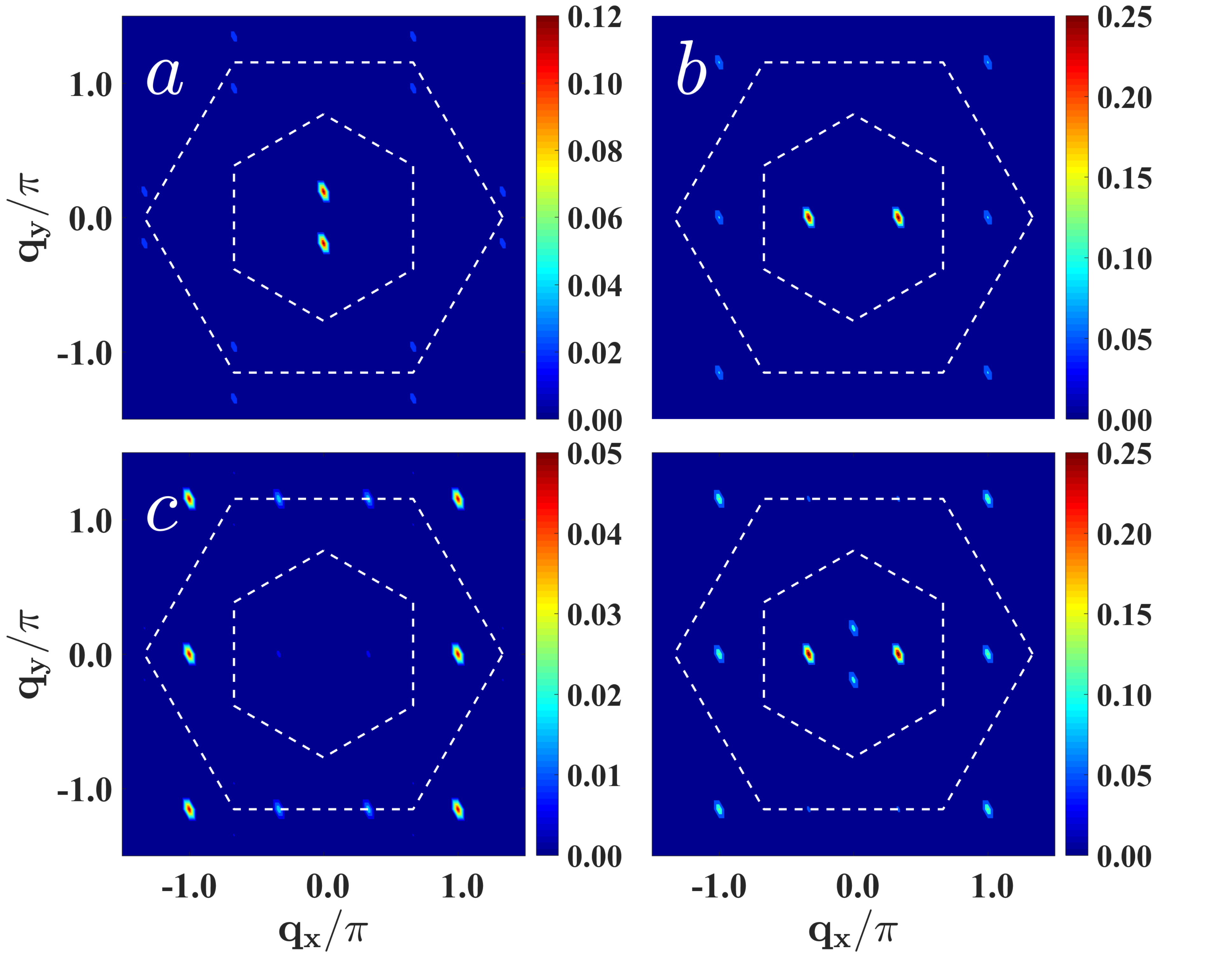}
			 }
			 \caption{\textcolor{red}{$\rm 48_B$} phase ($\theta=0.35\pi$, $\phi=0.83\pi$)}
			 \label{FigSMP-20}
		\end{minipage}
	\end{tabular}
\end{figure*}

\begin{figure*}[!h]
	\begin{tabular}{cc}
		\hspace{-.2in}
		\begin{minipage}[t]{0.6\textwidth}
			\includegraphics[width=\textwidth]{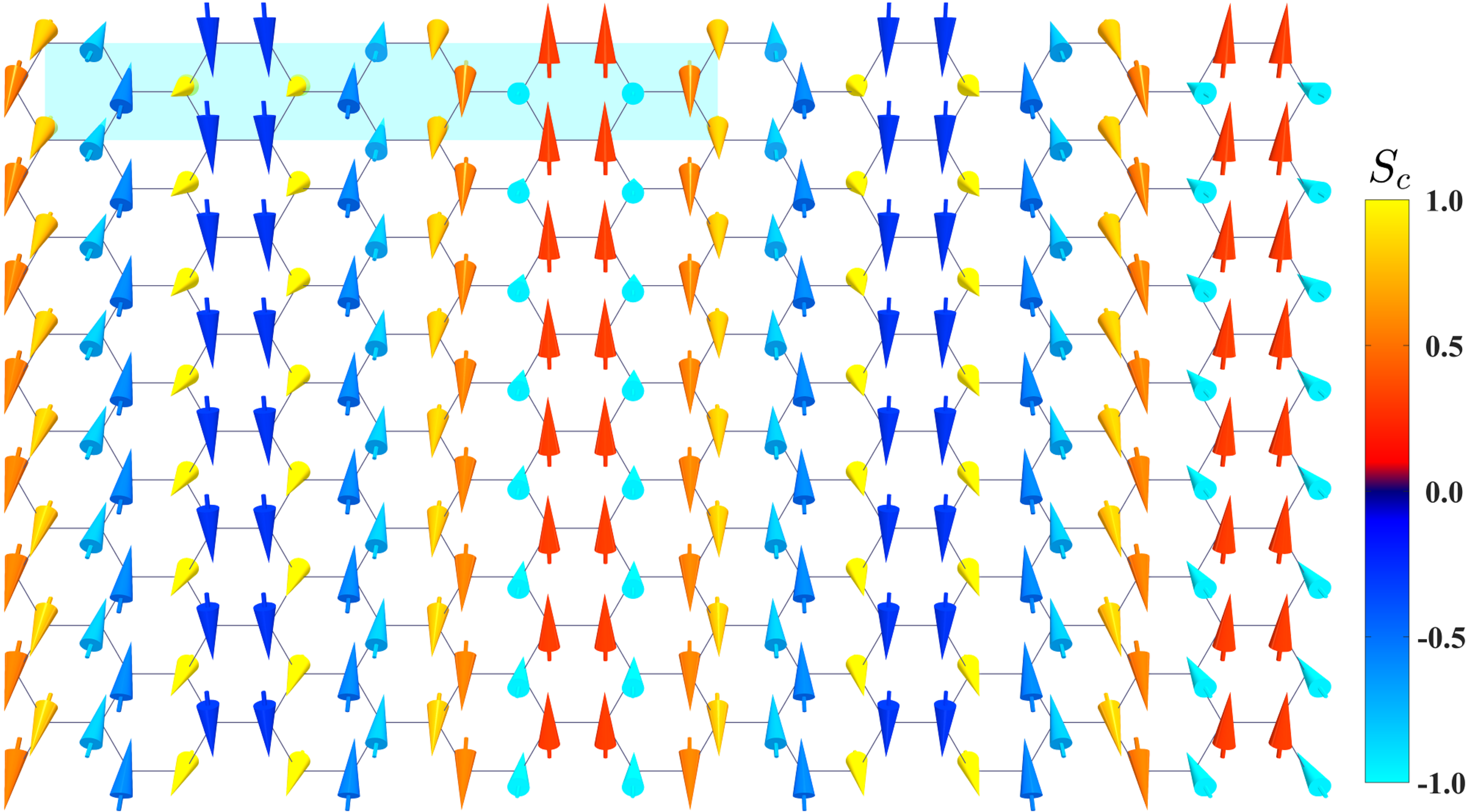}
		\end{minipage}
		\begin{minipage}[t]{0.35\textwidth}
			\includegraphics[width=\textwidth]{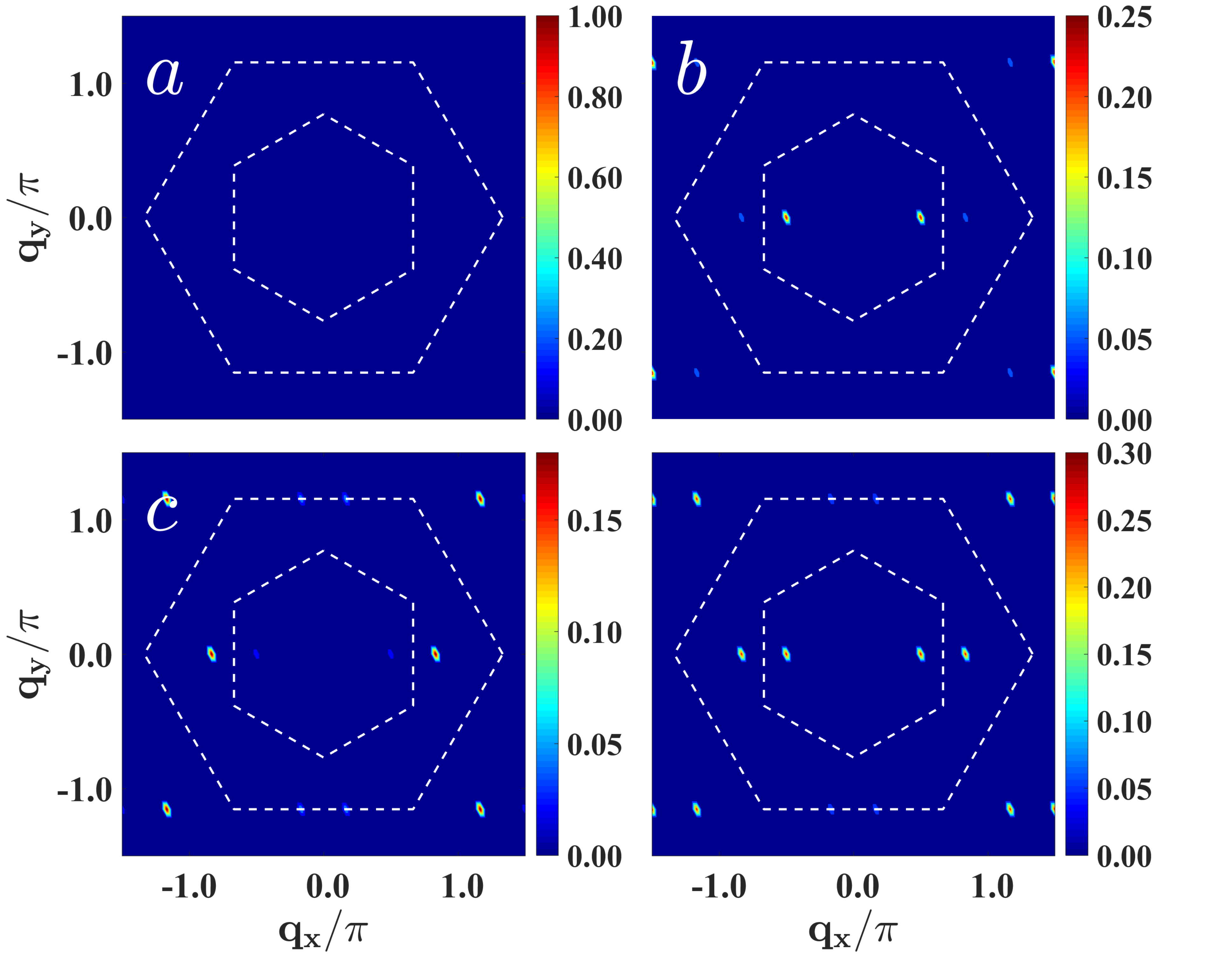}
		\end{minipage}   		
	\end{tabular}
	\caption{\textcolor{red}{$\rm 16_A$} phase ($\theta=0.55\pi$, $\phi=0.85\pi$)}
	\label{FigSMP-21}
\end{figure*}

\begin{figure*}[!h]
	\begin{tabular}{cc}
		\hspace{-.2in}
		\begin{minipage}[t]{0.6\textwidth}
			\includegraphics[width=\textwidth]{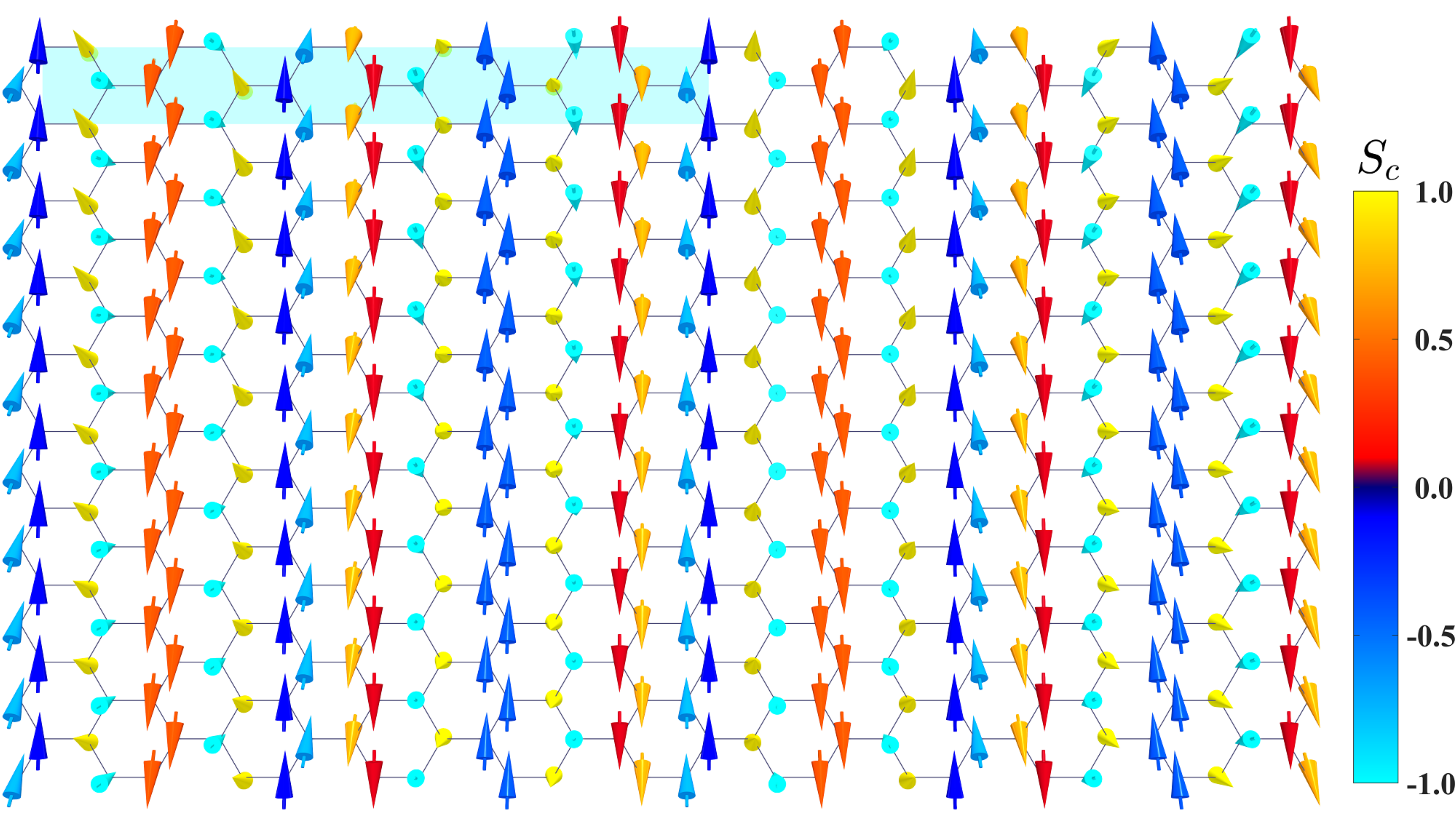}
		\end{minipage}
		\begin{minipage}[t]{0.35\textwidth}
			\includegraphics[width=\textwidth]{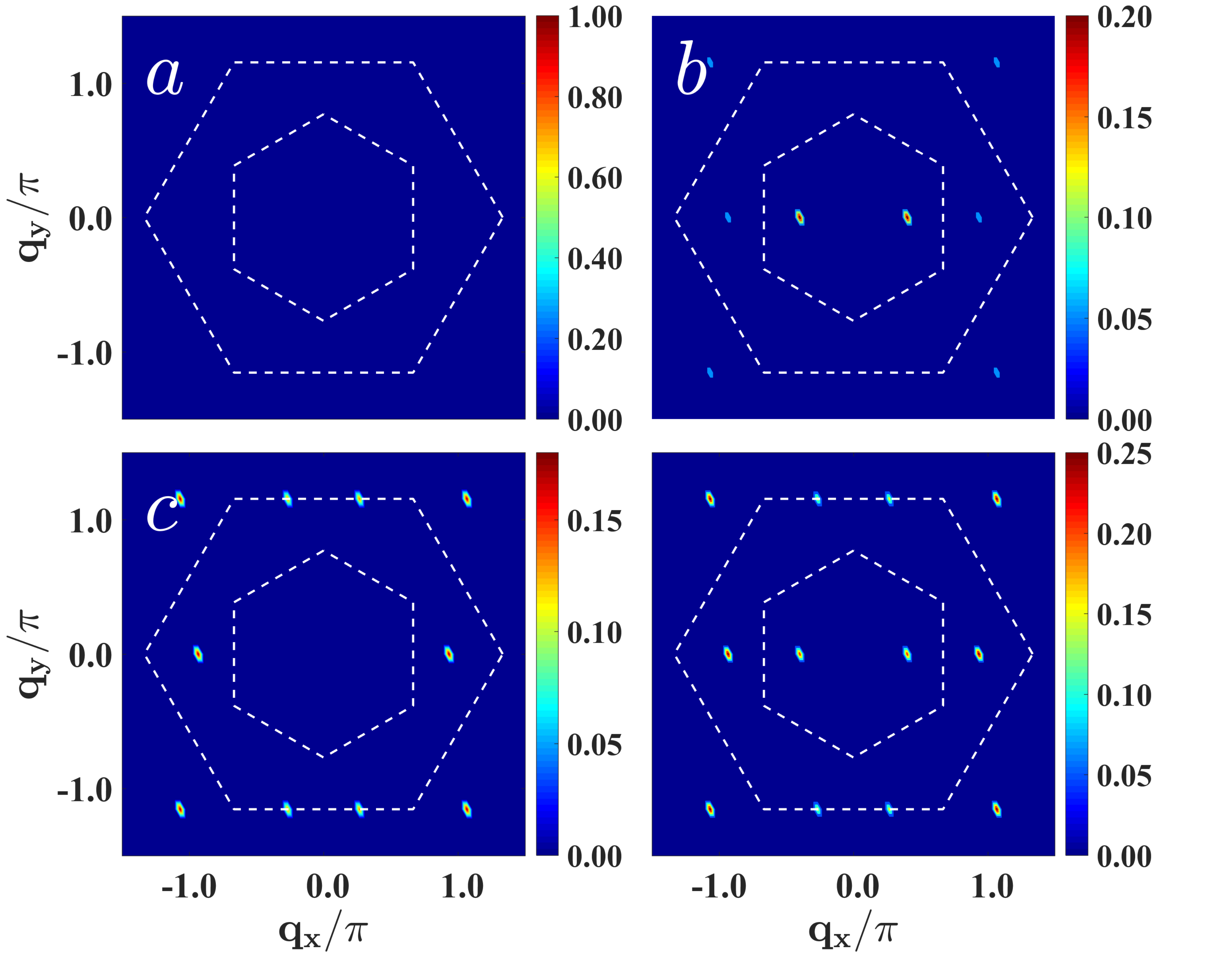}
		\end{minipage}   		
	\end{tabular}
	\caption{\textcolor{red}{20} phase~($\theta=0.55\pi$, $\phi=0.635\pi$)}
	\label{FigSMP-22}
\end{figure*}

\begin{figure*}[!htb]
	\begin{tabular}{cc}
		\hspace{-.2in}
		\begin{minipage}[t]{0.6\textwidth}   	
			\includegraphics[width=\textwidth]{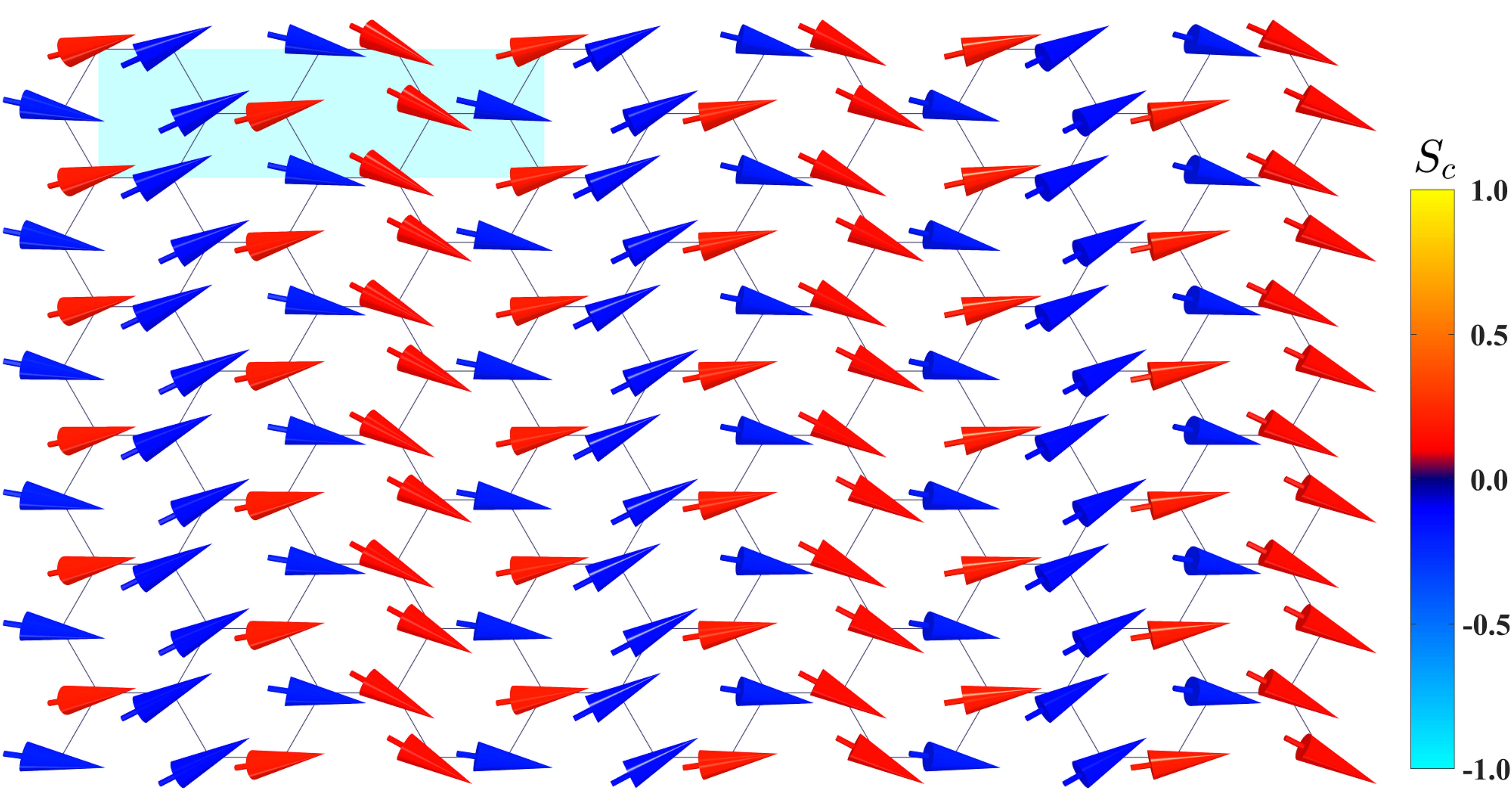}
		\end{minipage}
		\begin{minipage}[t]{0.35\textwidth}
			\includegraphics[width=\textwidth]{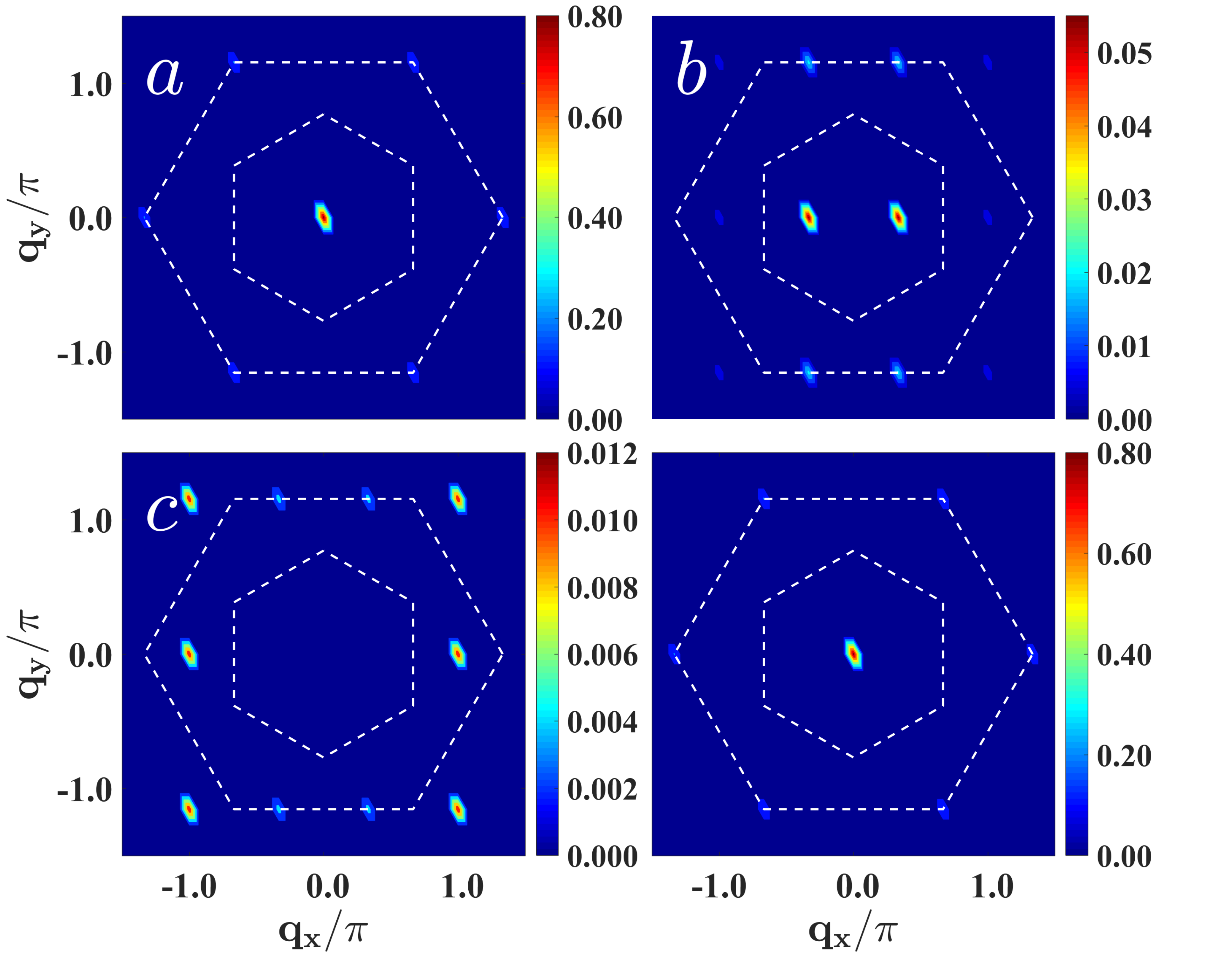}
		\end{minipage}   		
	\end{tabular}
	\caption{\textcolor{red}{\it{wave}} phase~($\theta=0.49\pi$, $\phi=0.96\pi$)}
	\label{FigSMP-23}
\end{figure*}

\begin{figure*}[!htb]
	\begin{tabular}{cc}
		\hspace{-.2in}
		\begin{minipage}[t]{0.6\textwidth}   	
			\includegraphics[width=\textwidth]{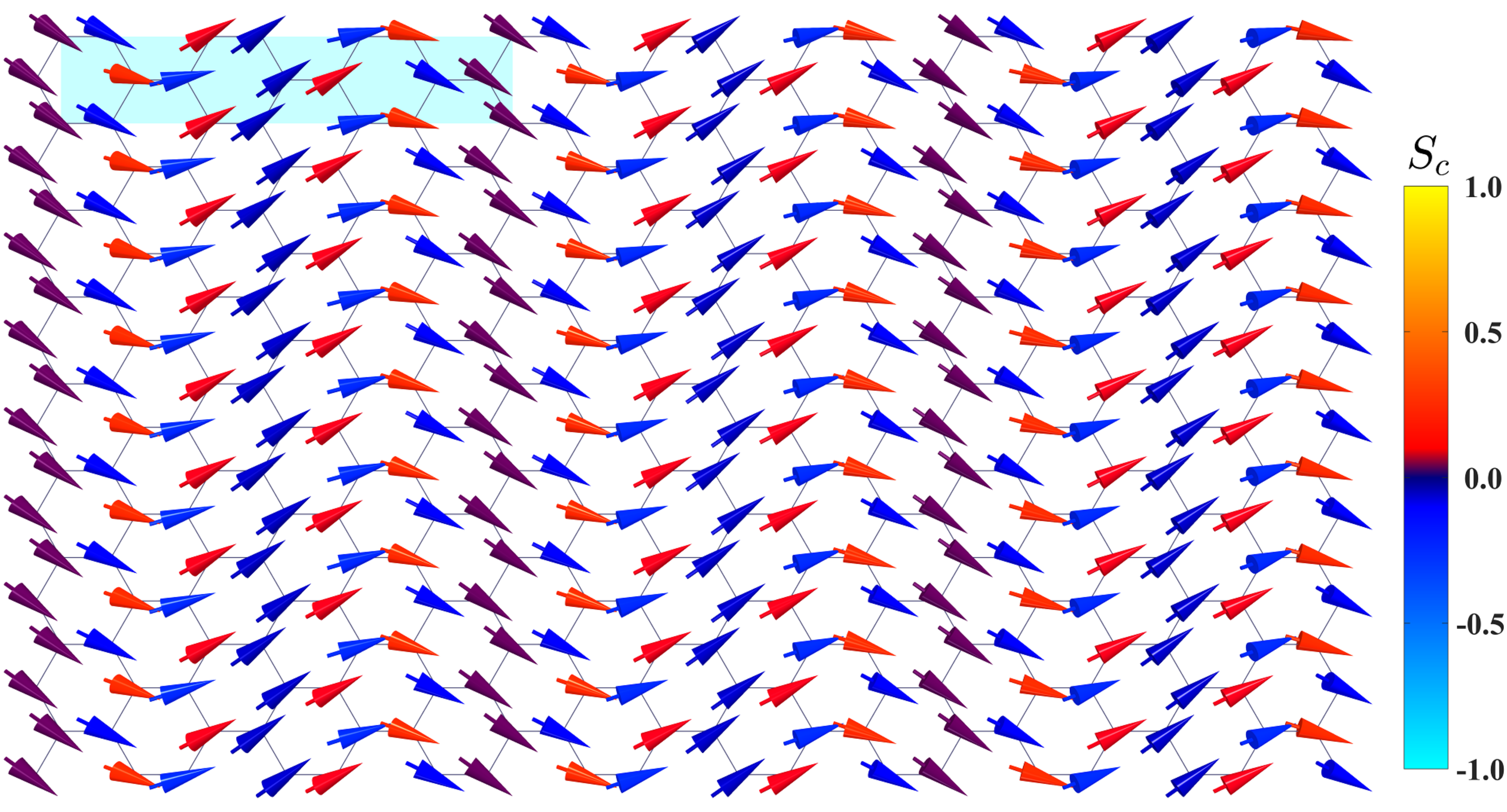}
		\end{minipage}
		\begin{minipage}[t]{0.35\textwidth}   		
			\includegraphics[width=\textwidth]{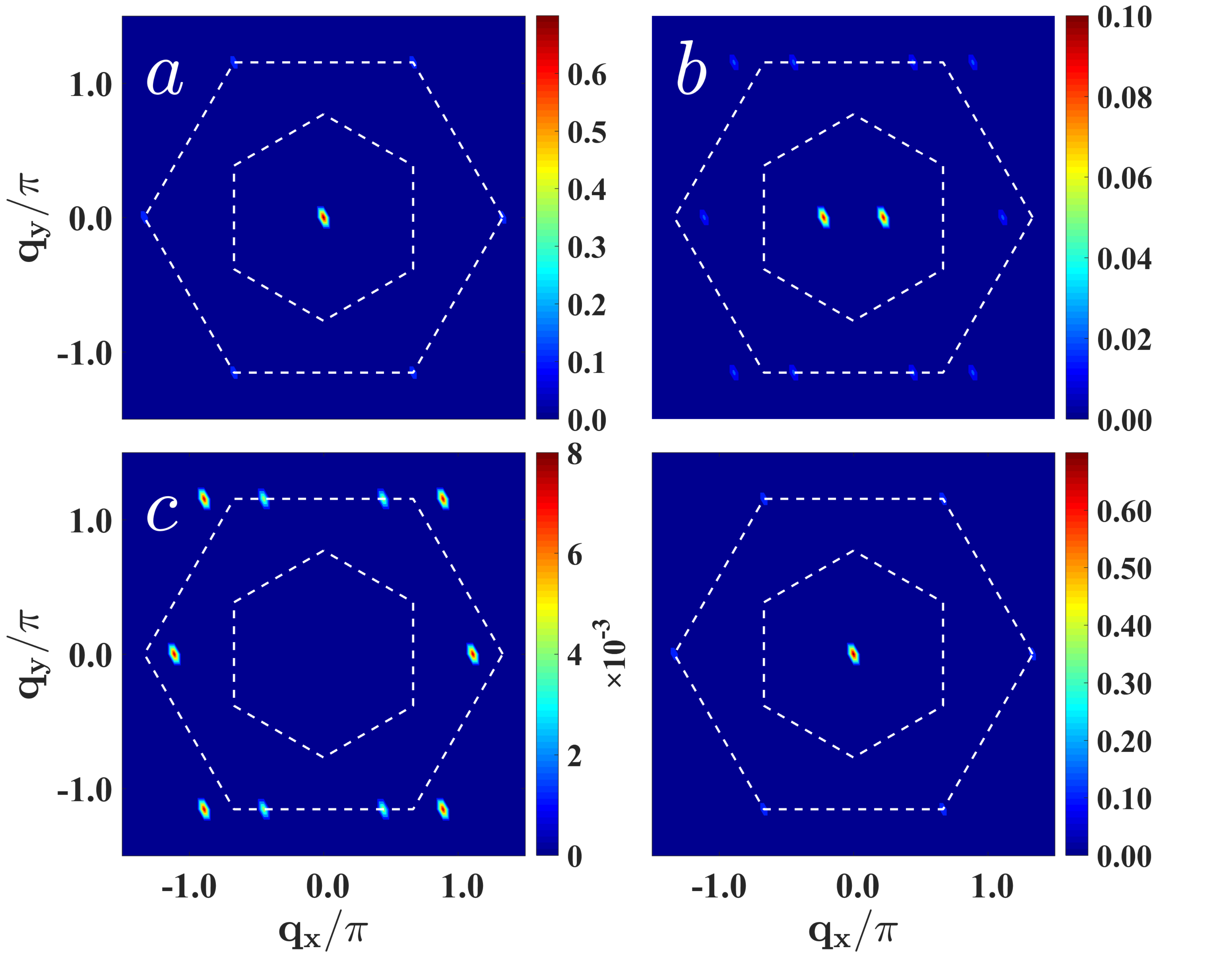}    			
		\end{minipage}   		
	\end{tabular}
	\caption{\textcolor{red}{\it{wave}} phase~($\theta=0.4\pi$, $\phi=0.875\pi$)}
	\label{FigSMP-24}
\end{figure*}

%\end{CJK*}

\begin{thebibliography}{99}
\bibitem{Braun2012}
    H.-B. Braun,
    Topological effects in nanomagnetism: From superparamagnetism to chiral quantum solitons,
    \href{https://doi.org/10.1080/00018732.2012.663070}{Adv. Phys. \textbf{61}, 1 (2012).}
    
\bibitem{Skyrme1962}
    T. H. R. Skyrme,
    A unified field theory of mesons and baryons,
    \href{https://doi.org/10.1016/0029-5582(62)90775-7}{Nucl. Phys. \textbf{31}, 556 (1962).}

\bibitem{NagaosaTokura2013}
    N. Nagaosa and Y. Tokura, 
    Topological properties and dynamics of magnetic skyrmions,
    \href{https://doi.org/10.1038/nnano.2013.243}{Nature Nanotech. \textbf{8}, 899 (2013).}

\bibitem{FertRC2017}
    A. Fert, N. Reyren, and V. Cros,
    Magnetic skyrmions: Advances in physics and potential applications,
    \href{https://doi.org/10.1038/natrevmats.2017.31}{Nat. Rev. Mater. \textbf{2}, 17031 (2017).}		

\bibitem{HellmanHTetal2017}
    F. Hellman, A. Hoffmann, Y. Tserkovnyak, G. S. D. Beach, E. E. Fullerton, C. Leighton, \textit{et al.}
    Interface-Induced Phenomena in Magnetism,
    \href{https://doi.org/10.1103/RevModPhys.89.025006}{Rev. Mod. Phys. \textbf{89}, 025006 (2017).}

\bibitem{Zhou2018}
    Y. Zhou, 
    Magnetic skyrmions: Intriguing physics and new spintronic device concepts,
    \href{https://doi.org/10.1093/nsr/nwy109}{Nat. Sci. Rev. \textbf{6}, 210 (2018).}

\bibitem{BogdanovPanagopoulos2020}
    A. N. Bogdanov and C. Panagopoulos, 
    Physical foundations and basic properties of magnetic skyrmions,
    \href{https://doi.org/10.1038/s42254-020-0203-7}{Nat. Rev. Phys. \textbf{2}, 492 (2020).}

\bibitem{GobelMT2021}
    B. G\"obel, I. Mertig, and O. A. Tretiakov,
    Beyond skyrmions: Review and perspectives of alternative magnetic quasiparticles,
    \href{https://doi.org/10.1016/j.physrep.2020.10.001}{Phys. Rep. \textbf{895}, 1 (2021).}

\bibitem{MuhlbauerBJetal2009}
    S. M\"uhlbauer, B. Binz, F. Jonietz, C. Pfleiderer, A. Rosch, A. Neubauer, R. Georgii,   and P. B\"oni,
    Skyrmion lattice in a chiral magnet,
    \href{https://doi.org/10.1126/science.1166767}{Science \textbf{323}, 915 (2009).}

\bibitem{YuOKetal2010}
    X. Z. Yu, Y. Onose, N. Kanazawa, J. H. Park, J. H. Han, Y. Matsui, N. Nagaosa, and Y. Tokura,
    Real-space observation of a two-dimensional skyrmion crystal,
    \href{https://doi.org/10.1038/nature09124}{Nature \textbf{465}, 901 (2010).}

\bibitem{SekiYIT2012}
    S. Seki, X. Z. Yu, S. Ishiwata,  and Y. Tokura, 
    Observation of skyrmions in a multiferroic material,
    \href{https://doi.org/10.1126/science.1214143}{Science \textbf{336}, 198 (2012).} 

\bibitem{NayakKMetal2017}
    A. K. Nayak, V. Kumar, T. Ma, P. Werner, E. Pippel, R. Sahoo, F. Damay, U. K. Rößler, C. Felser,  and S. S. P. Parkin, 
    Magnetic antiskyrmions aboveroom temperature in tetragonal Heusler materials,
    \href{http://dx.doi.org/10.1038/nature23466}{Nature \textbf{548}, 561 (2017).}

\bibitem{FujishiroKNetal2019}
    Y. Fujishiro, N. Kanazawa, T. Nakajima, X. Z. Yu, K. Ohishi, Y. Kawamura, K. Kakurai,
    Topological transitions among skyrmion- and hedgehog-lattice states in cubic chiral magnets,
    \href{https://doi.org/10.1038/s41467-019-08985-6}{Nat. Commun. \textbf{10}, 1059 (2019).}

\bibitem{HeinzeBMetal2011}
    S. Heinze, K. von Bergmann, M. Menzel, J. Brede, A. Kubetzka, R. Wiesendanger, G. Bihlmayer, and S. Bl{\"u}gel,
    Spontaneous atomic-scale magnetic skyrmion lattice in two dimensions,
    \href{https://doi.org/10.1038/nphys2045}{Nat. Phys. \textbf{7}, 713 (2011).}

\bibitem{RommingHMetal2013}
    N. Romming, C. Hanneken, M. Menzel, J. E. Bickel, B. Wolter, K. vonBergmann,  A. \'e Kubetzka, and  R.  Wiesendanger,
    Writing  and  deleting  single  magneticskyrmions,
    \href{https://dx.doi.org/10.1126/science.aau0968}{Science \textbf{341}, 636 (2013).}

\bibitem{TokunagaYRetal2015}
    Y. Tokunaga, X. Z. Yu, J. S. White, H. M. R$\phi$nnow, D. Morikawa, Y. Taguchi, and Y. Tokura,
    A new class of chiral materials hosting magnetic skyrmions beyond room temperature,
    \href{https://doi.org/10.1038/ncomms8638}{Nat. Commun. \textbf{6}, 7638 (2015).}

\bibitem{MoreauMRetal2016}
    C. Moreau-Luchaire, C. Moutafis, N. Reyren, J. Sampaio, C. A. F. Vaz, N. Van Horne, K. Bouzehouane, K. Garcia, C. Deranlot, and P. Warnicke et al.,
    Additive interfacial chiral interaction in multilayers for stabilization of small individual skyrmions at room temperature, 
    \href{https://dx.doi.org/10.1038/nnano.2015.313}{Nat. Nanotechnol. \textbf{11}, 444 (2016).}

\bibitem{KurumajiNHetal2019}    
    T. Kurumaji, T. Nakajima, M. Hirschberger, A. Kikkawa, Y. Yamasaki, H. Sagayama, H. Nakao, Y. Taguchi, T. Arima, and Y. Tokura, 
    Skyrmion lattice with a giant topological Hall effect in a frustrated triangular-lattice magnet, 
    \href{https://doi.org/10.1126/science.aau0968}{Science \textbf{365}, 914 (2019).}

\bibitem{HirschbergerNGetal2019}
    M. Hirschberger, T. Nakajima, S. Gao, L. Peng, A. Kikkawa, T. Kurumaji, M. Kriener, Y. Yamasaki, H. Sagayama, H. Nakao, K. Ohishi, K. Kakurai, Y. Taguchi, X. Yu, T. Arima, and Y. Tokura, 
    Skyrmion phaseand competing magnetic orders on a breathing kagom\'e lattice,	
    \href{https://dx.doi.org/10.1038/s41467-019-13675-4}{Nat. Commun. \textbf{10}, 5831 (2019).}

\bibitem{KhanhNYetal2020}
    N. D. Khanh, T. Nakajima, X. Yu, S. Gao, K. Shibata, M. Hirschberger, Y. Yamasaki, H. Sagayama, H. Nakao, L. Peng, K. Nakajima, R. Takagi,
    T. Arima, Y. Tokura, and S. Seki, 
    Nanometric square skyrmion lattice in acentrosymmetric tetragonal magnet,
    \href{https://dx.doi.org/10.1038/s41565-020-0684-7}{Nat. Nanotechnol. \textbf{16}, 444 (2020).}

\bibitem{LinSB2015}
    S. Z. Lin, A. Saxena, and C. D. Batista,
    Skyrmion fractionalization and merons in chiral magnets with easy-plane anisotropy,
    \href{https://doi.org/10.1103/PhysRevB.91.224407}{Phys. Rev. B \textbf{91}, 224407 (2015).}

\bibitem{YuKTetal2018}
    X. Z. Yu, W. Koshibae, Y. Tokunaga, K. Shibata, Y. Taguchi, N. Nagaosa, and Y. Tokura,
    Transformation between meron and skyrmion topological spin textures in a chiral magnet,
    \href{https://doi.org/10.1038/s41586-018-0745-3}{Nature (London) \textbf{564}, 95 (2018).}

\bibitem{GaoJIetal2019}
    N. Gao, S. G. Je, M. Y. Im, J. W. Choi, M. Yang, Q. Li, T. Y. Wang, S. Lee, H. S. Han, K. S. Lee, W. Chao, C. Hwang, J. Li, and Z. Q. Qiu,
    Creation and annihilation of topological meron pairs in in-plane magnetized films,
    \href{https://doi.org/10.1038/s41467-019-13642-z}{Nat. Commun. \textbf{10}, 5603 (2019).}

\bibitem{GaoRGetal2020}
    S. Gao, H. D. Rosales, F. A. G{\'o}mez Albarrac{\'i}n, V. Tsurkan, G. Kaur, T. Fennell, P. Steffens, M. Boehm, P. $\rm{\check{C}}$erm{\'a}k, A. Scheidewind, E. Ressouche, D. C. Cabra, C. R{\"uegg}, and O. Zaharko,
    Fractional antiferromagnetic skyrmion lattice induced by anisotropic couplings,
    \href{https://doi.org/10.1038/s41586-020-2716-8}{Nature (London) \textbf{586}, 37 (2020).}

\bibitem{Kitaev2006}
    A. Kitaev,
    Anyons in an exactly solved model and beyond,
    \href{http://dx.doi.org/10.1016/j.aop.2005.10.005}{Ann. Phys. \textbf{321}, 2 (2006).}

\bibitem{JackeliKhaliullin2009}
    G. Jackeli and G. Khaliullin, 
    Mott insulators in the strong spin-orbit coupling limit: From Heisenberg to a quantum compass and Kitaev models,
    \href{https://doi.org/10.1103/PhysRevLett.102.017205}{Phys. Rev. Lett. \textbf{102}, 017205 (2009).}

\bibitem{SinghMRetal2012}
    Y. Singh, S. Manni, J. Reuther, T. Berlijn, R. Thomale, W. Ku, S. Trebst, and P. Gegenwart,
    Relevance of the Heisenberg-Kitaev Model for the Honeycomb Lattice Iridates $\rm A_2IrO_3$,
    \href{https://doi.org/10.1103/PhysRevLett.108.127203}{Phys. Rev. Lett. \textbf{108}, 127203 (2012).}

\bibitem{PlumbCSetal2014}
    K. W. Plumb, J. P. Clancy, L. J. Sandilands, V. V. Shankar, Y. F. Hu, K. S. Burch, H.-Y. Kee, and Y.-J Kim, 
    $\rm \alpha{-RuCl_3}$ : A spin-orbit assisted Mott insulator on a honeycomb lattice,
    \href{https://doi.org/10.1103/PhysRevB.90.041112}{Phys. Rev. B \textbf{90}, 041112(R) (2014).}

\bibitem{RauLK2014}
    J. G. Rau, E. K. Lee, and H. Y. Kee,	
    Generic Spin Model for the Honeycomb Iridates beyond the Kitaev Limit,
    \href{https://doi.org/10.1103/PhysRevLett.112.077204}{Phys. Rev. Lett. \textbf{112}, 077204 (2014).}

\bibitem{KatukuriNYetal2014}
    V. M. Katukuri, S. Nishimoto, V. Yushankhai, A. Stoyanova, H. Kandpal, S. Choi, R. Coldea, I. Rousochatzakis, L. Hozoi, and J. van den Brink,	   Kitaev interactions between j = 1/2 moments in honeycomb $\rm Na_2IrO_3$ are large and ferromagnetic: insights from ab initio quantum chemistry calculations,
    \href{https://doi.org/10.1088/1367-2630/16/1/013056}{New J. Phys. \textbf{16}, 013056 (2014).}

\bibitem{WangDYetal2017}
    W. Wang, Z.-Y. Dong, S.-L. Yu, and J.-X. Li, 
    Theoretical investigation of magnetic dynamics in $\rm \alpha{-RuCl_3}$,
    \href{https://doi.org/10.1103/PhysRevB.96.115103}{Phys. Rev. B \textbf{96}, 115103 (2017).}

\bibitem{MaksimovChernyshev2020}
    P. A. Maksimov and A. L. Chernyshev, 
    Rethinking $\rm \alpha{-RuCl_3}$,
    \href{https://doi.org/10.1103/PhysRevResearch.2.033011}{Phys. Rev. Res. \textbf{2}, 033011 (2020).}

\bibitem{XuFXetal2018}
    C. Xu, J. Feng, H. Xiang, and L. Bellaiche, 
    Interplay between Kitaev interaction and single ion anisotropy in ferromagnetic $\rm CrI_3$ and $\rm CrGeTe_3$ monolayers,
    \href{https://doi.org/10.1038/s41524-018-0115-6}{npj Comput. Mater. \textbf{4}, 57 (2018).}

\bibitem{LeeUWetal2020}
    I. Lee, F. G. Utermohlen, D. Weber, K. Hwang, C. Zhang, J. van Tol, J. E. Goldberger, N. Trivedi, and P. C. Hammel,
    Fundamental spin interactions underlying the magnetic anisotropy in the Kitaev ferromagnet $\rm CrI_3$,
    \href{https://doi.org/10.1103/PhysRevLett.124.017201}{Phys. Rev. Lett. \textbf{124}, 017201 (2020).}

\bibitem{StavropoulosPLetal2021}
    P. P. Stavropoulos, X. Liu, and H. Y. Kee,
    Magnetic anisotropy in spin-3/2 with heavy ligand in honeycomb Mott insulators: Application to $\rm CrI_3$,
    \href{https://doi.org/10.1103/PhysRevResearch.3.013216}{Phys. Rev. Research \textbf{3}, 013216 (2021).}

\bibitem{ZhouCLetal2021}
    Z. Zhou, K. Chen, Q. Luo, H.-G. Luo, and J. Zhao,
    Strain-induced phase diagram of the $S=\frac{3}{2}$ Kitaev material CrSiTe$_3$,
    \href{https://doi.org/10.1103/PhysRevB.104.214425}{Phys. Rev. B \textbf{104}, 214425 (2021).}

\bibitem{OkuboCK2012}
    Tsuyoshi Okubo, Sungki Chung, and Hikaru Kawamura,
    Multiple-$q$ States and the Skyrmion Lattice of the Triangular-Lattice Heisenberg Antiferromagnet under Magnetic Fields,
    \href{https://doi.org/10.1103/PhysRevLett.108.017206}{Phys. Rev. Lett. \textbf{108}, 017206 (2012). }

\bibitem{LeonovMostovoy2015}
    A. Leonov and M. Mostovoy, 
    Multiply periodic states and isolated skyrmions in an anisotropic frustrated magnet. 
    \href{https://doi.org/10.1038/ncomms9275}{Nat. Commun. \textbf{6}, 8275 (2015).}

\bibitem{HayamiOM2017}
    Satoru Hayami, Ryo Ozawa, and Yukitoshi Motome,	
    Effective bilinear-biquadratic model for noncoplanar ordering in itinerant magnets,
    \href{https://doi.org/10.1103/PhysRevB.95.224424}{Phys. Rev. B \textbf{95}, 224424 (2017).}

\bibitem{KharkovSM2017}
    Y. A. Kharkov, O. P. Sushkov, and M. Mostovoy,
    Bound States of Skyrmions and Merons near the Lifshitz Point,
    \href{https://doi.org/10.1103/PhysRevLett.119.207201}{Phys. Rev. Lett. \textbf{119}, 207201 (2017).}

\bibitem{WangSLetal2021}
    Zhentao Wang, Ying Su, Shi-Zeng Lin, and Cristian D. Batista,
    Meron, skyrmion, and vortex crystals in centrosymmetric tetragonal magnets,
    \href{https://doi.org/10.1103/PhysRevB.103.104408}{Phys. Rev. B \textbf{103}, 104408 (2021).}

\bibitem{ChernKLetal2020}    
    L. E. Chern, R. Kaneko, H.-Y. Lee, and Y. B. Kim, 
    Magnetic field induced competing phases in spin-orbital entangled Kitaev magnets, 
    \href{https://doi.org/10.1103/PhysRevResearch.2.013014}{Phys. Rev. Res. \textbf{2}, 013014 (2020).}    

\bibitem{LiuSRetal2020}
    K. Liu, N. Sadoune, N. Rao, J. Greitemann, and L. Pollet, 
    Revealing the phase diagram of Kitaev materials by machine learning: Cooperation and competition between spin liquids,
    \href{https://doi.org/10.1103/PhysRevResearch.3.023016}{Phys. Rev. Res. \textbf{3}, 023016 (2021).}    

\bibitem{RayyanLK2021}
    A. Rayyan, Q. Luo, and H. Y. Kee,
    Extent of frustration in the classical Kitaev-$\Gamma$ model via bond anisotropy,
    \href{https://doi.org/10.1103/PhysRevB.104.094431}{Phys. Rev. B \textbf{104}, 094431 (2021).}

\bibitem{Hamamoto2015}
    K. Hamamoto, M. Ezawa, and N. Nagaosa,
    Quantized topological Hall effect in skyrmion crystal,
    \href{https://dx.doi.org/10.1103/PhysRevB.92.115417}{Phys. Rev. B \textbf{92}, 115417 (2015).}

\bibitem{HayamiOM2021}
    S. Hayami, T. Okubo, and Y. Motome,
    Phase shift in skyrmion crystals,
    \href{https://doi.org/10.1038/s41467-021-27083-0}{Nat. Commun. \textbf{12}, 6927 (2021).}

\bibitem{HuangCNetal2017}
    B. Huang, G. Clark, E. Navarro-Moratalla, D. R. Klein, R. Cheng, K. L. Seyler, D. Zhong, E. Schmidgall, M. A. McGuire, D. H. Cobden, W. Yao, D. Xiao, P. Jarillo-Herrero, and X. Xu,
    Layer-dependent ferromagnetism in a Van der Waals crystal down to the monolayer limit,
    \href{https://doi.org/10.1038/nature22391}{Nature~(London) \textbf{546}, 270 (2017).}

\bibitem{GongLLetal2017}
    C. Gong, L. Li, Z. Li, H. Ji, A. Stern, Y. Xia, T. Cao, W. Bao, C. Wang, Y. Wang, Z. Q. Qiu, R. J. Cava, S. G. Louie, J. Xia, and X. Zhang,
    Discovery of intrinsic ferromagnetism in two-dimensional Van der Waals crystals,
    \href{https://doi.org/10.1038/nature22060}{Nature~(London) \textbf{546}, 265 (2017).}

\bibitem{XingCOetal2017}
    W. Xing, Y. Chen, P. M. Odenthal, X. Zhang, W. Yuan, T. Su, Q. Song, T. Wang, J. Zhong, S. Jia, X. C. Xie, Y. Li, and W. Han,
    Electric field effect in multilayer $\rm Cr_2Ge_2Te_6$: A ferromagnetic 2D material,
    \href{https://doi.org/10.1088/2053-1583/aa7034}{2D Mater. \textbf{4}, 024009 (2017).}


\bibitem{XuFKetal2020}
    C. Xu, J. Feng, M. Kawamura, Y. Yamaji, Y. Nahas, S. Prokhorenko, Y. Qi, H. Xiang, and L. Bellaiche,
    Possible Kitaev quantum spin liquid state in 2D material with $S$ = 3/2,
    \href{https://doi.org/10.1103/PhysRevLett.124.087205}{Phys. Rev. Lett. \textbf{124}, 087205 (2020).}

\bibitem{HukushimaNemoto1996}
    K. Hukushima and K. Nemoto,
    Exchange Monte Carlo method and application to spin glass simulations,
    \href{https://doi.org/10.1143/JPSJ.65.1604}{J. Phys. Soc. Jpn. \textbf{65}, 1604 (1996).}

\bibitem{MetropolisRRetal1953}
    N. Metropolis, A. W. Rosenbluth, M. N. Rosenbluth, and A. H. Teller,
    Equation of state calculations by fast computing machines,
    \href{https://doi.org/10.1063/1.1699114}{J. Chem. Phys. \textbf{21}, 1087 (1953).}

\bibitem{MiyatakeYKetal1986}
    Y. Miyatake, M. Yamamoto, J. J. Kim, M. Toyonaga, and and O. Nagai,
    On the implementation of the 'heat bath' algorithms for Monte Carlo simulations of classical Heisenberg spin systems,
    \href{http://iopscience.iop.org/0022-3719/19/14/020}{J. Phys. C: Solid State Phys. \textbf{19}, 2539 (1986).}

\bibitem{JanssenAV2016}
    L. Janssen, E. C. Andrade, and M. Vojta,
    Honeycomb-Lattice Heisenberg-Kitaev Model in a Magnetic Field: Spin Canting, Metamagnetism, and Vortex Crystals,
    \href{https://doi.org/10.1103/PhysRevLett.117.277202}{Phys. Rev. Lett. \textbf{117}, 277202 (2016).}

\bibitem{Supp}
	See Supplemental Material at http://xxx, which includes Refs. \cite{Landau1935}, \cite{Gilbert2004},
    for the numerical methods of optimzating the ground-state energy~(Sec. I), determing the magnetic unit cell~(Sec. II), 
    the origin of the TmX phase~(Sec. III), the topological charges and the topological Hall effect in the TmX~(Sec. IV), atomistic
    spin-dynamics simulations~(Sec. V) as well as the complete phase diagram of the $\rm K\Gamma{A}$ model~(Sec. VI).

\bibitem{BergLuscher1981}
    B. Berg and M. L\"uscher,
    Definition and statistical distributions of a topological number in the lattice O(3) $\sigma$-model,
    \href{https://doi.org/10.1016/0550-3213(81)90568-X}{Nucl. Phys. B \textbf{190}, 412 (1981).}

\bibitem{ZhangZhouEzawa2016}
    X. Zhang, Y. Zhou, and M. Ezawa,
    High-topological-number magnetic skyrmions and topologically protected dissipative structure,
    \href{https://doi.org/10.1103/PhysRevB.93.024415}{Phys. Rev. B \textbf{93}, 024415 (2016).}

\bibitem{ZhouEzawa2014}
    Y. Zhou and M. Ezawa,
    A reversible conversion between a skyrmion and a domain-wall pair in a junction geometry,
    \href{https://doi.org/10.1038/ncomms5652}{Nat. Commun. \textbf{5}, 4652 (2014).}	

\bibitem{AugustinJEetal2021}
    M. Augustin, S. Jenkins, R. F. L. Evans, K. S. Novoselov and E. J. G. Santos, 
    Properties and dynamics of meron topological spin textures in the two-dimensional magnet CrCl$_{\rm{3}}$, 
    \href{https://doi.org/10.1038/s41467-020-20497-2}{Nat. Commun. \textbf{12}, 185 (2021).}

\bibitem{HayamiYambe2021}
    Satoru Hayami and Ryota Yambe,
    Meron-antimeron crystals in noncentrosymmetric itinerant magnets on a triangular lattice,
    \href{https://doi.org/10.1103/PhysRevB.104.094425}{Phys. Rev. B \textbf{104}, 094425 (2021).}

\bibitem{RousochatzakisPerkins2017}
    Ioannis Rousochatzakis and Natalia B. Perkins,
    Classical Spin Liquid Instability Driven By Off-Diagonal Exchange in Strong Spin-Orbit Magnets,
    \href{https://doi.org/10.1103/PhysRevLett.118.147204}{Phys. Rev. Lett. \textbf{118}, 147204 (2017).}

\bibitem{Actor1979}
    A. Actor,
    Classical solutions of SU(2) Yang-Mills theories,
    \href{https://doi.org/10.1103/RevModPhys.51.461}{Rev. Mod. Phys. \textbf{51}, 461 (1979).}

\bibitem{Ezawa2011}
    Motohiko Ezawa,
    Compact merons and skyrmions in thin chiral magnetic films,
    \href{https://doi.org/10.1103/PhysRevB.83.100408}{Phys. Rev. B 83, 100408(R) (2011).}

\bibitem{PhatakLH2012}
    C. Phatak, A. K. Petford-Long, and O. Heinonen,
    Direct Observation of Unconventional Topological Spin Structure in Coupled Magnetic Discs,
    \href{https://doi.org/10.1103/PhysRevLett.108.067205}{Phys. Rev. Lett. \textbf{108}, 067205 (2012).}

\bibitem{GuoXGetal2020}
    Cheng Guo, Meng Xiao, Yu Guo, Luqi Yuan, and Shanhui Fan,
    Meron spin textures in momentum space,
    \href{https://doi.org/10.1103/PhysRevLett.124.106103}{Phys. Rev. Lett. \textbf{124}, 106103 (2020).}

\bibitem{SenthilVBetal2004}
    T. Senthil, A. Vishwanath, L. Balents, S. Sachdev, and M. P. A. Fisher,
    Deconfined quantum critical points,
    \href{https://doi.org/10.1126/science.1091806}{Science \textbf{303}, 1490 (2004).}

\bibitem{Landau1935}
    L. D. Landau and E. Lifschitz,
    On the theory of magnetic permeability in ferromagnetic bodies,
    Phys. Z. Sowjetunion \textbf{8}, 153 (1935).

\bibitem{Gilbert2004}
   T. L. Gilbert,
   A phenomenological theory of damping in ferromagnetic materials,
   \href{http://dx.doi.org/10.1109/TMAG.2004.836740}{IEEE Trans. Magn. \textbf{40}, 3443 (2004).}

%%%%%%%%%%%%%%%%%%%%%%%%%%%%%%%%%%%%%%%%%%%%%%%%%%%%%%%%%%%%%%%%%%%%%%%%%%%%%%%%%%%%%%%%%%%%%%%%%%%%%%%%%%%%%%%%%%%%%%%%%%%%%%%%%%%%%%%%%%%%%%%%
\end{thebibliography}

\begin{thebibliography}{99}

\bibitem{RousochatzakisPerkins2017}
    Ioannis Rousochatzakis and Natalia B. Perkins,
    Classical Spin Liquid Instability Driven By Off-Diagonal Exchange in Strong Spin-Orbit Magnets,
    \href{https://doi.org/10.1103/PhysRevLett.118.147204}{Phys. Rev. Lett. \textbf{118}, 147204 (2017).}

\bibitem{BergLuscher1981}
    B. Berg and M. L\"uscher,
    Definition and statistical distributions of a topological number in the lattice O(3) $\sigma$-model.
    \href{https://doi.org/10.1016/0550-3213(81)90568-X}{Nucl. Phys. B \textbf{190}, 412-424 (1981).}

\bibitem{HayamiOM2021}
    S. Hayami, T. Okubo, and Y. Motome,
    Phase shift in skyrmion crystals,
    \href{https://doi.org/10.1038/s41467-021-27083-0}{Nat. Commun. \textbf{12}, 6927 (2021).}

\bibitem{Hamamoto2015}    
K. Hamamoto, M. Ezawa, and N. Nagaosa,
Quantized topological Hall effect in skyrmion crystal,
\href{https://dx.doi.org/10.1103/PhysRevB.92.115417}{Phys. Rev. B \textbf{92}, 115417 (2015).}


\bibitem{Landau1935}
    L. D. Landau and E. Lifschitz,
    On the theory of magnetic permeability in ferromagnetic bodies.
    Phys. Z. Sowjetunion \textbf{8}, 153 (1935).

\bibitem{Gilbert2004}
   T. L. Gilbert,
   A phenomenological theory of damping in ferromagnetic materials.
   \href{http://dx.doi.org/10.1109/TMAG.2004.836740}{IEEE Trans. Magn. \textbf{40}, 3443 (2004).}

\end{thebibliography}
\end{document}